\documentclass{aa}
\usepackage{psfig,times}
\newcommand{\PutWin}[4]{
\put(#1,#2){\parbox{#3}{#4}}}
\usepackage{graphicx}

\newcommand\ra[4]{~~#1$^{\rm h}$#2$^{\rm m}$#3$^{\rm s}$.#4 }
\newcommand\dec[4] {$#1$$^{\circ}$#2$^{\rm '}$#3$^{\rm ''}$.#4}

\newcommand{\sbb}{mag/$\sq\arcsec$}
\def\ha{H$\alpha$}
\def\ergsec{${\rm erg\,\,s^{-1}}$}
\def\h2{H{\small II}}
\newcounter{qub}
\def\sbu{mag/arcsec$^2$}

\def\p23{{\it P$^{\tiny \cal J}_{23}$}}
\def\e25{{\it E$^{\cal J}_{23}$}}

\def\ew{{\sl EW}}
\def\msun{$M_{\odot}$}
\def\zsun{$Z_{\odot}$}

\def\ha{H$\alpha$}
\def\cg{mag kpc$^{-1}$}
\def\hst{\sl HST\rm }
\def\bg{$^{\ddag}$}
\def\rr{$R^*$}
\def\rt{$R^*_{\rm T}$}
\def\kmsec{${\rm km\,\,s^{-1}}$}
\def\cgg{mag kpc$^{-1}$}
\def\flat{type\,V}
\def\med{{\sf med}}
\def\ser{S\'ersic }
\def\zsun{$Z_{\odot}$}

\begin{document}

\title{New insights to the photometric structure of Blue Compact Dwarf 
Galaxies from deep Near-Infrared studies}
\subtitle{I. Observations, surface photometry and decomposition of surface brightness profiles}

\author{K.G. Noeske\inst{ }
\and P. Papaderos\inst{ }
\and L.M. Cair\'os \inst{ }
\and K.J. Fricke \inst{ }}
\offprints{knoeske@uni-sw.gwdg.de}
\institute{      Universit\"ats--Sternwarte, Geismarlandstra\ss e 11,
                 D--37083 G\"ottingen, Germany
}
\date{Received \hskip 2cm; Accepted}

\abstract{ We have analyzed deep Near Infrared (NIR) broad band images
for a sample of Blue Compact Dwarf Galaxies (BCDs), observed with the 
ESO NTT\thanks{European Southern Observatory, program ID 65.N-0318(A)} 
and Calar Alto\thanks{German--Spanish Astronomical Center, Calar Alto,
operated by the Max--Planck--Institute for Astronomy, Heidelberg,
jointly with the Spanish National Commission for Astronomy.} 3.6m
telescopes.
The data presented here allows for the detection and quantitative
study of the extended stellar low-surface brightness (LSB) host 
galaxy in all sample BCDs.
NIR surface brightness profiles (SBPs) of the LSB host galaxies 
agree at large galactocentric radii with those from optical studies, 
showing also an exponential intensity decrease and compatible scale lengths.  
At small to intermediate radii (within 1--3 exponential scale lengths), however,  
the NIR data reveals for more than one half of our sample BCDs evidence for 
a significant flattening of the exponential profile of the LSB component.
Such profiles (\flat\ SBPs, Binggeli \& Cameron 1991) have rarely been 
detected in the LSB component of BCDs at optical wavelengths, where 
the relative flux contribution of the starburst, being stronger than in 
the NIR, can readily hide a possible central intensity depression in
the underlying LSB host.
The structural properties, frequency and physical origin of \flat\ LSB profiles in 
BCDs and dwarf galaxies in general have not yet been subject 
to systematic studies. 
Nevertheless, the occurrence of such profiles 
in an appreciable fraction of BCDs would impose important 
new observational constraints to the radial mass distribution 
of the stellar LSB component, as well as to the photometric 
fading of these systems after the termination of star-forming 
activities.
We test the suitability of two empirical fitting functions, 
a modified exponential distribution (Papaderos et al. 1996a)
and the S\'ersic law, for the systematization of the structural 
properties of BCD host galaxies which show a \flat\ intensity 
distribution. Either function has been found to satisfactorily 
fit a \flat\ distribution. 
However, it is argued that the practical applicability 
of \ser\ fits to the LSB emission of BCDs is limited by the 
extreme sensitivity of the achieved solutions to, 
e.g., small uncertainties in the sky subtraction 
and SBP derivation.
We find that most of the sample BCDs show in their stellar LSB host galaxy optical-NIR
colors indicative of an evolved stellar population with subsolar metallicity.
Unsharp-masked NIR maps reveal numerous morphological details and
indicate in some cases, in combination with optical data,
appreciable non-uniform dust absorption 
on a spatial scale as large as $\sim$1 kpc.}

\authorrunning{Noeske et al.}
\titlerunning{NIR observations of Blue Compact Dwarf Galaxies}

\maketitle
\keywords{galaxies: dwarf --- galaxies: evolution --- galaxies: structure --- 
galaxies: starburst}

\section{Introduction}                                                
\label{introduction}
The star-formation history and chemodynamic evolution of 
Blue Compact Dwarf (BCD) galaxies are central issues in 
the contemporary dwarf galaxy research. In spite of being old 
in their vast majority, BCDs resemble in many aspects unevolved 
low-mass galaxies in the early Universe. 
They are gas--rich (\ion{H}{i} mass fraction of typically $>$30\%) and
metal--deficient (\zsun/50$\la Z\la$\zsun/3) extragalactic systems,
undergoing intense, spatially extended star-forming (SF) activity.
Such properties are believed to have been common among young low-mass
objects at high to intermediate redshift, such as pre--galactic
building blocks (Lowenthal et al. \cite{lowenthal97}, Hirashita et
al. \cite{hirashita00}, Fujita et al. \cite{fujita01}) or the
progenitors of the present-day dwarf spheroidals (e.g. Babul \& Rees
\cite{babul92}; Guzm\'an et al. \cite{guzman98}).  BCDs are therefore
convenient nearby laboratories to study at high spatial resolution the
impact of collective star formation on the spectrophotometric and chemodynamic
properties of these distant and faint extragalactic sources. Moreover,
they are important testbeds for deducing constraints to cosmological parameters,
such as the primordial $^4$He abundance ratio, and to monitor the
synthesis and dispersal of heavy elements in a nearly pristine
environment (Peimbert \& Torres--Peimbert \cite{peimbert74}, 
Pagel et al. \cite{pagel92}, Izotov et al. \cite{izotov97}).

The understanding of the origin and implications of the starburst
phenomenon in BCDs is necessary for elucidating evolutionary pathways
of dwarf galaxies (DGs) in general. Are BCDs active phases in the
lifetime of dormant dwarf irregulars (dIs) and do the latter fade to
dwarf ellipticals (dEs) once their gas reservoir has been depleted
(see e.g. Lin \& Faber \cite{lin83}, Thuan \cite{thuan85}, Silk et
al. \cite{silk87}, Davies \& Phillipps \cite{davies88})?  What is the
role of the environment (e.g., Babul \& Rees \cite{babul92},
Pustil'nik et al. \cite{pustilnik01}) and of Dark Matter (DM; Dekel \&
Silk \cite{dekel86}, Ferrara \& Tolstoy \cite{ferrara00}), and does
the latter invariably dominate the mass \emph{within} the Holmberg
radius of a BCD (Papaderos et al. 1996b; hereafter \cite{papaderos96b})?
Do most BCDs undergo intermittent bursts or rather prolonged periods 
of elevated star formation (Vallenari \& Bomans \cite{vallenari96},
Noguchi \cite{noguchi01}, Rieschick \& Hensler \cite{rieschick01}, 
Schulte-Ladbeck et al. \cite{schulte01})?
Despite much previous effort, the observational evidence available 
thus far is still too fragmentary to allow for unambiguous answers to 
the aforementioned questions.

Recent studies suggest, however, that key information for assessing DG
evolution can be inferred from studies of the stellar low-surface
brightness (LSB) host galaxy of these systems.  In BCDs, the LSB
component, underlying the SF regions, has first been disclosed through
deep CCD imaging by Loose \& Thuan (1986, hereafter \cite{loose86}),
and has in the following been confirmed and further studied by various
authors (cf., e.g., the list given later in this Section).  This
extended stellar host was found to account for $\sim$1/2 of the light
inside the 25 $B$ \sbb\ isophote (\cite{papaderos96b}, Salzer \&
Norton \cite{salzer99}), and 
to typically dominate 
the intensity and color distribution of BCDs for $\mu\ga$24.5 $B$ \sbb. Such an
evenly distributed, evolved stellar component is observed in all
types of DGs, except for the extremely rare type of i0 BCDs in the
classification scheme of \cite{loose86}.  Its red colors and smooth
appearance in the main class of iE/nE BCDs (\cite{loose86}) indicate
that these systems are several Gyr old, gas-rich DGs, having not been
forming stars at the presently large rate throughout their lifetime.
Different lines of evidence, outlined in the following, suggest that 
elaborate studies of the structural and kinematic properties of the 
LSB component are fundamental to assess at least two
central issues of DG research: the evolutionary connections between
DGs and the regulation of the SF process in these systems.

According to the standard evolutionary hypothesis for dwarf galaxies,
dIs, dEs and BCDs differ basically by their gas content and the
amplitude of their ongoing SF activity (Thuan \cite{thuan85}, Davies \&
Phillipps \cite{davies88}). One would therefore expect that, on
average, the evolved stellar LSB host in all these three main DG
classes is indistinguishable from one another with respect to its
structural properties.  However, \cite{papaderos96b} and subsequent
authors (Patterson \& Thuan \cite{patterson96}, Marlowe et
al. \cite{marlowe97}, Salzer \& Norton \cite{salzer99}, see also
Papaderos et al. \cite{papaderos02}) found that, at equal absolute magnitude,
the stellar LSB component of iE/nE BCDs is systematically more compact 
than other types of DGs.
\cite{papaderos96b} interpreted this structural disparity as the
result of adiabatic contraction of the stellar LSB component of BCDs
in response to a large-scale gas inflow, preceding the ignition of a
starburst in a BCD. Quite interestingly, subsequent 
interferometric \ion{H}{i} studies have shown that the gaseous 
halo of BCDs is by at least a factor of two more centrally concentrated 
than in dIs (van Zee et al. \cite{vanzee98},\cite{vanzee01}; see also Simpson et
al. \cite{simpson00}), lending circumstantial support to the latter hypothesis.
Whether or not such observational constraints can be accounted for by the 
dynamical evolution of the stellar and gaseous matter in DGs (\cite{papaderos96b})
or, alternatively, an extraordinarily dense Dark Matter halo, being
particular to BCD galaxies (Meurer et al. \cite{Meurer98}) awaits to
be investigated by numerical 3D-simulations. These are also needed to
explore a possible connection between the LSB morphology and the 
evolutionary state of BCDs, as proposed by Noeske et al. (\cite{noeske00}).

The evolved LSB component contains, due to its high $M/L$, the bulk of
the stellar matter in a BCD. Thus, provided that DM does not dominate
within the optical radius, it forms, together with the gaseous
component, the graviational potential within which SF activity occurs.
It is therefore worth exploring whether the structural properties of
the LSB component influence the SF process in a BCD.  %
\cite{papaderos96b} found a trend for the fractional surface area of
the SF component of BCDs to decrease with increasing LSB
luminosity. For a constant $M/L$, this trend translates into a
mass--morphology connection for BCDs: SF activity in more massive BCDs
occurs mainly in the inner part of the LSB component, leading to a nE
morphology and surface brightness profiles (SBPs) sometimes
superficially resembling a de Vaucouleurs law. Conversely, SF activity
in low-mass BCDs is spread over a larger portion of the LSB host,
resulting in an iE morphology and SBPs possessing a conspicuous {\it
plateau} feature (cf. Papaderos et al. 1996a, hereafter
\cite{papaderos96a}) at intermediate
intensity levels.

Our current knowledge of the nature of the underlying
LSB component relies mainly on optical surface photometry studies
of its faint periphery
(e.g. \cite{loose86}, Kunth et
al. \cite{kunth88}, \cite{papaderos96a}, Telles et al.
\cite{telles97}, Marlowe et al.  \cite{marlowe97}, Doublier et
al. \cite{doublier97}, \cite{doublier99}, Salzer \& Norton \cite{salzer99}, 
Vennik et al. \cite{vennik00}, Cair\'os et al. \cite{cairos01a}, 
\cite{cairos01b}, Makarova et al. \cite{makarova02}).
In optical wavelengths, extended stellar and gaseous starburst
emission overshines the LSB component out to a galactocentric
radius of typically $\sim$2 exponential scale lengths.
Only at larger radii, where the starburst emission is in most cases 
negligible, the LSB host can be studied directly.
However, to explore its possible dynamical influence on the global SF
process, it is essential to pin down its intensity distribution at
smaller radii, if possible just beneath the SF regions. Deprojection
of SBPs would then allow one to put constraints on the density
profile and the gravitational potential of the evolved stellar
host (cf., e.g., \cite{papaderos96a}). In addition, one would be able
to correct optical colors inside the SF component
for the line-of-sight contribution of the underlying stellar background
(cf. Cair\'os et al. \cite{cairos02a}, Papaderos et
al. \cite{papaderos02}), thus better constrain the SF history of these systems.

One way to alleviate the problem of the extended starburst emission 
is to extend studies to the Near Infrared (NIR). 
At these wavelengths, the young stellar populations and the ionized gas contribute a smaller 
fraction of the total light of the galaxy than in the optical. 
For instance, evolutionary synthesis models by Kr\"uger et
al. (\cite{krueger95}) predict
that a moderately strong burst during its peak luminosity accounts 
for only $\sim$20\% of a BCD's emission in the $K$, but for
$\ga$80\% in the $B$ band. 
Observations of BCDs also show that starburst emission is weaker
in the NIR, and becomes negligible at a smaller galactocentric 
distance than in the optical
(e.g. Vanzi et al. \cite{vanzi96}, \cite{vanzi02}; Beck et
al. \cite{beck97}; Alton et al. \cite{alton94}; James \cite{james94}).
NIR data allow therefore to study
the stellar LSB component at smaller radii and, using optical--NIR colors 
(e.g. $B-J$), better constrain its formation history.
To achieve these objectives one needs to unambiguously detect the 
LSB component and study its intensity over a sufficient span 
($\Delta \mu \ga 2$ mag).
Empirical estimates, based on published optical and NIR data, suggest
that these requirements are met by extending NIR studies to
$\sim$22--24 $J$ \sbb.  So far, only a few BCDs have been studied at
those surface brightness levels with an accuracy high enough to
pinpoint structural properties and colors (Mkn 86, Gil de Paz et
al. \cite{gildepaz00a}; Tol 0645--376, Doublier et
al. \cite{doublier01}; Tol3, Vanzi et al. \cite{vanzi02}).

The present analysis is the first part of an observational project (see also Cair\'os et
al. 2003; hereafter \cite{c02b}), aiming at a systematic study of the NIR 
properties of nearby BCDs with large array detectors on 4m-class telescopes.
For this purpose, we take advantage of a large set of imaging 
data, homogeneous with respect to its limiting surface
brightness and the methods used in its processing. The observations
have been conducted so as to permit NIR surface photometry out to a
comparable radius as in optical wavelengths, with the purpose of a
multiwavelength investigation of the LSB component. 

This paper is structured as follows: In Sect. 2, we describe the
sample selection and data acquisition, and discuss the data reduction,
photometric transformations and extinction corrections.  Section 3
focusses on the derivation of SBPs and their decomposition into the
luminosity components owing to the old LSB host and the younger
stellar populations and SF regions. In Sect. 4, individual objects are
presented in detail. Our results are discussed in Sect. 5. Sect. 6
summarizes this work.

\section{Observations and data reduction 
\label{SectObs}}
\subsection{Sample selection}
\label{sample}
Our sample 
covers the whole morphological spectrum of BCDs, containing both
examples of the main morphological class (iE/nE systems according to
\cite{loose86}), and of the less frequent iI/iI,C/i0 BCD classes.  The
latter subset includes the metal--poor galaxies Tol\,65 and
Tol\,1214--277 ($Z <$\zsun /20). The sample also includes one 
intrinsically luminous
blue compact galaxy (BCG), UM\,448.

Table \ref{tab_sample} lists the adopted distance to each BCD.
Whenever available, literature distances based on standard candles
were preferred for nearby objects. Those literature distances that 
rely on redshifts have been corrected for peculiar motions within 
the local supercluster and assume a Hubble constant $H_0=$75\,km\,s$^{-1}$\,Mpc$^{-1}$.
When no literature data was available, the distance was inferred from 
the $H_0$ and the heliocentric velocity listed in the 
NASA Extragalactic Database (NED). 
The latter was first transformed to the velocity relative to the 
local group (LG) centroid using the NED velocity calculator, and 
subsequently corrected for a LG infall towards the Virgo Cluster 
center ($l=284\degr, b=74\degr$) at 200\,km\,s$^{-1}$ 
(Tammann \& Sandage \cite{tammann85}).

\subsection{Data acquisition}
\label{observations}
The NIR images were observed with the ESO 3.6m NTT telescope, La
Silla/Chile, during three consecutive nights from April 21st to
24th, 2000. The seeing ranged from 0\farcs 4 to 1\farcs 0 FWHM and 
the transparency variations were $\la$ 1\% during the first and third night, 
and $\la$ 2\% during the second night.
Additional NIR images were taken at the 3.6m telescope of the
German--Spanish Astronomical Center, Calar Alto, Spain, during several
observing runs. The seeing conditions were generally average (December
26th, 1999: FWHM 1\farcs 3 -- 3\farcs 5, transparency fair; May
10th--15th, 2000: FWHM 0\farcs 8 -- 1\farcs 4, transparency good to
average; October 6th--10th, 2000: FWHM 1\farcs 2 -- 2\farcs 0,
transparency good to fair).
Both cameras used, the SOFI at the NTT Nasmyth focus and the OMEGA
PRIME at the Calar Alto 3.6m telescope's prime focus, were equipped
with 1024$\times$1024 pixel Rockwell HAWAII detectors. The pixel
scales for SOFI, using the large field objective, and OMEGA PRIME were
0\farcs 292 and 0\farcs 396, yielding a FOV of 4\farcm 94 and 6\farcm
76, respectively.
The data was taken through the $J$ and $H$ broad band filters, as well
as the modified $K$ filters $K_s$ at the NTT and $K$\arcmin\ at Calar
Alto, both selected to extenuate the contribution of thermal background.

To achieve an adequate sampling of the background variations, the
telescope was offset between exposures typically each 60
seconds. While compact galaxies were jittered within the FOV, larger
objects required to alternately observe the object and the sky. To
avoid detector saturation, images were obtained by summing up
subexposures of few seconds each.
The total on-object exposure times, after rejection of 
subexposures affected by unstable readout electronics 
or strong background gradients, are listed in Table \ref{tab_sample}.
%
\begin{table*}
\caption{Sample galaxies }
\label{tab_sample}
\tabcolsep1.8mm
\begin{tabular}{llcccccccl}\hline\hline
Object       &   RA(J2000)      & t$_{J, \rm NTT}$ & t$_{H,\rm NTT}$ &t$_{K_s,\rm NTT}$& M$_B$& A$_B$&  $D$            &FWHM     & other \\
             &   DEC(J2000)     & t$_{J,\rm CA}$  & t$_{H,\rm CA}$  & t$_{K',\rm CA} $& (ref.) &    &   (ref.) &(final)$^{\spadesuit}$  & names \\
             &                  &  [s]        & [s]         & [s]         & [mag] & [mag]&   [Mpc]  &  [\arcsec]      &                  \\
             &   (2)            & (3)    & (4)    & (5)          & (6)  &   (7)     &    (8)         & (9)  &(10)        \\ \hline
Tol 3        &   \ra{10}{06}{33}{6}    & 1440   & 1560   & 2100  &--18.0& 0.33 &   13.8    &   1.2          & NGC\,3125; ESO\,435-G041;   \\
(iE)         &   \dec{-29}{56}{09}{0}     & ---    & ---    & ---& (a)  &      & (a)       &                & Tol1004-296          \\ \hline
Haro 14      &   \ra{00}{45}{46}{4}    & ---    & ---    & ---   &--17.0& 0.09 &   12.5    &   1.6          & NGC\,0244; UGCA\,010;       \\
(iE/nE)      &   \dec{-15}{35}{49}{0}   & 1440   & 1680   & 1920 & (a)  &      & (a)       &                & VV\,728               \\ \hline
UM 461       &   \ra{11}{51}{33}{0}    & 1500   & 1500   & 2100  &--14.9& 0.08 &   14.3    &   0.7          & SCHG\,1148-020            \\
(iI)         &   \dec{-02}{22}{23}{0}     & ---    & ---    & ---& (j)  &      & (e)       &                &                     \\ \hline 
He 2-10      &   \ra{08}{36}{15}{0}    & 1140   & 1500   & 2040  &--18.7& 0.48 &    8.7    &   0.9          & ESO\,495-G021            \\
(iE/nE)      &   \dec{-26}{24}{34}{0}     & ---    & ---    & ---& (k)  &      & (f)       &                                      \\ \hline
Tol 1400-411 &   \ra{14}{03}{21}{0}    & 1140   & 1500   & 2100  &--16.5& 0.30 &   4.8     &   0.9          & NGC\,5408; Tol\,116;         \\
(iI,C)       &   \dec{-41}{22}{44}{0}     & ---    & ---    & ---& (l)  &      & (c)       &                & ESO\,325-G047        \\ \hline
Pox 4/Pox 4B &   \ra{11}{51}{11}{6}    & 1500   & 1500   & 2100  &--18.8& 0.17 &   46.7    &   0.6         & CAM\,1148-2020             \\
(iI + comp.) &   \dec{-20}{36}{02}{0}     & ---    & ---    & ---& (m)  &      & (b)       &                &                    \\ \hline
Tol 65       &   \ra{12}{25}{46}{9}    & 2100   & 2100   & 3000  &--15.3& 0.32 &   34.2    &   0.6          & ESO\,380-G027;           \\
(i0/iI,C)    &   \dec{-36}{14}{01}{0}s  & ---    & ---    & ---  &(n)   &      & (b)       &                & Tol\,1223-359      \\ \hline
Tol 1214-277 &   \ra{12}{17}{17}{1}    & 1200   & 1500   & 2100  &--16.9& 0.28 &   102.6   &    0.9         & SCHG\,1214-277;          \\
(i0/iI,C)    &   \dec{-28}{02}{33}{0}     & ---    & ---    & ---&(o)   &      & (b)       &                & Tol\,21              \\ \hline
Mkn 178      &   \ra{11}{33}{29}{1}    & 240    & 1380   & 1380  &--13.9& 0.08 &    4.2    &   1.6          & UGC\,06541; HOLMBERG\,263A;\\
(iE)         &   \dec{+49}{14}{17}{0}     & ---    & ---    & ---&(p)   &      & (h)       &                & CGCG\,242-046; SBS\,1130+495 \\ \hline
Mkn 1329     &   \ra{12}{37}{03}{0}    & 300    & 360    & 480   &--16.8& 0.10 &   16.0    &   0.7          & VCC\,1699; IC\,3589/91;    \\
(iI,C)       &   \dec{+06}{55}{36}{0}     & 1200   & 360    & 960& (i)  &      & (i)       &                & UGC\,7790                \\ \hline
IC 4662      &   \ra{17}{47}{06}{4}    & 1980   & 1620   & 660   &--14.9& 0.30 &    2.0    &   0.8          & ESO\,102-G014;               \\
(dI/iE)      &   \dec{-64}{38}{25}{0}  & ---    & ---    & ---   & (l)  &      & (g)       &                & He 2-269                  \\ \hline
UM 448       &   \ra{11}{42}{12}{4}    & 300    & 360    & 480   &--19.8& 0.11 &   76.1    &   0.7          & UGC\,06665; SCHG\,1139+006;\\
(iI/BCG)     &   \dec{+00}{20}{03}{0}   & 840    & 900    & 1200 & (l)  &      & (d)       &                & Ark\,312; ARP\,161; Mkn\,1304 \\ \hline\hline
\end{tabular}\\
$\spadesuit$: Resolution of the best image set available for the respective galaxy, after reduction and combination 
(a): Marlowe et al. (\cite{marlowe99}), (b): inferred from the heliocentric
velocity $v_{\rm hel}$ listed in the NED, corrected for solar motion with respect to 
the center of the Virgo Cluster and adopting 
H$_0$=75\,km\,s$^{-1}$\,Mpc\,$^{-1}$ (cf. Sect. \ref{sample}).
(c): Karachentsev et al. (\cite{karachentsev02}), 
(d): Mirabel \& Sanders \cite{mirabel88}, 
(e): $v_{LG}$ from Smoker et al. (\cite{smoker00}), corrected for Local Group 
infall to the Virgo Cluster like b); (f): Tully (\cite{tully88}),
(g): Heydari-Malayeri \cite{heydari90}, (h): Schulte-Ladbeck et al. (\cite{schulte00}), 
(i): Yasuda et al. (\cite{yasuda97}), (j): m$_B$ from M\'endez \& Esteban (\cite{mendez00}), 
(k): m$_B$ from Papaderos (\cite{papaderosphd}), (l): m$_B$ from the RC3 
(de Vaucouleurs et al. \cite{devaucouleurs91}), 
(m): m$_B$ for Pox 4 from M\'endez \& Esteban (\cite{mendez99}); 
no published m$_B$ for Pox 4B (n): m$_B$ from \cite{papaderos99}, 
(o): m$_B$ from \cite{fricke01}, (p): m$_B$ from Papaderos et al. (\cite{papaderos02})\\
\hrule
\end{table*}

\subsection{Data reduction}
\label{reduction}
The exposures have been processed semi--interactively, employing our
NIR image reduction software based on ESO MIDAS\footnote{Munich Image
data Analysis System, provided by the European Southern Observatory
(ESO).}.  The procedures used for basic reduction, background
subtraction, image alignment and coaddition follow the recipes which
are detailed in, e.g., the SOFI User Manual\footnote{Issue 1.2,
available online at {\tt http://www.eso.org}}.

Since surface photometry in the LSB regime depends sensitively on the
quality of the background subtraction, care has been exercised to 
eliminate residuals of this correction on both, small and large
spatial scales. For this purpose we have implemented a number of 
additional corrective steps into our software package. 
For each single science frame of an exposure sequence we first computed an 
individual master background image, using subexposures that were taken 
closest in time, but at a sufficiently large distance from the target.
Prior to the calculation of the master background frame, the input 
background images were cleaned from bright contaminating sources and 
normalized to the same mean intensity (see the SOFI User Manual for details).
The master background frame computed this way was in turn scaled to 
the background level of the respective science frame, and subtracted.
Each of the resulting subexposures was subsequently checked for a possible
residual background gradient and, whenever necessary, rectified using a 
first order polynomial fit. The result image was obtained through 
combination of all background--corrected and coregistered subexposures. 
In this procedure, out of all pixels with identical {\em sky} coordinates,
only those at sound {\em detector} coordinates have been included in
the calculations.

Small residual background variations in the final images needed to be
interactively corrected by fitting twodimensional polynomials.
Alternatively, regions still affected by small--scale residuals in the
background subtraction were extracted as subimages, corrected in the
latter manner, and inserted back into the original frame.  Likewise,
bleeding artifacts, caused by bright sources in the NTT/SOFI FOV, or
blooming of bright, overexposed foreground stars
in the very
outskirts of IC\,4662 (northern edge) and Tol\,1400-411 (east and
west) were modeled and subtracted out. SBPs of the latter two galaxies
were only analyzed well above intensity levels where slight residuals
from the replaced bright stars could possibly contribute.

Final images taken in the same filter during different nights or at
different telescopes were aligned to each other, transformed to equal
pixel scales and resolutions, and coadded after weighting each one by
its $(S/N)^2$.  The FWHM of the result images is listed in Table
\ref{tab_sample}.  
Images used for aperture or surface photometry were
manually cleaned for 
fore-- and background sources.
%
\subsection{Flux Calibration \label{SectFluxCal}}
\label{calibration}
The SOFI data was calibrated by observing at different airmasses
standard stars from Persson et al. (\cite{persson98}) six times each
night. The excellent photometric stability throughout the NTT
observing run has allowed for the derivation of an airmass--dependent
calibration with a scatter of $\la$ 0.01 mag during nights 1 and 3 and
$\la$ 0.02 mag during night 2. Zero points and airmass-dependent
calibration coefficients both agree well with the average values
supplied by the NTT/SOFI support team.
Images of UM\,448 and Mkn\,1329, obtained by combining Calar Alto
and NTT data, were calibrated using aperture photometry of bright
stars in the SOFI FOV. Despite different $K$ filters used at those 
telescopes, the integral $K_s$ fluxes are essentially preserved, as 
color terms in the filter transformations (Eqs. \ref{trafos_k}) are 
not exceeding a few 0.01 mag.

The BCDs Haro\,14 and Mkn\,178, for which no NTT observations 
are available, were calibrated using the 
{\em Two Micron All Sky Survey} (2MASS) catalogue\footnote{\tt
http://www.ipac.caltech.edu/2mass/} (Cutri et
al. \cite{cutri00}, Jarrett et al. \cite{jarrett00}). 
As 2MASS data (cf. Andreon \cite{andreon02}) may, due to their 
limited sensitivity, slightly underestimate the flux within the extended 
LSB component, we computed calibration terms using 2MASS field stars in the close vicinity 
of Haro\,14 and Mkn\,178.
\subsubsection{Transformation to other NIR photometric systems}
\label{transformations}

Unless stated otherwise, all magnitudes and colors given in this paper
refer to the calibrations described in the previous section, and to
the photometric systems defined by the instrumental setup of each
telescope (see Sect. \ref{SectObs}).  However, whenever photometric
quantities of the sample BCDs are compared either among each other, or
with model predictions and data from the literature, they are first
transformed to the 2MASS photometric system, using the relations
described below.

The calibration obtained at the NTT with the SOFI instrument is based
on standards by Persson et al. (\cite{persson98}), and can therefore
be transformed to the Persson et al. (\cite{persson98}) Las Campanas
Observatory (LCO) system using the color transformations given by the
SOFI user manual (Issue 1.3, 16/08/2000). A transformation from the
latter system to the 2MASS system is described by Carpenter
(\cite{carpenter01}).  Combination of both latter transformations
yields the relations:
\begin{equation}
K_{s,\rm 2M} = K_{s,\rm S} + 0.021(J-K)-0.010\\
\label{trafos_s2m}
\end{equation}
\[
(J-H)_{\rm 2M} = 0.995 (J-H)_{\rm S} + 0.015 (J-K) + 0.002\\
\]
\[
(H-K_s)_{\rm 2M} = 1.029 (H-K_s)_{\rm S} - 0.046 (J-K) + 0.005,
\]
where the indices ``2M'' and ``S'' denote the 2MASS and SOFI systems,
respectively. Cumulative uncertainties introduced by this
transformation are $\la$ 0.01 mag for $K_s$ and $\la$ 0.015 mag for
$(J-H)$ and $(H-K_s)$, given the typical range of values expected for
NIR colors of BCDs.

No transformations to other standard NIR systems are available
for the photometric system defined by the Calar Alto 3.6m telescope
and the OMEGA PRIME camera. As stated in Sect. \ref{SectFluxCal},
the calibration of the Calar Alto frames is tied to the 2MASS zero
points. The remaining uncertainties introduced by the unknown color
terms can be estimated from Carpenter (\cite{carpenter01}) to be 
\begin{equation}
|({\rm color})_{2M}-({\rm color})_{\Omega '}|\la |0.08 ({\rm
 color})_{\Omega '}|
\end{equation}
per transformed color, i.e., the approximate upper limit of the color
terms entering into the transformation between standard NIR
photometric systems.

The $K$\arcmin\ and $K_s$ magnitudes are related by the 
following equations (cf. Wainscoat \& Cowie \cite{wainscoat92} 
and the SOFI user manual):
\begin{equation}
K_s = K + 0.005 (J-K)
\label{trafos_k}
\end{equation}
\[
K = K\arcmin\ - (0.22 \pm 0.03) (H-K)
\]

\subsubsection{Extinction correction:}
\label{extinction}

Magnitudes and colors given in this paper are corrected for Galactic
extinction, adopting values derived from the $B$ band extinction 
maps by Schlegel et al. (\cite{schlegel98}) (cf. Table \ref{tab_sample}) and 
the standard (R$_V=3.1$) extinction law 
(Cardelli et al. \cite{cardelli89}) implemented into the NED.
No attempt was made to correct for internal extinction, since this is
known to vary spatially even in the most metal-deficient BCDs 
(cf., e.g., Guseva et al. \cite{guseva01}, Cannon et al. \cite{cannon02}, 
Hunt et al. \cite{hunt03}) and can be reliably constrained in the SF regions only.

\section{Data analysis and SBP derivation \label{SectDataSBP}}
\subsection{Derivation of surface brightness and color profiles}
\label{sbps}
Surface photometry aims at a standardized one-dimensional 
representation of a galaxy's two-dimensional flux pattern.
One technique to compute surface brightness profiles (SBPs) 
requires the determination of the size $A(\mu)$ of the galaxy in 
$\sq\arcsec$ for a series of surface brightness levels $\mu$ (\sbb). 
By this definition, the equivalent radius \rr=$\sqrt{(A(\mu)/\pi)}$ 
is a monotonic function of the surface brightness $\mu$ and {\sl vice versa}. 
In order to derive SBPs this way, one has to keep track of the morphology and
angular extent of a BCD throughout its intensity span, i.e. in general to be 
able to interpolate an isophote down to the faintest measured level 
$\mu$ of an SBP. By this condition one can visually check for and screen-out
fore-- or background sources in the periphery of the galaxy, thus make sure 
that source confusion does not affect SBPs at faint levels.
This task is more difficult to achieve when computing SBPs 
employing photon statistics inside circular or elliptical 
annuli, extending out to a user-defined maximal radius $r_{\rm max}$. 
SBPs derived for an irregular system by such techniques may vary
from case to case, depending on, e.g., the adopted $r_{\rm max}$ or
``center'' of the galaxy. It should be borne in mind
that techniques of this kind, when applied to galaxies with a 
morphology significantly departing from the assumed circular 
symmetry, can strongly overestimate the exponential scale 
length and underestimate the central intensity of the LSB component 
(cf., e.g., Marlowe et al. \cite{marlowe97}). 

In the present analysis, we compute SBPs using method iv), 
described in Papaderos et al. (\cite{papaderos02}). This is a 
hybrid technique, incorporating features of both aforementioned
approaches (determination of the \rr\ corresponding to a user-defined
$\mu$, as opposed to the determination of the mean surface brightness
$\overline{\mu}$ inside a circular annulus with a user-defined 
radius \rr$\pm\Delta$\rr).

For a set of $n$ intensity intervals $I_n$, with a mean intensity 
decreasing as $n$ increases, masks $M_n$ are generated, each of which 
extracts from a smoothed image of the galaxy the areas with intensities 
within $I_n$. The equivalent radius $R^{\star}_n$ corresponding to the
mask area $A_n$ is:
\begin{equation}
R^{\star}_n=\left(
\frac{1}{\pi}
\left[ \sum_{i=1}^{n-1}A_i+\frac{1}{2}A_n \right] \right)^{\frac{1}{2}}
\label{eq_lazy}
\end{equation}
The surface brightness $\mu (R^{\star}_n)$ is calculated from the mean
intensity of the {\em original} image within the area $M_n$.  Method
iv) overcomes the problem of the artificial SBP flattening for low
$S/N$ levels (see discussion in e.g. Noeske \cite{noeske99}, Noeske et
al. \cite{noeske00} and Cair\'os \cite{cairos00}) and proves reliable
down to very faint surface brightness levels ($I_n<$ sky noise).

As a check for consistency, we also derived SBPs through
ellipse fitting to isophotes or methods ii) and iii) in 
\cite{papaderos96a}, applying in all cases a minimum of 
image filtering to moderate photon noise.  
The resulting SBPs have been compared with deep optical surface 
photometry to ensure that the underlying LSB component has been 
detected and modelled (see Sect. \ref{decomposition}) over a sufficient 
radius range. 

Color profiles were derived by subtracting the SBPs from each other,
after the latter had been smoothed to equal FWHM. 
Surface brightness and color profiles, corrected for Galactic 
extinction (Sect. \ref{extinction}), are shown in Figs. \ref{ftol3} 
through \ref{fum448}.
\subsubsection{Notes on the errors}

The uncertainties of each data point include the fully propagated
effects of {\sf (i)} Poisson photon noise and {\sf (ii)} residual
background variations on various spatial scales. The latter originate
both from small residuals from the background reduction, and from
undetected fore-- and background sources.
The treatment of {\sf (i)} is explained in, e.g., \cite{papaderos96a} or
Cair\'os et al. \cite{cairos01a}. The resulting uncertainties are in
most cases small, owing to the large number of pixels typically
involved in the calculation of each profile point at small $S/N$
levels, i.e. in the outer LSB region of a galaxy.
In these low $S/N$ regions, the uncertainties from 
{\sf (ii)} are typically dominant: background variations on spatial
scales which are not small compared to the size of the considered mask
segment or aperture do not average out.

The amplitude of such variations was estimated as the standard
deviation of the mean intensities measured in quadratic apertures, 
placed on the background well away from the object, with sizes
$\lambda$ of the order $1/10$ of the object's diameter. 
The resulting uncertainties of each SBP point might be decreased by a
factor of $\sqrt{A_n/\lambda ^2}$, accounting for the averaging of
background variations over the area $A_n$ (Eq. \ref{eq_lazy}) of the
respective mask segment. This first-order correction was however
discarded, since the background variations cannot be determined at the
position of the galaxy's LSB component, and since the outer mask
segments cut through large areas of the image, possibly sampling
individual stronger background anomalies.  Error bars shown in the
SBPs and color profiles (Figs. \ref{ftol3} through \ref{fum448})
therefore represent upper limits, i.e. typically overestimates, to the
true errors, which are difficult to quantify. Nevertheless, the error
constraints shown here give a better estimate of the true errors than
the, typically too small, pure Poisson noise errors.

The reliability limits of the SBPs were not estimated from these error
constraints, but determined at a surface brightness where an isophote
could still be robustly interpolated on a mildly smoothed image.

\begin{table*} 
\tabcolsep1.9mm
\caption{Structural properties of the dwarfs$^a$; see also the discussion of individual objects}
\label{tab_phot}
\begin{tabular}{llccccccccccc}
\hline
\hline
Name & Band & $\mu_{E,0}$ & $\alpha $ & $m_{\rm LSB}^{\rm fit}$ &
$P_{\rm iso}$  & $m_{P_{\rm iso}}$ & $E_{\rm iso}$ & $m_{\rm E_{\rm iso}}$  & $m_{\rm SBP}$ &
$m_{\rm tot}$ & $r_{\rm eff}$,$r_{80}$ & $\eta _{\rm\,SBP}$  \\
($b$,$q$)$^b$&     & \sbb\ &  kpc   & mag    & kpc       &  mag       &  kpc
          &  mag             &    mag & mag & kpc & \\
$\eta _{\rm LSB}^c$ &   &       &       & & & & & & & & & \\ 
   (1) &   (2)         &   (3)     &  (4)      &   (5)            &
 (6)    &  (7)             &  (8)     &  (9)  & (10) & (11) & (12) & (13)  \\
\hline
Tol\,3$^e$      &$J$ & 18.24$\pm$0.05 & 0.52$\pm$0.01 & 11.78 & 1.20 & 12.42 & 2.29 & 11.86 & 11.31 & 11.29 & 0.61,1.23 & 1.45 \\
$\star$         &$H$ & 17.74$\pm$0.06 & 0.53$\pm$0.01 & 11.26 & 1.18 & 11.76 & 2.07 & 11.37 & 10.73 & 10.69 & 0.61,1.22 & 1.43 \\
                &$K_s$& 17.45$\pm$0.09 & 0.51$\pm$0.01 & 11.06 & 1.18 & 11.48 & 2.12 & 11.15 & 10.50 & 10.46 & 0.57,1.15 & 1.49 \\
\hline
Haro\,14$^f$    &J & 17.46$\pm$0.12 & 0.37$\pm$0.01 & 12.70 & 1.39 & 12.45 & 1.88 & 12.82 & 11.81 & 11.78 & 0.61,1.21 & 1.18 \\
{\sl 3.6,0.94} &H & 16.97$\pm$0.26 & 0.39$\pm$0.02 & 12.11 & 1.37 & 11.69 & 1.76 & 12.31 & 11.12 & 11.07 & 0.61,1.22 & 1.25 \\
0.49           &K'& 16.88$\pm$0.94 & 0.39$\pm$0.08 & 11.99 & 1.32 & 11.54 & 1.83 & 12.17 & 10.98 & 10.97 & 0.60,1.21 & 1.45 \\
\hline
UM\,461$^e$     &J & 19.37$\pm$0.18 & 0.21$\pm$0.01 & 15.56 & 0.42 & 16.19 & 0.69 & 15.88 &15.09 & 15.04 & 0.34,0.57 & 1.96 \\
{\sl 2.3,0.85}&H & 18.82$\pm$0.25 & 0.21$\pm$0.02 & 14.98 & 0.35 & 16.12 & 0.60 & 15.46 &14.66 & 14.57 & 0.37,0.61 & 1.82 \\
0.49            &K$_s$& 18.47$\pm$0.79 & 0.19$\pm$0.04 & 14.79 & 0.36 & 15.75 & 0.63 & 15.14 &14.46 & 14.37 & 0.32,0.54 & 2.04 \\
\hline
Henize\,2--10$^e$&J & 18.46$\pm$0.03 & 0.67$\pm$0.01 & 10.46 & 0.87 & 10.57 & 2.80 & 10.55 & 9.80 & 9.70 & 0.41,1.26 & 2.50 \\
$\star$         &H & 17.75$\pm$0.04 & 0.64$\pm$0.01 & 9.85  & 0.78 & 9.94  & 2.51 & 9.96  & 9.17 & 9.13 & 0.40,1.22 & 2.44 \\
                &K$_s$& 17.51$\pm$0.03 & 0.62$\pm$0.01 & 9.68  & 0.79 & 9.62  & 2.56 & 9.78  & 8.95 & 8.89 & 0.34,1.04 & 2.44 \\
\hline
Tol\,1400--411$^e$ & J & 18.29$\pm$0.11 & 0.33$\pm$0.01 & 11.18 & 0.69 & 12.67 & 1.41 & 11.32 &10.99 & 10.91 & 0.60,0.98 & 2.04\\
{\sl 3.0,0.82}& H$^d$ & 17.30$\pm$0.14 & 0.27$\pm$0.01 & 10.57 & 0.31 & 13.65 & 1.18 & 10.71 &10.57 & 10.55 & 0.57,0.93 & 1.33\\
0.63       &K$_s$& 17.39$\pm$0.09 & 0.30$\pm$0.01 & 10.43 & 0.50 & 12.49 & 1.28 & 10.58 &10.32 & 10.31 & 0.59,0.97 & 1.75\\
\hline
Pox\,4$^e$      &J & 19.26$\pm$0.15 & 0.86$\pm$0.03 & 15.15 & 2.27 & 14.80 & 2.87 & 15.54 & 14.24 & 14.20 & 1.24,2.17 & 1.75\\
{\sl 2.7,0.90}&H$^d$& 19.10$\pm$0.31 & 0.88$\pm$0.07 & 14.95 & 2.18 & 14.35 & 1.79 & 16.10 &13.85 & 13.81 & 1.27,2.15 & 1.45\\
0.42             &K$_s$$^d$& ---           & ---     & ---   & ---  & ---   & ---  & ---   &13.54 & 13.52 & 1.18,2.09 & 1.96\\
\hline
Pox\,4\,B$^e$   &J & 19.47$\pm$0.40 & 0.29$\pm$0.04 & 17.37 & 0.51 & 19.29 & 0.93 & 17.67 & 17.21 & 17.12 & 0.54,0.87 & 0.75\\
{\sl 2.0,0.80}&H$^d$ & ---            & ---         & ---   & ---  & ---   & ---  & ---   &16.83 & 16.72 & 0.51,0.80 & 0.77 \\
0.49             &K$_s$$^d$& ---       & ---            & ---   & ---  & ---   & ---  & ---   &16.75 & 16.52 & 0.47,0.69 & 0.75 \\
\hline
Tol\,65$^e$     &$J$ & 19.58$\pm$0.18 & 0.28$\pm$0.02 & 17.25 & 0.43 & 17.47 & 0.82 & 17.79 &16.57 & 16.55 & 0.37,0.80 & 1.59 \\
{\sl 2.7,0.90}  &$H^d$ & ---            & ---         & ---   & ---  & ---   & ---  & ---   &   16.23 & 16.30 & 0.42,0.84 & --- \\
0.46            &$K_s$$^d$ & ---      & ---           & ---   & ---  & ---   & ---  & ---   &16.03 & 15.99 & 0.24,0.61 & --- \\      \hline
Tol\,1214--277$^{e,g}$ & J$^d$ & 19.16$\pm$0.24 & 0.53$\pm$0.03 & 18.05 & 0.75 & 18.56 & 1.69 & 18.55 &17.46 & 17.41 & 0.86,1.83 & 1.89\\
{\sl 3.3,0.92}&H$^d$ & ---            & ---         & ---   & ---  & ---   & ---  & ---   & 17.22 & 17.26 & 1.15,1.90 & --- \\
0.48             &K$_s$$^d$ & ---     & ---           & ---   & ---  & ---   & ---  & ---   & 16.94 & 17.07 & 0.81,1.67 & --- \\ 
\hline
Mkn\,178$^f$    &J & 20.62$\pm$0.11 & 0.28$\pm$0.01 & 12.94 & 0.39 & 14.36 & 0.62 & 13.41 & 12.78 & 12.70 & 0.34,0.57 & 1.27 \\
$\star$         &H$^d$& 19.87$\pm$0.18 & 0.26$\pm$0.02 & 12.35 & 0.26 & 14.85 & 0.51 & 12.92 & 12.29 & 12.25 & 0.34,0.58 & 1.12 \\
                &K'$^d$& 19.92$\pm$0.43 & 0.27$\pm$0.04& 12.31 & 0.36 & 14.00 & 0.52 & 12.91 & 12.13 & 12.01 & 0.36,0.58 & 0.97 \\
\hline
Mkn\,1329$^e$   &J & 19.25$\pm$0.04 & 0.67$\pm$0.01 & 12.81 & 0.99 & 14.61 & 2.32 & 13.01 & 12.64 & 12.60 & 1.12,1.97 & 1.01 \\
{\sl 1.6,0.70}&H$^d$& 18.81$\pm$0.01 & 0.71$\pm$0.01 & 12.25 & 1.01 & 13.85 & 2.09 & 12.57 & 12.00 & 11.94 & 1.17,2.08 & 1.11 \\
0.68             &K$_s$& 18.24$\pm$0.06 & 0.60$\pm$0.01 & 12.03 & 0.79 & 14.56 & 2.09 & 12.23 & 11.95 & 11.93 & 1.09,1.87 & 0.72 \\
\hline
IC\,4662$^e$    &J & 16.79$\pm$0.24 & 0.15$\pm$0.01 & 10.60 & 0.55 & 10.32 & 0.83 & 10.73 & 9.68 & 9.64 & 0.29,0.53 & 1.41 \\
{\sl 4.8,0.97}&H & 16.61$\pm$0.73 & 0.15$\pm$0.02 & 10.36 & 0.58 & 9.64  & 0.69 & 10.66 & 9.15 & 9.05 & 0.29,0.52 & 1.37 \\
0.48             &K$_s$$^d$&---            & ---     & ---   & ---  & ---   & ---  & ---   & 9.04 & 8.93 & 0.27,0.48 & 1.75\\
\hline
UM\,448$^e$     &J & 20.55$\pm$0.07 & 3.83$\pm$0.10 & 13.48 & 4.26 & 12.50 & 8.66 & 13.92 & 12.17 & 12.15 & 1.28,3.63 & 2.33 \\
$\star$         &H & 19.95$\pm$0.17 & 4.02$\pm$0.31  & 12.77 & 4.08 & 11.88 & 7.60 & 13.39 & 11.52 & 11.50 & 1.32,3.94 & 2.40 \\
               &K$_s$& 19.55$\pm$0.15 & 3.61$\pm$0.19 & 12.61 & 4.01 & 11.53 & 8.16 & 13.05 & 11.22 & 11.30 & 1.21,3.24 & 2.44 \\
\hline
\hline
\end{tabular}
\parbox{17.2cm}{$a$: All values are corrected for Galactic extinction, 
adopting the $A_B$ from Table \ref{tab_sample} (cf. Sect. \ref{extinction}).}
\parbox{17.2cm}{$b$,$c$: See Sect. \ref{decomposition} for details. Objects whose 
LSB component has been modelled by a pure exponential fit (Eq. \ref{exponential})
are marked with an asterisk.}
\parbox{17.2cm}{$d$: Decomposition uncertainty higher (faint object, or data affected by nearby bright stars).}
\parbox{17.2cm}{$e$: Calibration obtained at the ESO NTT, cf. Sect. \ref{calibration}.}
\parbox{17.2cm}{$f$: Calibrated using field stars from the 2MASS 
catalogue, cf. Sect. \ref{calibration}}
\parbox{17.2cm}{$g$: Fit radius interval and ($b$,$q$) adopted from \cite{fricke01}.}
\end{table*}


\subsection{Profile Decomposition \label{decomposition} }
%
The stellar emission of a BCD is due to the superposition of two
distinct populations with respect to their ${\cal M/L}$ ratio and
spatial extent: (i) the underlying LSB host galaxy which, owing to 
its high ${\cal M/L}$, contains the bulk of the system's stellar
mass, and (ii) the younger stellar population, attributable to the
ongoing and recent SF activity.  In the main class of iE/nE BCDs, the
younger stellar component dominates the optical emission inside $\sim$2
exponential scale lengths of the underlying LSB host and contributes,
together with ionized gas emission, $\sim$1/2 of the $B$ light within
the 25 $B$ \sbb\ isophote (\cite{papaderos96b}, Noeske
\cite{noeske99}, Salzer \& Norton \cite{salzer99}).
Evidently, a meaningful study of the structural properties 
of a BCD requires the decomposition of SBPs into these
two main photometric components. 
In the following, we employ a simple SBP decomposition scheme, 
in which we fit only the LSB emission. Subtraction of the best fit
LSB model from the SBP allows us to deduce the luminosity fraction 
and spatial extent of the superimposed SF component. 
In order to make sure that extended starburst emission does not
affect the decomposition results, we fitted the LSB component 
beyond a
\emph{transition radius} \rt\ (cf. \cite{papaderos96a}, \cite{c02b}), 
where optical and optical--NIR color
gradients vanish and isophotes become more regular.  In addition,
whenever available, we used \ha\ maps to trace the size of the SF
component, and get from it an additional constraint to \rt.  The LSB
emission has been fit out to the radius where SBPs became increasingly
uncertain as a result of background noise (typically for
$\la$0.5$\sigma _{\rm bgr}$ for method iv). The fitted SBP range in
$J$ is indicated with the dashed-gray line at the bottom of each
profile (Figs. \ref{ftol3}--\ref{fum448}).

An exponential fitting law
\begin{equation}
\label{exponential}
I(R^{\star})=I_{\rm E,0}\exp \left(-\frac{R^{\star}}{\alpha}\right)
\end{equation}
with a central intensity $I_{\rm E,0}$ and an exponential scale length
$\alpha$ has been found to approximate well the intensity distribution
of our sample galaxies in their LSB component. An exponential law is
also known to fit well the outer parts of dwarf irregulars (dIs, Patterson \&
Thuan \cite{patterson96}, van Zee \cite{vanzee00}) and dwarf
ellipticals (dEs, Vigroux et al. \cite{vigroux88}, Binggeli \& Cameron
\cite{binggeli91}) over a few scale lengths.

However, for several of our sample BCDs, an extrapolation of the
exponential LSB slope to smaller radii yields for \rr$\sim$1\dots
3$\alpha$ a higher intensity than the observed value. Thus, a
meaningful decomposition of such SBPs cannot be achieved on the usual
assumption that the exponential law is valid in the LSB component all
the way to \rr=0\arcsec. Instead, one has to adopt an alternative
fitting formula, approaching the exponential law for large radii and
flattening in its inner part. Such a distribution should be compatible
to the ``type II'' profiles of disc galaxies (Freeman
\cite{freeman70}, see also MacArthur et al. \cite{macarthur02}), or
the ``\flat'' profiles described for spheroidal early type DGs by
Binggeli \& Cameron (\cite{binggeli91}).

Previous optical studies have allowed for the detection and modelling
of such an inwards flattening exponential LSB profile in a few BCDs
only (e.g. I Zw 115, \cite{papaderos96a}; Tol 65, Papaderos et
al. 1999, hereafter \cite{papaderos99}; Tol 1214-277, Fricke et
al. 2001, hereafter \cite{fricke01}; SBS 0940+544, Guseva et
al. \cite{guseva01}).  However, SBPs of this kind do not appear to be
rare among intrinsically faint dEs and dIs. They have been observed in
several dEs with and without a central nucleus (cf. Binggeli \&
Cameron \cite{binggeli91}, Cellone et al. \cite{cellone94}, Young \&
Currie \cite{young94}, \cite{young95}), and Vennik et
al. (\cite{vennik00}) deduce a fraction of $>$10\% for the SBPs of
late-type dwarf galaxies falling into this category.  In fact, a
casual inspection of the sample of Patterson \& Thuan
(\cite{patterson96}), Makarova et al. (\cite{makarova98}), Vennik et
al. (\cite{vennik00}) and van Zee (\cite{vanzee00}) reveals several
examples of dIs showing an exponential outer LSB slope and a
pronounced flattening for intermediate to small radii (e.g., UGC 2034,
UGC 2053, UGC 5423, UGC 9128, UGC 10669 in the sample of Patterson \&
Thuan \cite{patterson96}, KKH35 and KKSG 19 in Makarova et
al. \cite{makarova98}, or UGCA 9, UGCA 15, UGC 2345,UGC 9240, UGC
10445 in the sample by van Zee \cite{vanzee00} ).  Other examples are
the dIs Holmberg~I (Ott et al. \cite{Jurgen01}), Holmberg~II (Noeske
\cite{noeske99}), H\,1032-2722 (Duc et al. \cite{duc99}) and Kar~50
(Davidge \cite{davidge02}).  Following the nomenclature of Binggeli \&
Cameron (\cite{binggeli91}) we shall henceforth refer to this type of
SBPs as to \flat.

The physical origin and exact form of \flat\ profiles in puffed-up 
stellar systems has not yet been studied in detail, neither 
observationally nor theoretically. Therefore it is difficult to say 
which fitting formula approximates best their intensity distribution 
(see discussion in Sect. \ref{dis3}). An empirical fitting function 
that yields by deprojection a finite luminosity density for \rr=0\arcsec\
has been proposed in P96a, as
\begin{equation}
\label{med}
I(R^{\star})=I_{\rm E,0}\exp
\left(-\frac{R^{\star}}{\alpha}\right)\{1-q\exp (-P_3(R^{\star}))\}
\end{equation}
with
\begin{equation}
P_3(R^{\star})=\left(\frac{R^{\star}}{b\alpha}\right)^3+
\left(\frac{R^{\star}}{\alpha}\cdot \frac{1-q}{q}\right).
\end{equation}
The modified exponential distribution Eq. (\ref{med}), in the
following referred to as \med, flattens with respect to a pure
exponential law inside of a cutoff radius $b\alpha$, and attains at
\rr=0\arcsec\ an intensity given by the relative depression parameter
$q=\Delta I/I_{\rm E,0}< 1$.  An advantage of the \med\ is that its
exponential part (left part of Eq. \ref{med}), depending on $I_{\rm
E,0}$ and $\alpha$ only, can be 
separately constrained by fitting Eq. \ref{exponential} to the outer
exponential part of a \flat\ profile. This has been done by means of
unweighted fits, after SBPs were splined to equidistant radius steps,
in order to ensure that the fit solution is not biased through a
clustering of points to a specific radius range.  Once the scale
length $\alpha$ is fixed, the multiplicative right-hand side part of
Eq. (\ref{med}) is a function of $b$ and $q$, only.
Because the ($b$,$q$) parameter space has not been explored so far,
e.g., by fitting Eq. (\ref{med}) to a sample of dEs with \flat\ SBPs,
it is difficult to judge which values are appropriate for dwarf
galaxies falling into this category.  P96a inferred a tentative ratio
of $b/q\sim$3, they noted, however, that this could be
luminosity-dependent.

In order to deduce plausible constraints to ($b$,$q$), we first
subtracted from the $J$ image of each sample BCD most of the irregular 
starburst emission. The latter has been approximated by an unsharp-mask 
version of the original image (Sect. \ref{hb}), adjusted such that no
regular emission from the LSB component was removed.
\ha\ exposures and color maps were used to further ascertain that the
subtracted emission was not part of the underlying host galaxy. 
$J$ band exposures processed this way were then used to compute 
SBPs for the LSB component.
These profiles, denoted $J_{\rm LSB}$, allow to better trace the LSB 
component down to smaller \rr\ than the ones derived prior to partial
subtraction of the starburst light.
By fitting Eq. (\ref{med}) to the $J_{\rm LSB}$ SBPs we derived $b$,$q$
(first column in Table \ref{tab_phot}). In most cases, fit
uncertainties are 0.1 and 0.05 for $b$ and $q$, respectively.  No
reliable $J_{\rm LSB}$ profiles could be computed for Tol
1214-277. For this system we fixed ($b$,$q$) to values inferred by
\cite{fricke01} from optical VLT data.

Alternatively, \flat\ $J_{\rm LSB}$ profiles were fit with 
a S\'ersic model (S\'ersic \cite{sersic68}) of the form
\begin{equation}
\label{sersic}
I(R^{\star})=I_{\rm S,0}\exp
\left(-\frac{R^{\star}}{\beta}\right)^{1/\eta}.
\end{equation}
The exponent\footnote{For consistency with, e.g., Caon et
al. (\cite{caon93}) and \cite{c02b} we define
here the S\'ersic shape parameter as $\eta$, i.e. the reciprocal 
exponent in Eq. (\ref{sersic}). Note that in other studies (Cellone et
al. \cite{cellone94}, Young \& Currie \cite{young94}, \cite{young95};
\cite{papaderos96a}) the S\'ersic shape parameter is referred to as 
1/$\eta$.} $\eta$ in Eq. (\ref{sersic}) is a handy indicator 
of systematic deviations of the observed SBP from an exponential 
distribution.  
An $\eta<$1 corresponds to a convex profile, which might be
compatible to a \med\ distribution with $q>0$ (see discussion in 
Sect. \ref{dis3}), whereas an $\eta>1$ indicates a concave profile
which, in the special case of $\eta=4$, translates into the 
de Vaucouleurs law (see, e.g., Caon et al. \cite{caon93}). 

Note that Eq. (\ref{sersic}) approximates well the projected light of
a variety of stellar populations with both a nearly constant ${\cal M/L}$
(ellipticals or bulges; Caon et al. \cite{caon93}, 
Andredakis et al. \cite{andredakis95}, Graham et
al. \cite{graham96}, early-type dwarfs; Cellone et 
al. \cite{cellone94}, Young \& Currie 1994) or 
strongly varying
${\cal M/L}$ (for instance, nE BCDs or the \emph{plateau} component 
in the SBP of iE BCDs; \cite{papaderos96a}).
Therefore a \ser\ exponent $\eta>1$ gives no strong indication for a stellar 
system being similar to an intrinsically luminous early-type galaxy 
(see also discussion in \cite{c02b}).
%

The \ser exponents of the LSB components ($\eta_{\rm LSB}$) were 
obtained from non-weighted fits to the $J_{\rm LSB}$ SBPs (see above) 
of \flat\ profiles.
This limits $\eta_{\rm LSB}$ to values $<$1.
We avoided to fit Eq. (\ref{sersic}) to the outer part 
of SBPs alone, i.e. for \rr$\geq$\rt, as solutions obtained 
this way are very uncertain and depend strongly on the accuracy
of the sky subtraction (see Sect. \ref{dis3} and detailed 
discussion in \cite{c02b}).
Notwithstanding the fact that Eqs. (\ref{med}) and (\ref{sersic}) give
comparably good fits in terms of $\chi^2$, we decided not to
include the full ($I_{\rm S,0}$,$\beta$,$\eta$) \ser solutions for the
$J_{\rm LSB}$ SBPs in Table \ref{tab_phot}.  Instead, in its Col. 1,
we quote only the \ser\ exponent $\eta_{\rm LSB}$, the uncertainties 
of which are estimated to be of the order of 20\%.

The photometric quantities of the sample BCDs are summarized in Table
\ref{tab_phot}. BCDs without signatures of a \flat\ profile in their 
underlying LSB component 
are marked with an asterisk in column 1. 
For the remaining systems we list the ($b$,$q$) and $\eta_{\rm LSB}$ parameters, as 
obtained respectively by fitting Eqs. (\ref{med}) and (\ref{sersic}) to $J_{\rm LSB}$ SBPs.  
Columns 3 and 4 list, respectively, the \emph{extrapolated} central surface brightness
$\mu_{\rm E,0}$ (\sbb) and exponential scale length (kpc), obtained
by fitting Eq. (\ref{exponential}) to the outer exponential LSB part
of each SBP. Column 5 lists the total apparent magnitude of the LSB
component, computed by extrapolating the fitted model 
(i.e. Eq. \ref{exponential} or Eq. \ref{med}) to $R^{\star}=\infty$.
Columns 6 through 9 list the radii and magnitudes of the star-forming
(P) and underlying stellar LSB component (E), as obtained by profile
decomposition. Following P96a, we measure the respective radial extent
($P_{\rm iso}$, $E_{\rm iso}$) and encircled magnitude ($m_{P_{\rm
iso}}$,$m_{E_{\rm iso}}$) of each component at an isophotal level
$iso$, taken to be 23 \sbb\ for $J$ and 22 \sbb\ for $H$ and $K$. The
isophotal radii determined for the sample BCDs at 23 $J$ \sbb\ turn
out to be comparable to those obtained from optical SBPs at 25 $B$
\sbb\ ($P_{25}$ and $E_{25}$ in P96a).
Column 10 lists the magnitude from an SBP integration out to 
the last data point, and total magnitudes from aperture measurements 
(cf. Sect. \ref{apphot}) are listed in column 11. The radii
$r_{\rm eff}$ and $r_{80}$, enclosing 50\% and 80\% of the SBP's
flux are included in column 12.
Finally, a {\em formal} \ser exponent for the whole SBP 
($\eta _{\rm SBP}$), for later comparison with literature data, 
is listed in column 13 of Table \ref{tab_phot}.

Errors of the fitting law parameters determined from unweighted fits
are an underestimation of the true errors. LSB slope differences
between $J$, $H$ and $K$ (Table \ref{tab_phot}) should therefore not be
considered significant, but rather to reflect the true exponential
slope uncertainties, which are typically up to 10\% in $J$, 
and may become somewhat larger in $H$ and $K$, depending on the data 
quality.

\subsection{Aperture photometry and total magnitudes}
\label{apphot}
Since a SBP can only be accurately derived in regions with a 
sufficiently high $S/N$ (i.e., in general, down to the minimum 
intensity level in which the morphology of a BCD can still be 
visually checked and the problem of source confusion can be 
handled; cf. Sect. \ref{sbps}), profile integration out to the 
last measured data point may, in some cases, underestimate the 
object's total flux.
For extended LSB sources, or when NIR SBPs do not go sufficiently
deep (e.g., the $K_s$ SBP of Tol\,1214-277), the fractional flux 
missed can easily exceed 10\%. 
Similar problems may affect growth curve flux determinations,  
since these basically rely on a crude SBP derivation and can 
sensitively depend on the quality of background subtraction.

We therefore measured total magnitudes within polygonal apertures
which extend typically out to 1.5 Holmberg radii (Col. 11 of Table
\ref{tab_phot}), after removal of fore- and background sources from
the area of interest (cf. Sect. \ref{reduction}).  Errors take into
account the Poisson noise and small--scale variations of the local
background.

\begin{figure}[!ht]
\includegraphics[angle=270,width=8cm,clip=]{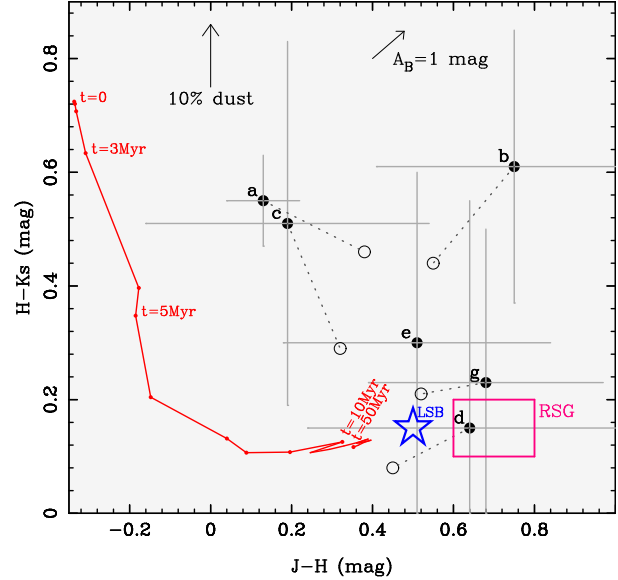}
\caption{NIR two-color diagram for the brightest compact regions 
in the iI BCD UM\,461. 
Filled circles show the colors of the regions {\sf a} 
through {\sf g} indicated in Fig. \ref{fum461}, after correction
for the flux contribution of the underlying LSB host galaxy.
Open circles, connected with dotted lines, indicate the 
color of the respective region prior to that correction.
The star marks the color of the stellar host galaxy (``LSB'').
The color range covered by red supergiants in the SMC
(Elias et al. \cite{elias85}) is shown by the box labelled {\sf RSG} 
to the lower right. The temporal evolution of the $J-H$ vs. $H-K_s$ color 
for an instantaneously formed stellar population with \zsun/10
for an age between $t$=0 and $t$=50\,Myr (solid line) has been calculated 
with the PEGASE code (Fioc \& Rocca--Volmerange \cite{fioc97}).
The colors for $t=$0 correspond to purely gaseous emission. 
Arrows depict the effect of a 10\% contribution of warm
($\sim$200\,K) dust to the $K$ band (cf. Campbell \& Terlevich
\cite{campbell84}), as well as the extinction vector. All colors are
transformed to the 2MASS photometric system.}
\label{knots461}
\end{figure}

Magnitudes and colors of selected features, such as stellar clusters
or \ion{H}{ii} regions, have been corrected for the flux contribution
of the underlying LSB component, by interpolating the mean surface
brightness of adjacent regions.  Values computed this way are marked
with the superscript $\ddag$. As pointed out in Cair\'os et
al. (\cite{cairos02a}) and Papaderos et al. \cite{papaderos02}, corrections 
for the LSB emission are generally
not negligible in optical wavelengths. That this statement is also true in
the NIR domain is illustrated on the example of the iE BCD UM 461
(Fig. \ref{knots461}); correction for the flux contribution of the
LSB background shifts even the brightest SF region ({\sf a}, $m_J\ddag$=17.26 
mag, open circle) of this system by +0.25 mag and --0.1 mag in the $J-H$ 
and $H-Ks$ color diagram.
A census and photometric study of compact stellar clusters 
in the extended BCD sample included in this project
after correction for the LSB- and ionized gas emission 
will be presented in a subsequent paper.

\subsection{Unsharp masking \label{hb}}
The considerable intensity range of a BCD, from its faint LSB
outskirts to the brightest nuclear starburst region, renders the
detection of fine coherent morphological features in the central
portion of the galaxy difficult. We use therefore a modified unsharp
masking technique (cf., e.g., Papaderos \cite{papaderosphd}), 
referred to in the following as {\em hierarchical binning 
({\rm hb}) -- transformation}. This contrast-enhancing procedure 
is stable against noise at low intensities and allows for the 
flux determination of faint sources within the bright 
background of a BCD, with an angular size smaller than a 
user-defined value. Morphological features of interest revealed 
using this procedure are displayed in the grayscale/isophote 
insets of the sample galaxies, and described in Sect. \ref{sample_galaxies}.

\subsection{Colors of the underlying LSB host galaxy}
\label{lsbcolors}
Colors of the LSB host galaxy were derived 
as the error-weighted mean of the color profiles 
for \rr$>$\rt, after rejection of deviant points, being probably 
affected by uncertainties in the sky determination, and local
residuals in the subtraction of background sources.
For Tol 1400-411, Pox 4, Mkn 178 and IC 4662, uncertainties in the LSB
colors are larger, due to extended starburst emission or crowding with
nearby bright stars. 
As for the very metal-deficient systems Tol 65 and Tol 1214-277, 
the faintness of their LSB component in $K_s$ has not permitted, 
despite generous exposure times, to pin down their $H-K_s$ colors.

Whenever calibrated optical data were available, optical--NIR colors
were derived. Because the quality of the SBPs was typically better in
$J$ than in $H$, and the $J-H$ color shows little evolution with time (few
0.1 mag) for old ($\ga$ 1Gyr) stellar populations, we derived 
$B-J$ colors instead of the more commonly used $B-H$ colors. 

The mean colors of the host galaxies are shown at the right edge of
each color profile (Figs. \ref{ftol3} -- \ref{fum448}), in the
photometric system in which the respective galaxy was observed and
calibrated (see Sect. \ref{transformations}).  
Table \ref{tab_lsbcolors} lists the NIR colors, transformed to the 2MASS
photometric system to facilitate comparisons and the $B-J$ color, where available. 
Errors give cumulative uncertainties in the calibration,
transformation to the 2MASS system, and the scatter and systematic
uncertainty of each color profile.
\begin{table}
\caption{Colors of the host galaxy$^a$}
\label{tab_lsbcolors}
\begin{tabular}{lccc}\hline
Object          & $J-H$         &  $H-K_s$      & $B-J$          \\
                & [mag]         & [mag]         & [mag]         \\ \hline
Tol\,3          & 0.55$\pm$0.07 & 0.10$\pm$0.06 & ---   \\
Tol\,65$^c$     & 0.48$\pm$0.17$^b$& ---        & 1.3$\pm$0.15$^b$      \\
Tol\,1214--277$^c$&0.39$\pm$0.16$^b$& ---       & 1.0$\pm$0.13$^b$      \\
Tol\,1400--411$^c$&0.43$\pm$0.08& 0.26$\pm$0.09 & 1.45$\pm$0.16\\
Pox\,4$^{c}$    & 0.42$\pm$0.13 & 0.12$\pm$0.13 & ---   \\
Pox\,4B         & 0.43$\pm$0.15$^b$& 0.1$\pm$0.15$^b$   & ---   \\
UM\,448         & 0.69$\pm$0.10$^b$& 0.14$\pm$0.11$^b$  & 2.09$\pm$0.15 \\
UM\,461         & 0.50$\pm$0.06 & 0.15$\pm$0.06 & 1.92$\pm$0.16 \\
He\,2-10        & 0.59$\pm$0.06 & 0.11$\pm$0.07 & ---   \\
IC\,4662$^{c}$  & 0.57$\pm$0.12$^b$& 0.03$\pm$0.12$^b$  & ---   \\
Mkn\,178        & 0.53$\pm$0.15 & 0.25$\pm$0.16$^b$ & 2.00$\pm$0.15 \\
Mkn\,1329       & 0.65$\pm$0.09 & 0.03$\pm$0.10 & ---   \\
Haro\,14        & 0.69$\pm$0.12$^b$& 0.20$\pm$0.12$^b$  & 2.02$\pm$0.15 \\ \hline 
\end{tabular}\\[0.5em]
$^a$: Corrected for galactic extinction (see Sect. \ref{extinction});
NIR colors are transformed to the 2MASS system (cf. Sect.
\ref{transformations}). The $B$ band denotes the Johnson $B$. 
Errors include cumulative uncertainties in the determination of 
color profiles, the calibration and the transformation to the 
2MASS system.\\
$^b$: possible local instabilities in one SBP at low $S/N$ levels.\\ 
$^c$: possible contamination by gas emission over 
a large portion of the LSB host galaxy\\
\end{table}

\section{Results and discussion of individual objects \label{sample_galaxies}}
\subsection{Tol 3 (NGC 3125)}                                         
\label{tol3}

\begin{figure*}[!ht]
\begin{picture}(18,10)
\put(0,0.1){{\psfig{figure=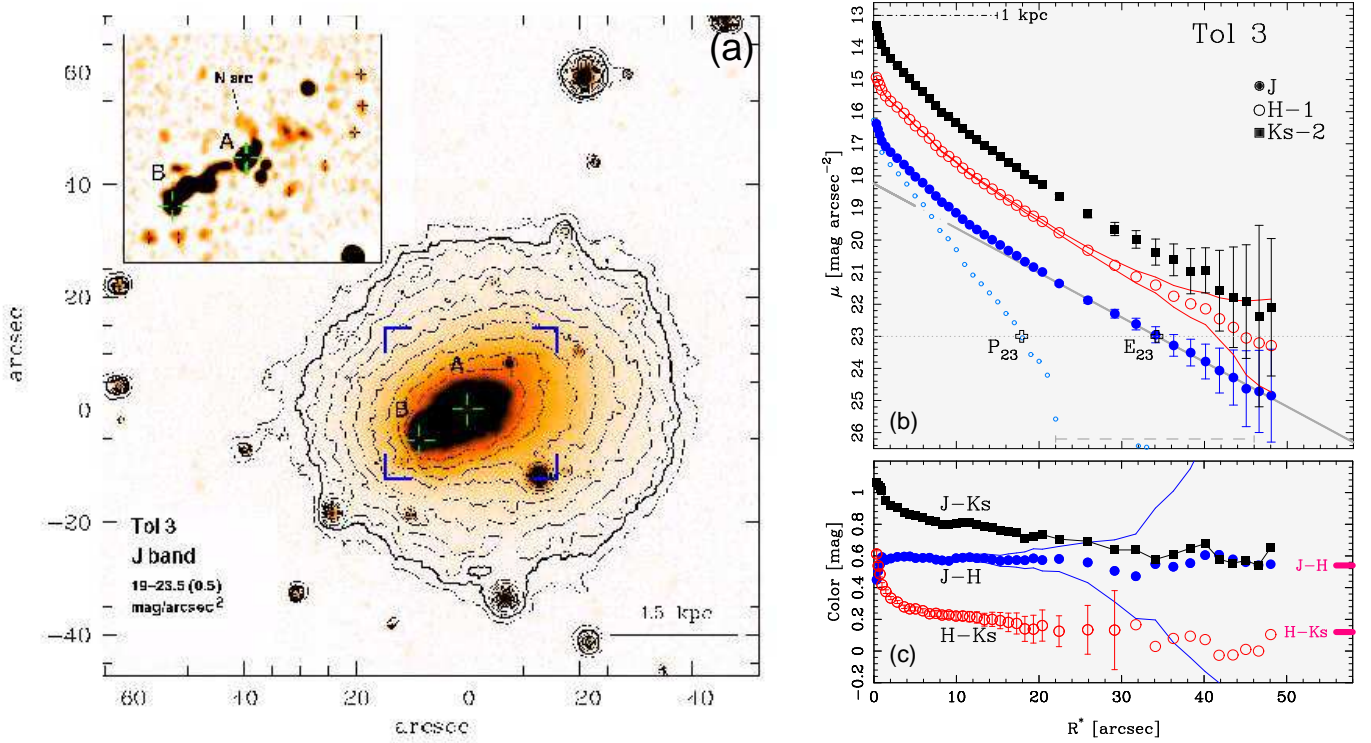,height=9.9cm,angle=0,clip=}}}
\put(10.9,4.01){{\psfig{figure=H4029F3.ps,width=7.1cm,angle=-90,clip=}}}
\put(10.95,0.19){{\psfig{figure=H4029F4.ps,width=7.1cm,angle=-90,clip=}}}
\put(9.4,9.2){\Large\sf (a)}
\put(11.8,4.3){\sf (b)}
\put(11.8,1.2){\sf (c)}
\end{picture}

\caption[]{{\bf a):} Contours overlaid with a $J$ image of Tol
3 ($D$=13.8 Mpc). North is up, east to the left. Contours, corrected for Galactic
extinction, go from 19 to 23.5 $J$ \sbb\ in increments of 0.5 mag. The
23\ $J$ \sbb\ isophote is illustrated by the thick contour. The
brighter star-forming region A and the fainter knot B (following the
denomination by Schaerer et al. \cite{schaerer99}) are marked with
crosses.
The inset shows a contrast-enhanced (see Sect. \ref{hb}) and magnified 
version of the central region of the BCD (indicated by brackets in 
the contour map image). Compact sources arranged along the 
southwestern arc--like chain are marked with small crosses.
{\bf b):} Surface brightness profiles (SBPs) of Tol 3 
in the $J$, $H$ and $K_s$, corrected for galactic extinction. 
For a better visualization, the $H$ and $K_s$ SBPs are shifted 
by --1 and --2 mag, respectively. The thick solid line 
illustrates an exponential fit to the stellar LSB component 
in $J$ (cf. Sect. \ref{decomposition}), computed in the 
radius range indicated by the light gray, long--dashed bar
at the bottom of the figure. 
The emission in excess to the fit (small open circles) is 
attributable to the starburst component, which dominates the light
in the inner part of Tol 3.
The isophotal radii $P_{23}$ and $E_{23}$ of the star-forming 
and LSB component at the surface brightness level of 
23 $J$ \sbb\ (horizontal dotted line) are indicated.
The bar at the upper left of the figure corresponds to
a galactocentric distance of 1 kpc.
{\bf (c):} Color profiles, computed by subtraction 
of the SBPs shown in the upper right panel. The thick  
lines at the rightmost part of the diagram indicate the 
mean $J-H$ and $H-K_s$ colors of the LSB component (see Sect.
\ref{lsbcolors}).}
\label{ftol3}
\end{figure*}
\begin{figure}[!h]
\includegraphics[width=8.5cm,angle=0,clip=]{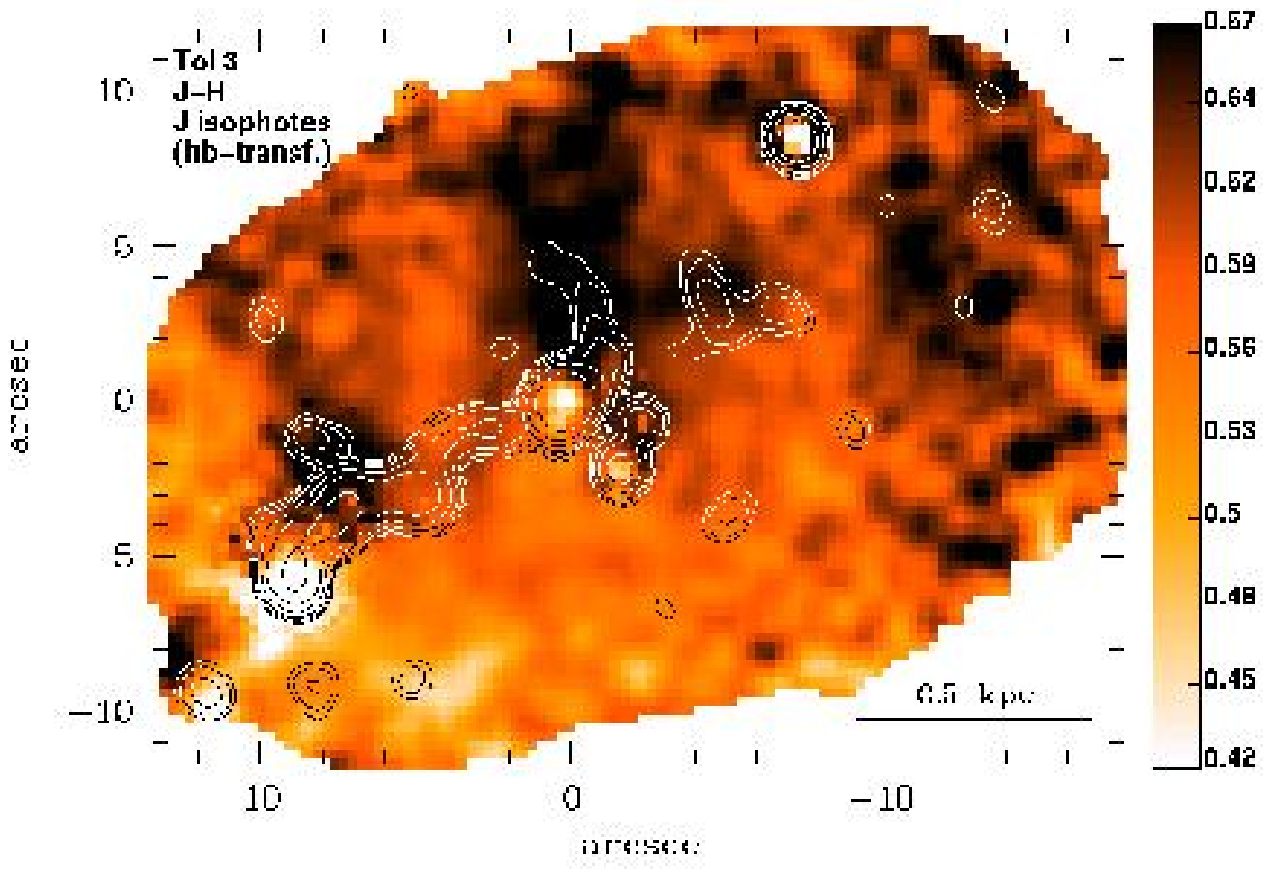}
\hspace*{0.7mm}\includegraphics[width=8.34cm,angle=0,clip=]{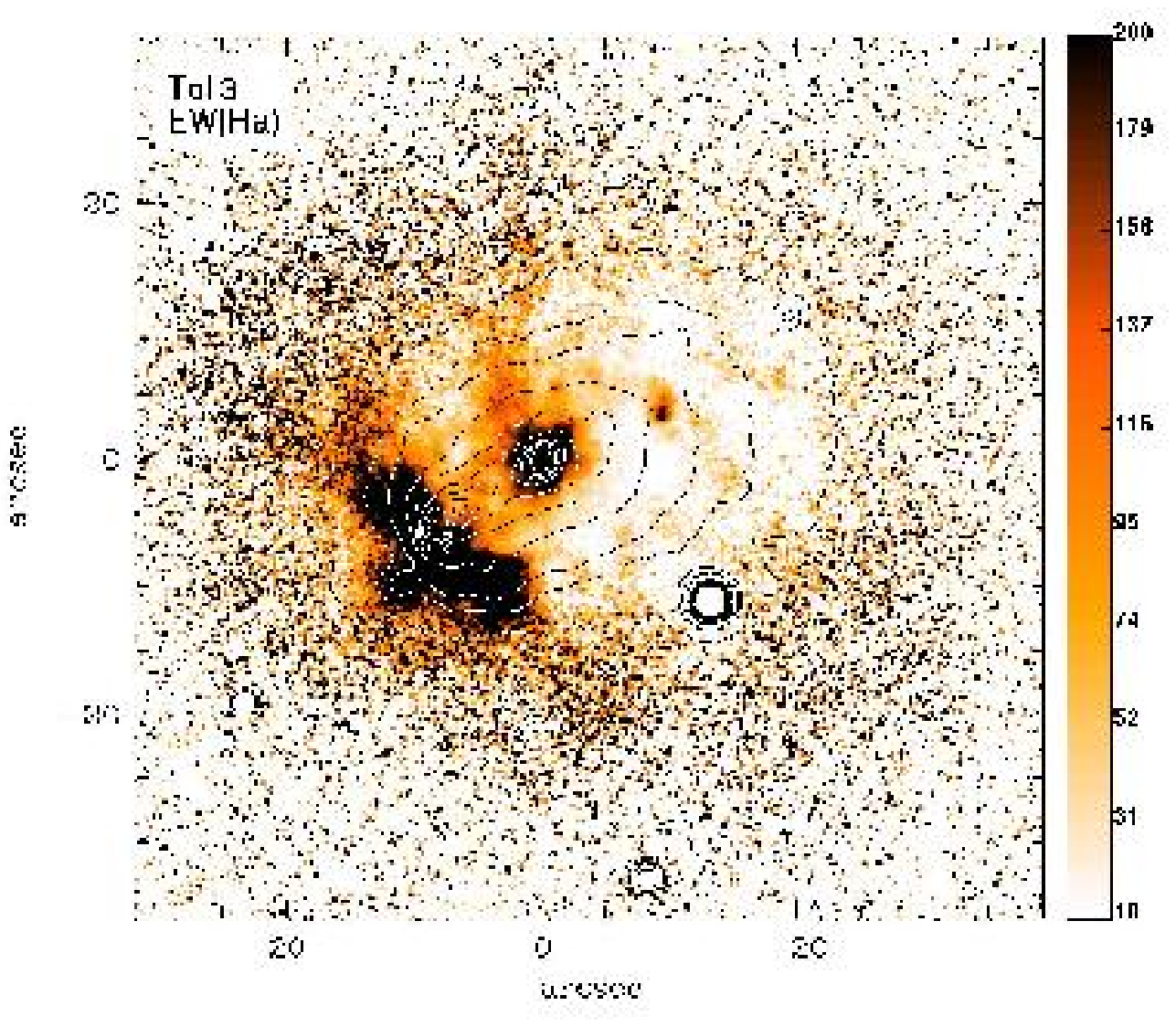}
\caption[]{{\bf top:} $J-H$ color map of the central region of Tol 3,
corrected for Galactic extinction.  The overlaid $J$ contours are
computed from the contrast--enhanced blow-up included in
Fig. \ref{ftol3} (see text for details).  {\bf bottom:} EW(\ha) map of
Tol 3, overlaid with $R$ contours. }
\label{ftol3b}
\end{figure}
%
This luminous ($M_B \sim$ --18.0; Marlowe et al. \cite{marlowe99}) and
relatively metal-rich (\zsun/6\dots \zsun/3; Kobulnicky et
al. \cite{kobulnicky99}, Schaerer et al. \cite{schaerer99}, Marlowe et
al. \cite{marlowe99} and references therein) BCD is known to be a member
of the NGC 3175 galaxy group (Garc\'{\i}a \cite{garcia93}).
NIR and optical images reveal two compact ($\la$ 0.3 kpc) 
high-surface brightness (HSB) regions: 
the brighter northwestern knot A, roughly coinciding with 
the geometrical center of the smooth LSB host galaxy, and the fainter
knot B, located $\sim$1 kpc southeast of A (see Fig. \ref{ftol3}a). 
Either knot is the locus of ongoing star formation, as witnessed by 
the detection of Wolf-Rayet features (Kunth \& Sargent \cite{kunth81}, 
Vacca \& Conti \cite{VC92}, Schaerer et al. \cite{schaerer99}) 
and red supergiants from CO absorption studies (Campbell \& Terlevich \cite{campbell84}). 
The intense SF activity in Tol 3 is also reflected on relatively 
blue optical colors ($B-V$=0.24, $V-I$=0.28; Marlowe
et al. \cite{marlowe97}) in its nuclear region, as 
well as on copious \ha\ emission ($1.1\times10^{41}$ \ergsec; 
Marlowe et al. \cite{marlowe97}). 
Deep \ha\ imaging by Marlowe et al. (\cite{marlowe95}) revealed a
bipolar outflow roughly perpendicular to the major axis of
the BCD, extending out to $\sim$2.8 kpc from its nuclear region.

Unsharp-masked NIR images reveal a complex morphology in the 
HSB regime of the BCD (inset in Fig. \ref{ftol3}a).
Regions A and B are immersed in an extended ``S-shaped'' pattern,
$\sim$0.9 kpc in length, ending at its NW tip with a curved 
feature ({\sf N arc}).  There is some evidence for propagation
of SF activities, as both archival optical NTT and NIR data reveal 
signatures of a younger age and stronger ionized gas emission 
towards region B (cf. Fig. \ref{ftol3b}, bottom).
For region A we determine within a rectangular 4\arcsec$\times$4\arcsec\ 
aperture colors of $J-H$=0.61 ($J-H^{\ddag}$=0.56) and 
$H-K$=0.41 ($H-K^{\ddag}$=0.58). For knot B we infer 
a $J-H$=0.39 ($J-H^{\ddag}$=0.14) and $H-K$=0.48 ($H-K^{\ddag}$=0.71).
Such colors
suggest a younger stellar age together with an appreciable ionized gas 
contribution towards the SE part of the SF component. 
The latter is verified from the 
\ha\ equivalent width (EW) map in Fig. \ref{ftol3b} (bottom) 
(see also Gil de Paz et al. 2002) which shows that 
SE of knot B, and all over an extended rim perpendicular to the 
major axis of the BCD, the EW(\ha) rises to $>$200 \AA.  

Interestingly, unsharp-masked NIR and optical images reveal on
larger scales a chain of faint knots 
(depicted with crosses in the inset
of Fig. \ref{ftol3}a), arranged over $\sim$2.5 kpc SW of regions A
and B. Their typical $J$\bg\ magnitudes of $\sim$20.6\dots 19 mag
translate into absolute magnitudes of --10\dots --12 mag.  The nature
and formation history of this extended feature is intriguing. One
interpretation is that it delineates the approaching side of an oblate
star-forming shell triggered by the burst, or that it may be
associated with an inclined circumnuclear disk of $\sim$1 kpc in
radius. The first hypothesis is consistent with the findings by Alton
et al. (\cite{alton94}), who suggested from imaging-polarimetry the
presence of a large-scale bipolar reflection dust-nebula illuminated
by the central starburst region. This could provide effective
UV-shielding, thereby allowing for secondary SF activity.  The
systematically redder colors in the NE half of Tol 3
(Fig. \ref{ftol3b}, top) are also in line with the same hypothesis, if
they originate from the more strongly absorbed far side of the
galaxy.

SBPs in the NIR (Fig. \ref{ftol3}b) show in the radius range
22\arcsec$\leq R^*\leq$48\arcsec\ an exponential intensity fall-off
with a scale length $\alpha=0.52$ kpc. This value is close to the $B$
scale length of $\alpha=$0.48 kpc, inferred for the LSB component by
Marlowe et al. (\cite{marlowe97}) within $\la$30\arcsec. By contrast,
our surface photometry does not appear to be compatible to that of
Kunth et al. (\cite{kunth88}).  These authors show optical SBPs out to
a radius \rr=80\arcsec, by a factor of 1.6 larger than the study here
or in Doublier et al. (\cite{doublier99}) and up to 2.7 times larger
than in Marlowe et al. (\cite{marlowe97}).  
NIR SBPs show no evidence for a dominant $R^{1/4}$ profile in Tol 3 
(cf. Kunth et al. \cite{kunth88}, Doublier et al. \cite{doublier99}). 
A de Vaucouleurs profile is neither supported by the Sersic index 
$\eta=1.4$ we derive for the {\em entire} $J$ profile.

The NIR color profiles (Fig. \ref{ftol3}c) show minor gradients ($<$0.15 \cg) 
and level off to $J-H$=0.54 mag and $H-K$=0.12 mag for \rr $>$ 20\arcsec. 
These results are not compatible to Doublier et al. (\cite{doublier01}), who report for
\rr$>$10\arcsec\ a roughly linear $J-H$ color increase from $\sim$0.5 mag to $\sim$2.0 mag.
A good agreement is found with Vanzi et al. (\cite{vanzi02}), who derive for the
LSB component of Tol~3 colors of $J-H$=0.6 mag and $H-K$=0.25 mag.

\subsection{Haro 14 (NGC 244)}
\label{haro14}
\begin{figure*}[!ht]
\begin{picture}(18,10)
\put(0,0.1){{\psfig{figure=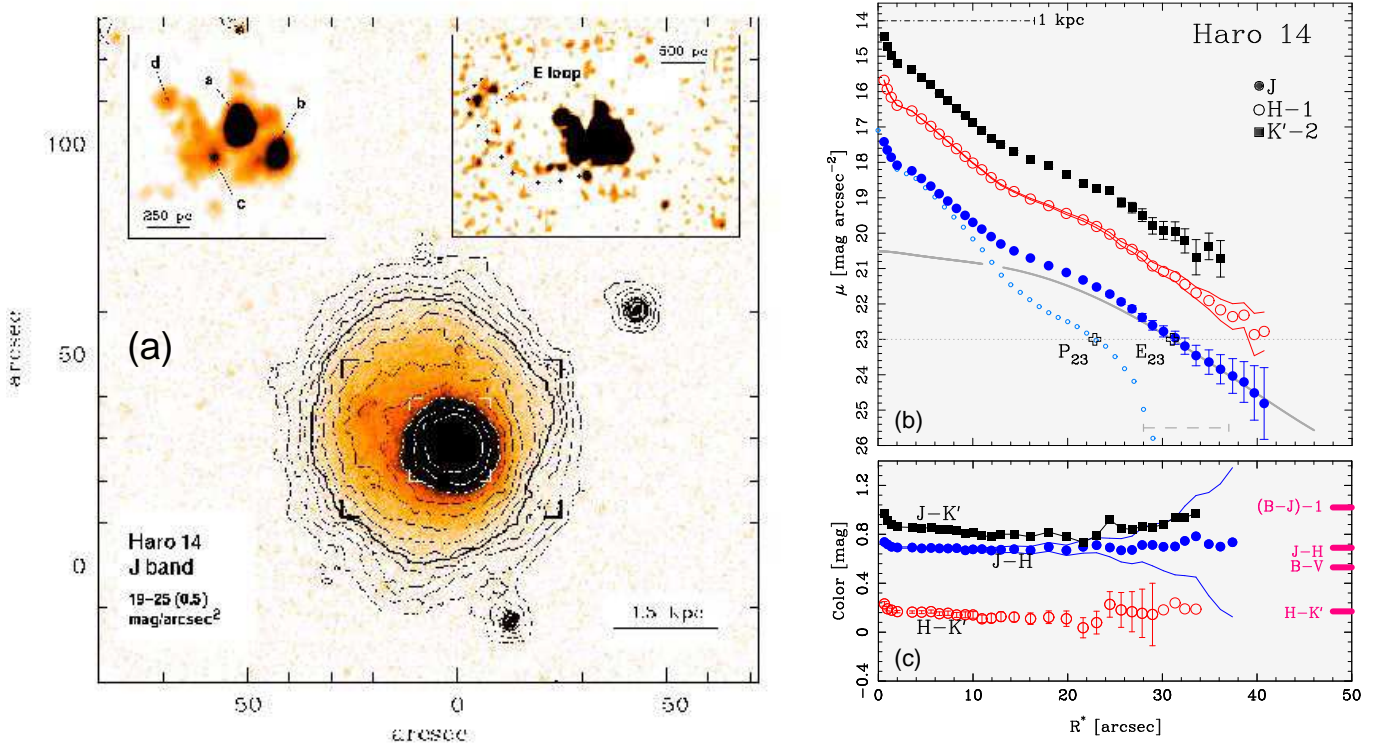,height=9.9cm,angle=0,clip=}}}
\put(10.9,3.92){{\psfig{figure=H4029F8.ps,width=7.0cm,angle=-90,clip=}}}
\put(10.95,0.19){{\psfig{figure=H4029F9.ps,width=7.05cm,angle=-90,clip=}}}
\put(1.6,5.2){\Large\sf (a)}
\put(11.8,4.28){\sf (b)}
\put(11.8,1.12){\sf (c)}
\end{picture}
\caption[]{Haro 14 ($D$=12.5 Mpc). For explanations of symbols and labels, refer to
Fig. \ref{ftol3}.  {\bf a):} $J$ band image and isophotes. Note that
the center of the star--forming regions is offset by $\sim$ 0.5 kpc
from the center of the outer circular isophotes.
The upper-left and upper-right insets show unsharp-masked versions
of the central part of the BCD, marked in the main image by the white 
and black brackets, respectively. In the upper-left inset we mark regions
{\sf a} through {\sf d}. The upper-right inset 
reveals an extended, curved chain of knots traceable up to $\sim$1.5 kpc 
from the brightest region {\sf a}.
{\bf b),c):} Surface brightness and color profiles. The thick grey
line shows a fit to the LSB component using a modified exponential
distribution Eq. (\ref{med}) with $b,q=3.6,0.94$\ .}
\label{fharo14}
\end{figure*}

This relatively metal--rich nE BCD (Z$\approx$\zsun/3, 
Hunter \& Hoffman \cite{hunter99}) 
shows SF activity ontop a smooth, nearly circular stellar LSB host galaxy
(Fig. \ref{fharo14}a). The intensity distribution of the latter
(Fig. \ref{fharo14}b) is approximated best by a modified
exponential distribution (Eq. \ref{med}), flattening for
\rr$\la$20\arcsec\ (see the detailed discussion in Sect. \ref{dis2}).

The central part of the BCD contains a massive complex of SF regions,
displaced by $\sim$0.5 kpc SW of the geometrical center of the outer
LSB isophotes. This SF component measures $\approx$1.4 kpc in diameter
in the $J$ band (this paper) and $\sim$2 kpc in the \ha\ line (Marlowe
et al. \cite{marlowe97}). Its composition out of several individual
regions, reported by Doublier et al. (\cite{doublier99}) from optical
images, is also evident from the contrast--enhanced NIR images
(Fig. \ref{fharo14}a, upper-left inset), which reveal a
wealth of individual regions, spanning a range of $\ga$3\,mag.
The upper-right inset of Fig. \ref{fharo14}a reveals a chain of faint
knots with a length of $\approx$ 1.5 kpc (labeled ``E loop''),
extending eastwards from the main SF complex. There is a hint for a 
similar feature in the southwestern direction.
Published data (e.g., the \ha\ image by Marlowe et
al. \cite{marlowe97}) do not allow to assess whether these
faint features may trace induced star formation 
along supergiant shells, as might be hypothesized from their morphology.

For the two brightest regions, denoted {\sf a} and {\sf b}, we obtain
respective absolute magnitudes of $M_{\rm J}$\bg=--14.7 mag and --13.3
mag, and effective radii $<$80 pc. 
If the mean $E(B-V)=$0.35 for Haro 14 (Hunter \& Hoffman \cite{hunter99}) 
applies to knots {\sf a} and {\sf b}, then their de-reddened colors would 
be $J-H$\bg $\sim$0.6 and $H-K_s$\bg $\sim$0.2. 
Such colors are reached at the earliest when the NIR emission becomes 
dominated by red supergiants (t$\sim$10$\dots$30 Myr),
and suggest no strong nebular line contamination.
This points against substantial ongoing star formation in regions {\sf a} and {\sf b}.

The large spatial extent of SF sources in the inner portion
of the BCD is further evidenced by the profile decomposition 
(Fig. \ref{fharo14}b), yielding a plateau radius \p23$\approx$1.4 kpc in the $J$ band.
The optical and NIR colors of the LSB host galaxy indicate
an age of several Gyr, in agreement with previous estimates 
by Marlowe et al. (\cite{marlowe99}).

\subsection{UM 461}                                                    
\label{um461}
\begin{figure*}[!ht]
\begin{picture}(18,10)
\put(.246,0.1){{\psfig{figure=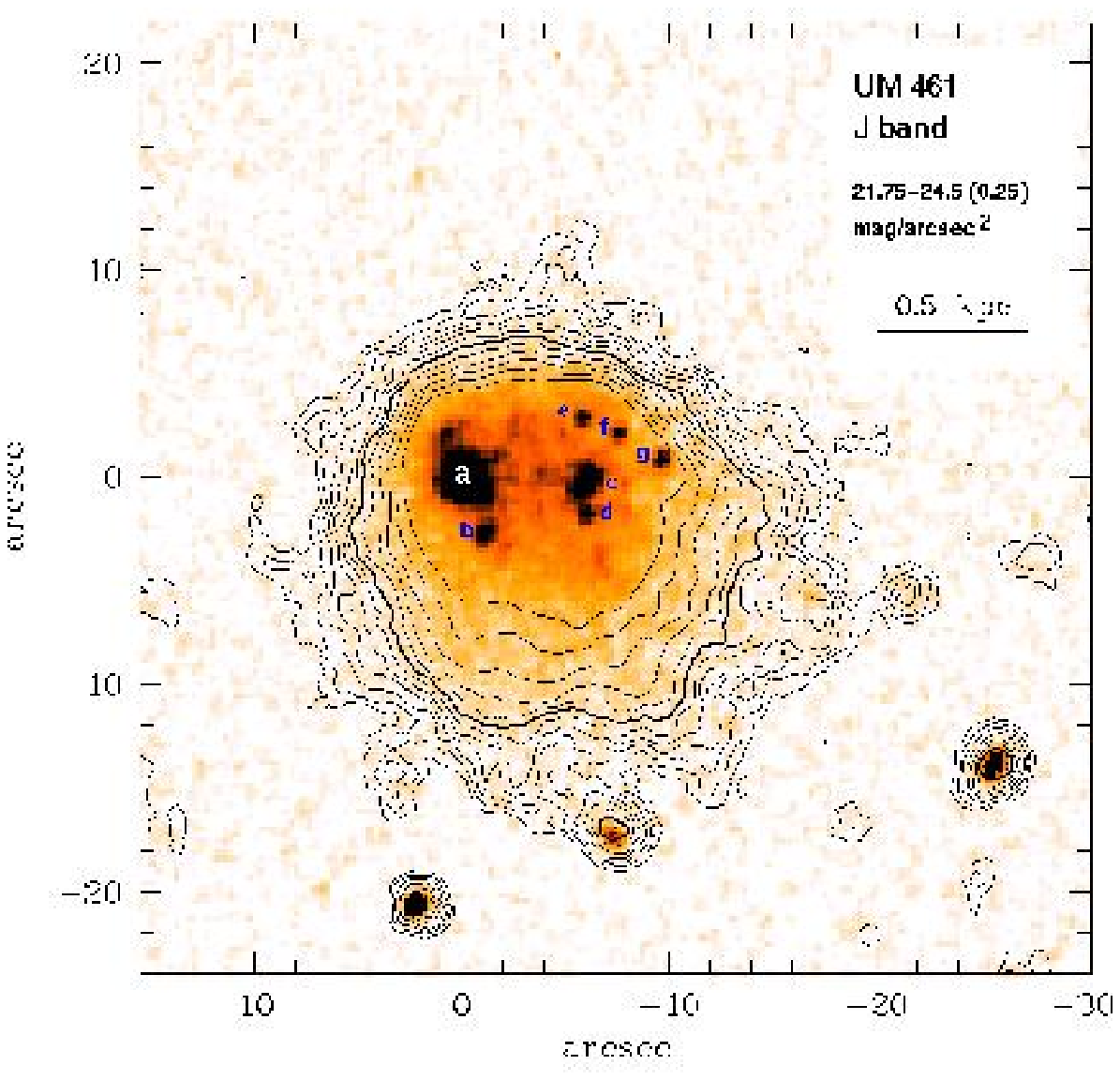,height=9.9cm,angle=0,clip=}}}
\put(10.9,3.99){{\psfig{figure=H4029F11.ps,width=7.1cm,angle=-90,clip=}}}
\put(10.95,0.21){{\psfig{figure=H4029F12.ps,width=7.12cm,angle=-90,clip=}}}
\put(1.7,1.5){\Large\sf (a)}
\put(11.8,4.3){\sf (b)}
\put(11.8,3.5){\sf (c)}
\end{picture}
   \caption[]{UM 461 ($D$=14.3 Mpc). 
For explanations of  symbols and labels, refer to Fig. \ref{ftol3}. 
{\bf a):} $J$ band image and isophotes. Bright stellar assemblies
in the central portion of the BCD are indicated.
{\bf b),c):} Surface brightness and color profiles. The thick grey
line shows a fit to the LSB host galaxy using Eq. (\ref{med}) with
parameters $b,q=2.3,0.85$.}
\label{fum461}
\end{figure*}

Telles \& Terlevich (\cite{telles95}) suggested that the iI BCD
UM 461 forms together with UM 463, UM 465 and UM 462 a 
loose group of dwarf galaxies. 

SF activity is confined to the whole northeastern part of the 
galaxy, i.e. within the 21.5 $J$ \sbb\ isophote, or on a spatial scale
of $\sim$0.9$\times$0.7 kpc. 
The two brightest SF regions {\sf a} (m$_J$\bg=17.3 mag; Fig. \ref{fum461}a) 
and {\sf c} (m$_J$\bg=19.5 mag) in UM 461 are separated by 20 \kmsec\ in the 
velocity space. This difference is of the order of the intrinsic \ion{H}{i} 
velocity dispersion within UM 461 ($\sim$30 \kmsec; van Zee et
al. \cite{vanzee98}).
The available data allow us to resolve 
a manifold of morphological features within the SF component, most 
notably a chain of compact sources northwest of region {\sf c} 
(labelled {\sf e}-{\sf g}) with a typical $M_J$\bg $\ga$ --10.4 mag. 
Their $J-H$\bg\ colors of 0.5-0.6 mag are consistent with 
the interpretation by M\'endez \& Esteban (\cite{mendez00}) that 
they are extinguished SF regions, formed not earlier than 100 Myr ago.
From profile decomposition we infer the summed up luminosity fraction of 
compact and diffuse SF sources to $\sim$37\% of the $J$ band light of UM 461.

UM 461 shows in its faint outskirts a slight asymmetry towards the SW direction. 
The intensity profile of the LSB host can be approximated by a \med\ model with 
a scale length $\alpha$=0.21 kpc and a depression parameter $q\approx$0.85 (Fig. \ref{fum461}b). 
The $J$ band scale length derived here is in excellent agreement with the value 
derived in the optical by Telles et al. (\cite{telles97}).
Color profiles reflect the ongoing SF for small radii
(\rr$\la$2\arcsec), and approach mean values of $J-H=$0.49 mag and
$H-K_s=$0.17 mag in the LSB periphery. Such colors, together with 
the $B-J\approx$1.9 mag determined from optical data, point 
consistently to a relatively evolved stellar LSB background. 
The NIR colors derived here do not appear to be compatible 
with those by Doublier et al. (\cite{doublier01}). 
These authors find within the radius range 
5\arcsec$\la$\rr$\la$10\arcsec\ the $J-H$ color to increase 
from 1.5 to $\sim$2.3 mag, whereas the $H-K$ color shows a continuous 
decrease from --0.5 mag to $<$--1 mag.
Also, the integrated $J-H$ and $H-K$ colors of 0.99 mag and --0.68 mag,
respectively, listed in Doublier et al. differ significantly
from the values of 0.47 mag and 0.2 mag derived in the present study.

\subsection{Henize 2-10 (ESO 495-G21)}                                                  
\label{he210}
\begin{figure*}[!ht]
\begin{picture}(18,10)
\put(0,0.1){{\psfig{figure=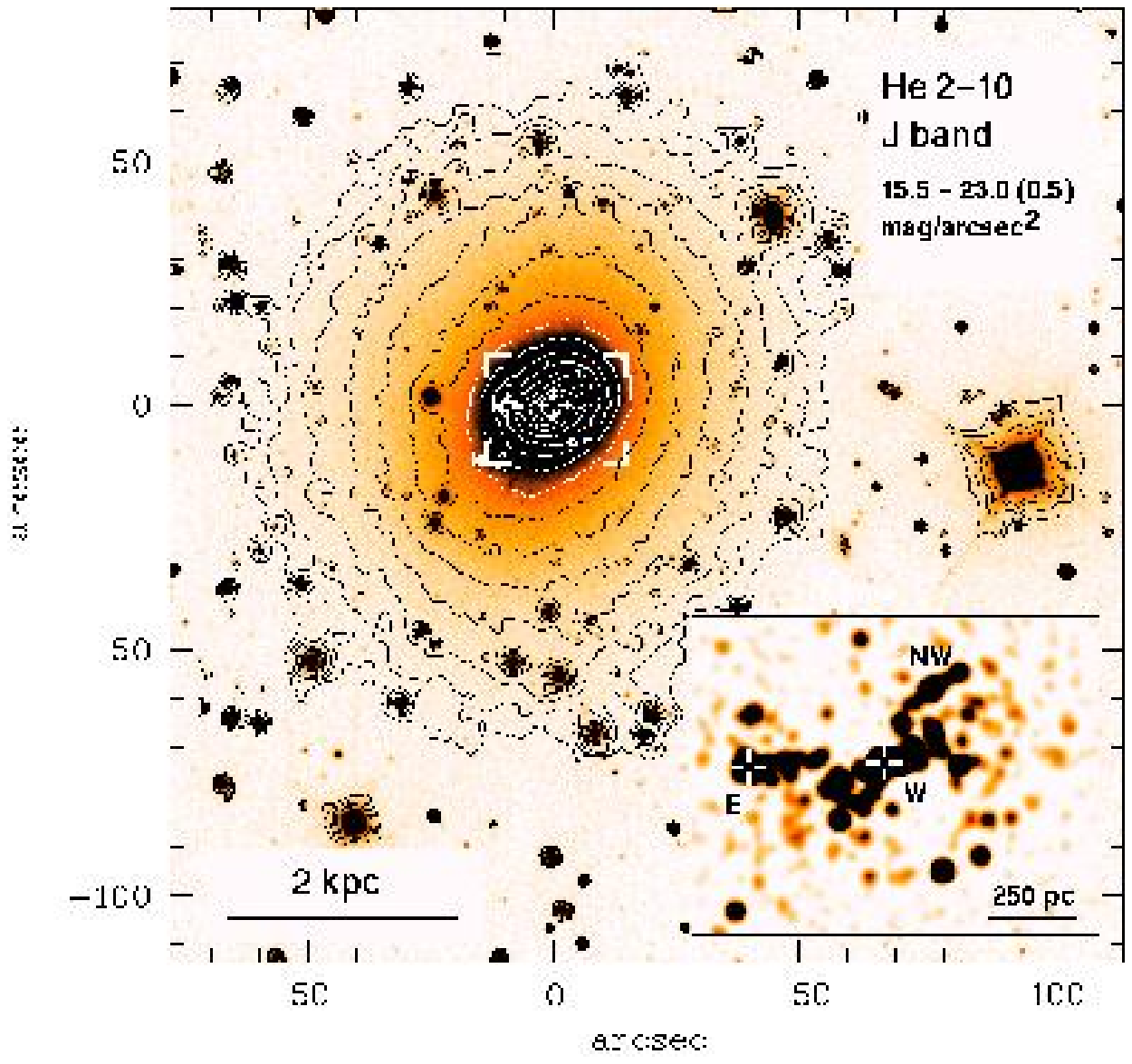,height=9.9cm,angle=0,clip=}}}
\put(10.9,4.01){{\psfig{figure=H4029F14.ps,width=7.1cm,angle=-90,clip=}}}
\put(10.95,0.21){{\psfig{figure=H4029F15.ps,width=7.12cm,angle=-90.,clip=}}}
\put(2.3,9.3){\Large\sf (a)}
\put(11.75,4.25){\sf (b)}
\put(11.75,1.15){\sf (c)}
\end{picture}
\caption[]{Henize 2--10 ($D$=8.7 Mpc). 
For explanations of  symbols and labels, refer to Fig. \ref{ftol3}.
{\bf a):} $J$ band image and isophotes. The inset shows an unsharp-masked
version of the nuclear region of the BCD (region marked by the brackets in the main image).
{\bf b),c):} Surface brightness and color profiles.}
\label{fhe210}
\end{figure*}

This relatively metal-rich iE BCD (\zsun/2.4\dots $\ga$\zsun; Schaerer et
al. \cite{schaerer99}, Kobulnicky et al. \cite{kobulnicky99}) is the
first extragalactic system in which the broad
\ion{He}{II}$\lambda$4686 line was detected. 
The starburst nature of He 2-10 is evidenced by a chain of bright
Super-Star Clusters (SSCs) in its brightest western SF region W 
(Conti \& Vacca \cite{conti94}), extended X-ray and \ha\ emission (Hensler et
al. \cite{hensler97}, Papaderos \& Fricke \cite{papaderos98}), as well
as a large bipolar outflow from the SF region (Papaderos \& Fricke
\cite{papaderos98}), with an expansion velocity between $\sim$250 and 
$\ga$360 \kmsec (M\'endez \& Esteban \cite{mendez99}, Johnson et
al. \cite{johnson00}).

Unsharp masking (inset in Fig. \ref{fhe210}a) reveals a
wealth of morphological features in the nuclear region of He 2-10,
most notably an extended feature protruding northwest of region W (labelled
NW) and an arc-like chain of compact sources connecting the tip of
region NW with the secondary SF knot E. Ground-based $B-R$ maps by
Papaderos \& Fricke (\cite{papaderos98}) indicate that region NW and
the concatenation of sources bending southwards of it are considerably
bluer than the underlying LSB host galaxy.

Our SBPs (Fig. \ref{fhe210}b) show an exponential intensity decrease 
in the LSB component (\rr$\ga$20\arcsec), with a $J$ scale length
identical to that obtained previously from optical data ($\alpha=0.67$
kpc, Papaderos \& Fricke \cite{papaderos98}).
\subsection{Tol 1400--411 (NGC 5408)}
\label{tol1400}
\begin{figure*}[!ht]
\begin{picture}(18,10)
\put(0.21,0.1){{\psfig{figure=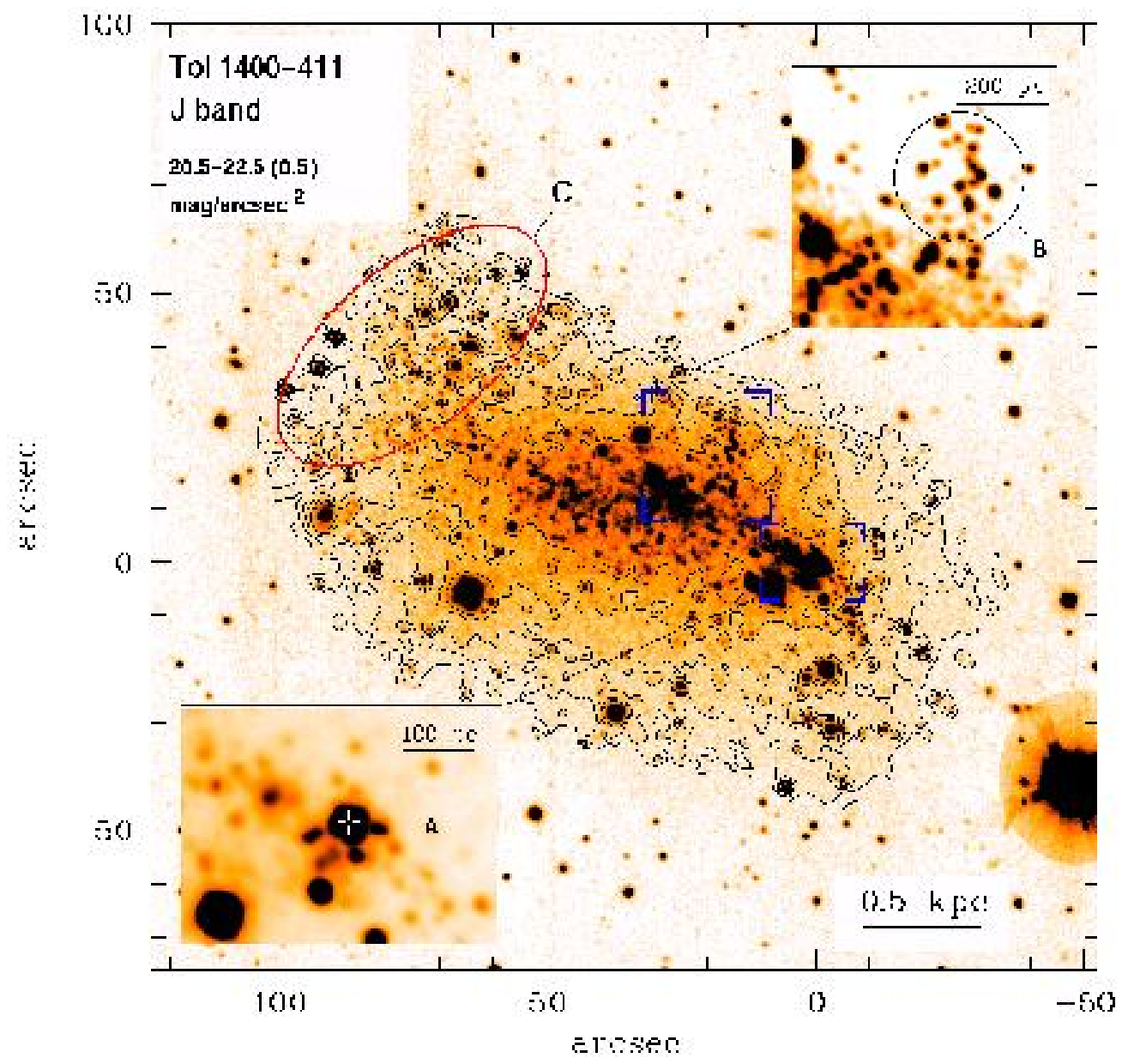,height=10cm,angle=0,clip=}}}
\put(10.9,3.99){{\psfig{figure=H4029F17.ps,width=7.08cm,angle=-90,clip=}}}
\put(10.94,0.17){{\psfig{figure=H4029F18.ps,width=7.10cm,angle=-90,clip=}}}
\put(5.4,9.3){\Large\sf (a)}
\put(11.8,4.3){\sf (b)}
\put(11.8,1.2){\sf (c)}
\end{picture}
\caption[]{Tol 1400--411 ($D$=4.8 Mpc). 
For explanations of symbols and labels, refer to Fig. \ref{ftol3}.
{\bf a):} $J$ band image and isophotes.  The insets show
magnifications of the bright SW star--forming knot {\sf A} as well as
the region {\sf B} extending northwestwards from the major axis.  The
bluer NE region {\sf C} is marked.
{\bf b),c):} Surface brightness and color profiles. The thick grey
line shows a fit to the host galaxy using Eq. (\ref{med}) with
$b,q=3.0,0.82$.}
\label{ftol1400}
\end{figure*}

This cometary iI BCD is a member of the Cen A group (van den Bergh
\cite{vandenbergh00}), situated at a projected distance of 7.2\degr\
($\sim$ 0.6 Mpc) from Cen A.  Intense SF activity is taking place mainly
at the SW part of its elongated stellar LSB host, where ground--based NIR
and archival \hst/WFPC2 optical data (PI: P. Seitzer, G0-8601) reveal
three bright (m$_V\la$16.2) stellar clusters spread over 8\arcsec\
($\sim$200 pc). The appreciable ionized gas contribution in the SF component
of Tol 1400-411 is reflected on the very blue $V-I$ color of $\la$--0.4 mag,
in the brightest cluster, {\sf A} (cf. l.h.s. blow-up of
Fig. \ref{ftol1400}a) and the
large EW(H$\beta$)$\ga$250 \AA\ determined for this system 
by Masegosa et al. (\cite{masegosa94}).

Interferometric 21cm line studies (Fritz \cite{fritz00}) show that 
the BCD is immersed within a large, rotationally supported \ion{H}{i} cloud of
about 8$\times$16 kpc in size. The \ion{H}{i} halo reveals two surface
density maxima, one coinciding with region {\sf A} and the other one
located $\sim$1\arcmin\ to the east.

The present data, as well as \ha\ images by Fritz (\cite{fritz00}),
reveal signatures of low-level SF activity all over the 22 $J$ \sbb\
isophote of the BCD, i.e 2.6 kpc across.  Interestingly, 30\arcsec\ NE
of region {\sf A} our data show a nearly circular region,
$\approx12$\arcsec\ in diameter, with relatively blue ($B-R$ of
$\la$0.75, $V-I\la$0.45 mag) colors in its {\sl unresolved} interior 
(denoted {\sf B} in Fig. \ref{ftol1400}a). Such colors, being significantly 
bluer than those in the LSB host ($B-R\approx$1, $V-I$=0.6--0.7), suggest 
that region {\sf B} is comparatively young. This is also suggested by the
NIR colors ($J-K$=0.7\dots 0.9) and luminosities ($M_J$\bg =--8.6\dots--9.7) 
of bright sources therein which are consistent with a population of young 
(10--20 Myr) red supergiants (cf. e.g. Elias et al. \cite{elias85}, 
Bertelli et al. \cite{bertelli94}).

Another conspicuous feature seen at the northeastern part of the BCD
is a comparatively blue ($B-R\approx 0.6$) curved strip 
(region {\sf C}), apparently bending from the NE tip of the 
LSB component to the north; a counter-feature of this region 
is probably present at the southern part of the BCD.

The large extent of the SF component, as well as several
bright foreground stars in the periphery of Tol 1400-411, 
render the determination of the structural properties and 
color of its LSB component difficult.
The SBPs (Fig. \ref{ftol1400}b) are approximated
best with Eq. (\ref{med}) with ($b$,$q$)=(3.0,0.8).
The mean LSB colors, $J-H$=0.42, $H-K_s$=0.28 and $V-I$=0.6--0.7
can be brought into rough agreement, given the uncertainties
discussed in Sect. \ref{lsbcolors}. 
The GALEV model (see Sect. \ref{dis4}) yields for a metal--poor stellar population
forming in a single burst or continuously an age between several
$10^8$ to a few $10^9$\,yr (cf. Fig. \ref{fig_lsbcolors}).  
However, these colors might be influenced by ionized gas emission 
and the younger stellar populations in the NE boundary.

\subsection{Pox 4}
\label{pox4}

\begin{figure*}[!ht]
\begin{picture}(18,21.7)
\put(0,11.7){{\psfig{figure=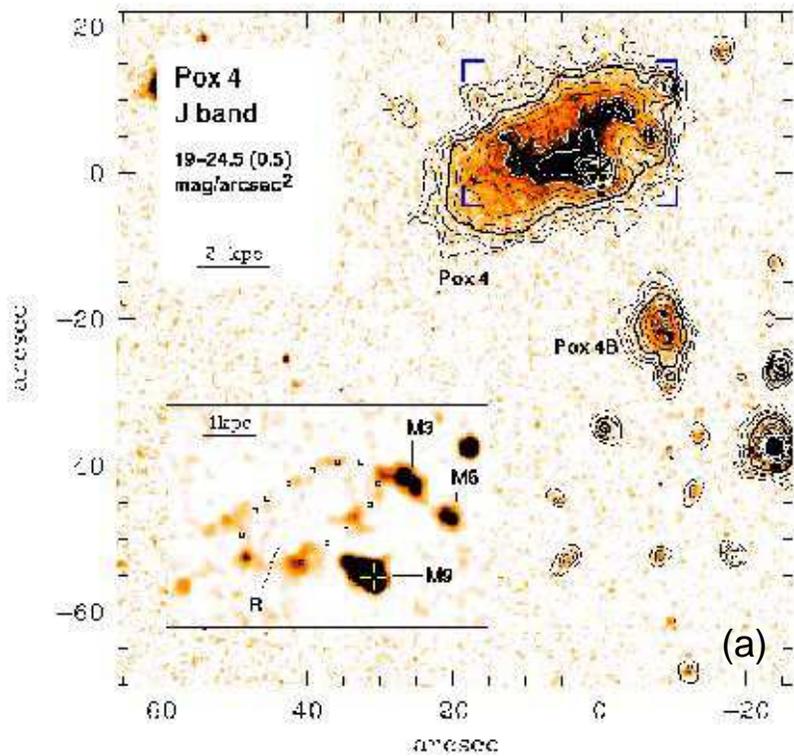,height=10cm,angle=0,clip=}}}
\put(0.7,4.2){{\psfig{figure=H4029F20.ps,width=7.8cm,angle=-90,clip=}}}
\put(0.75,0){{\psfig{figure=H4029F21.ps,width=7.8cm,angle=-90,clip=}}}
\put(9.7,4.2){{\psfig{figure=H4029F22.ps,width=7.8cm,angle=-90,clip=}}}
\put(9.75,0){{\psfig{figure=H4029F23.ps,width=7.8cm,angle=-90,clip=}}}
\put(9.5,13.0){\Large\sf (a)}
\put(1.7,4.6){\sf (b)}
\put(1.7,1.1){\sf (c)}
\put(10.7,4.6){\sf (d)}
\put(10.7,1.1){\sf (e)}
\PutWin{11}{18.1}{6.4cm}{
\caption[]{\label{fpox4}Pox 4 and its companion Pox
4B ($D$=46.7 Mpc). For explanations of symbols and labels, refer to Fig. \ref{ftol3}.
{\bf a):} $J$ band image and isophotes.  The inset shows an unsharp-masked
close--up of the SF regions in Pox 4 (area within
brackets in the main image). M3, M6 and M9 mark the brightest
\ion{H}{ii} regions referred to as 3, 6 and 9 by M\'endez \& Esteban 
(\cite{mendez99}). The ellipse of small boxes marks the ring of
SF regions described by the same authors. 
In our data, M9 resolves into 3 knots arranged over a projected 
length of 3\farcs2 (750 pc).
{\bf b)-e):} Surface brightness and color profiles of Pox 4 ({\bf
~b),c)~}) and Pox 4B ({\bf ~d),e)~}).  The thick grey lines show
decomposition fits to the stellar LSB host galaxies by means of
modified exponential distributions (see Sect.
\ref{decomposition}), with $b,q=2.7,0.90$\ for Pox 4 and $b,q=2.0,0.80$\
for Pox 4B.}}
\end{picture}
\end{figure*}

Despite its moderate metal deficiency (\zsun/9.3\dots \zsun/7.6;
Kunth \& Sargent \cite{kunth83}, Vacca \& Conti \cite{VC92}),
Pox 4 is a relatively luminous (M$_B$=--18.81) iI BCG
(M\'endez \& Esteban \cite{mendez99}).
It is accompanied by a faint SF dwarf galaxy, Pox 4\,B, at a projected 
distance of 5 kpc and a velocity difference of 130 km\,s$^{-1}$. 
On our NIR images, we do not detect any emission in between 
Pox 4 and Pox 4\,B down to an approximative surface brightness
level of 24 $J$ \sbb (Fig. \ref{fpox4}a).  

The optical morphology of Pox 4 is dominated by SF regions and
extended ionized gas emission on a spatial scale of 5$\times$2 kpc
(cf. Figs. 1 and 2 in M\'endez \& Esteban \cite{mendez99}).  
A substantial color shift due to intense (\ew(\ha)$=$1410 \AA) nebular
line emission has been observed in the brightest assembly of SF
sources M9 (inset of Fig. \ref{fpox4}a) for which
M\'endez \& Esteban (\cite{mendez99}) report a blue (--0.8 mag)
$U-B$ together with an extremely red (+0.7 mag) $B-V$ color. 
The NIR colors of region M9 ($J-H$\bg=0.28, $H-K_s$\bg=0.47) are
also suggestive of strong ionized gas contamination. 
The latter is also reflected on the color profiles (Fig. \ref{fpox4}c), 
approaching for \rr$\la$3\arcsec\ (0.7 kpc) colors as extreme 
as $J-H$=0.2 mag together with $H-K_s$=0.6 mag.

The SF morphology in Pox 4 is intriguing.  Unsharp masking reveals eastwards
of region M9 a ring-like distribution of compact sources with a projected 
size of 3.2$\times$1.5 kpc and a position angle of $\sim$124\degr. 
A similar morphology has been reported from optical data by
M\'endez \& Esteban (\cite{mendez99}), who interpreted Pox 4 
as a Cartwheel-like galaxy, downscaled by 1-2 orders of magnitude.

The available deep NIR data allow us to detect at faint intensities
($\approx$22.5 $J$ \sbb) a smooth underlying LSB host galaxy.  Its
$J_{\rm LSB}$ profile can be approximated with Eq. (\ref{med}) and a
depression parameter $q$=0.9 (Fig. \ref{fpox4}b).  From profile
decomposition we infer the absolute $J$ magnitude of this stellar host
to --~18.2 mag, which translates to $M_B>$--17.2 mag for a
$B-J>$1. Thus, despite its high total luminosity, Pox 4 qualifies 
by the absolute magnitude of its LSB host as a dwarf galaxy.

Our SBPs show some similarity to the uncalibrated profiles 
in Telles et al. (\cite{telles97}) out to $R^*=15$\arcsec. 
However, the scale length derived here for 10\arcsec$\leq$\rr$\leq$16\arcsec\ 
for the LSB component ($\alpha$=3\farcs 8=0.86 kpc) is twice as
large as that in Telles et al. (\cite{telles97}).
\subsection{Pox 4\,B} 
\label{pox4b}
This intrinsically faint ($M_J=$--16.2 mag) galaxy has been 
suggested by M\'endez \& Esteban (\cite{mendez99}) to have 
triggered the starburst activity in Pox 4 through a face-on 
collision. As evidenced by faint \ha\ emission 
(M\'endez \& Esteban \cite{mendez99}), Pox 4\,B still maintains a 
mild SF activity. The latter is probably taking place in
three compact ($\sim$0\farcs 5) sources, discernible in the central
part of the galaxy. Pox 4 B shows little morphological distortions,
except for a slight eastward extension of its LSB component
(Fig. \ref{fpox4}a).

The SBPs (Fig. \ref{fpox4}d) of Pox 4\,B show for
2\arcsec$\leq$\rr$\leq$5\arcsec\ an exponential fall-off with an
$\alpha\approx$0.29 kpc, and a flattening for \rr$<$2\arcsec.  We have
verified that the latter intensity regime is not due to seeing. For
this purpose, we convolved artificial exponential 2D models with 
the observed $\alpha$ and extrapolated central surface brightness 
($\mu_{\rm E,0}\approx$19.5 $J$ \sbb) with a Gaussian kernel 
matching the PSF of the coadded images ($\approx 0\farcs 7$ FWHM). 
The resulting SBPs deviate only slightly from a pure
exponential, even when the FWHM is further degraded by a factor of
2, and can by no means reproduce the strong flattening we detect in
the SBPs of Pox 4B. In addition, a perfect exponential distribution
with the $\mu_{\rm E,0}$ and $\alpha$ quoted above would yield a 16\%
larger luminosity than the one actually measured for Pox 4B.

The available data do not allow us to study the NIR colors in spatial
detail. The mean $J-H$ and $H-K_s$ of respectively $\approx$0.4 and
$\approx$0.2 are comparable to those of the LSB component of Pox 4.

\subsection{Tol 65}
\label{tol65}
\begin{figure*}[!ht]
\begin{picture}(18,10)
\put(0,0.1){{\psfig{figure=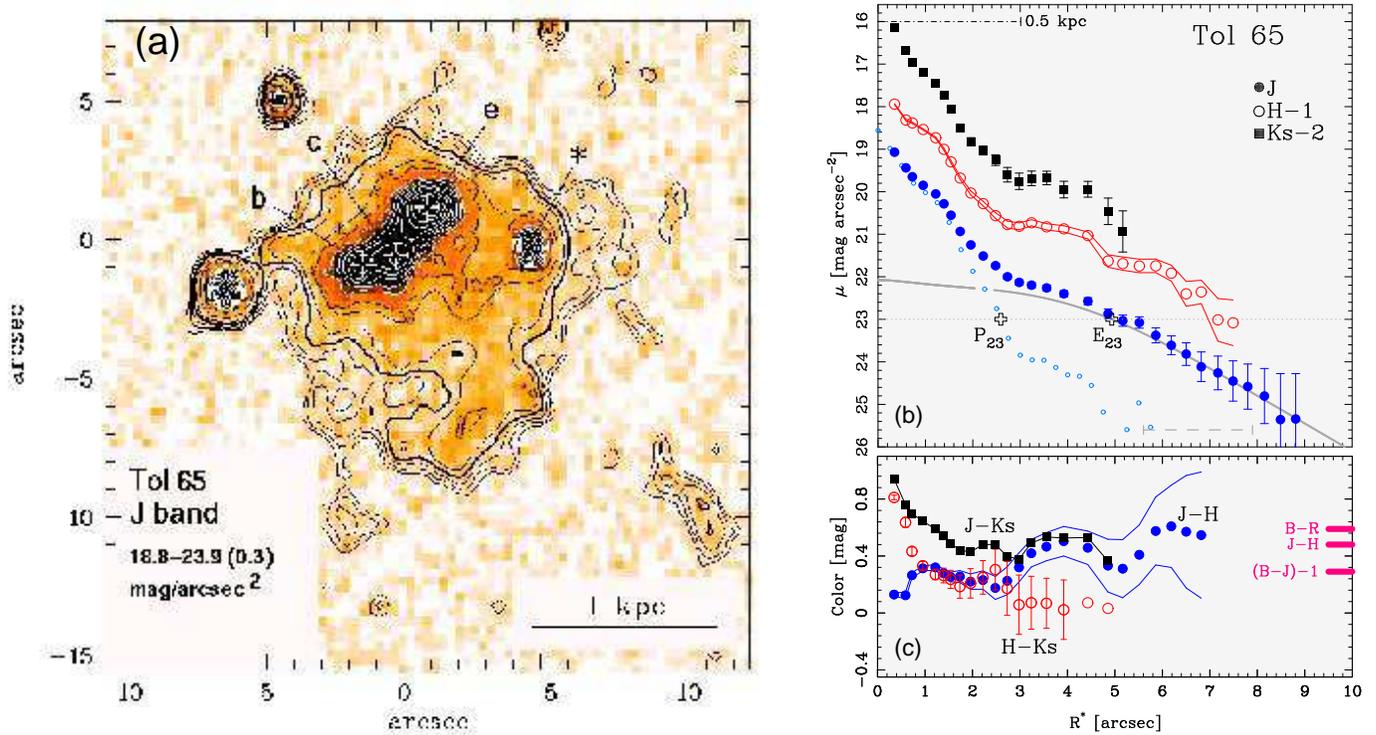,height=9.9cm,angle=0,clip=}}}
\put(10.9,3.85){{\psfig{figure=H4029F25.ps,width=7.0cm,angle=-90,clip=}}}
\put(10.94,.19){{\psfig{figure=H4029F26.ps,width=7.07cm,angle=-90,clip=}}}
\put(1.7,9.2){\Large\sf (a)}
\put(11.8,4.3){\sf (b)}
\put(11.8,1.2){\sf (c)}
\end{picture}
\caption[]{Tol 65 ($D$=34.2 Mpc). 
For explanations of symbols and labels, refer to Fig. \ref{ftol3}.
{\bf a):} $J$ band image and isophotes. The sources {\sf b, c and e}
along the NE--chain of SF regions are marked, following the
notation of \cite{papaderos99}. The feature marked by an asterisk
is a red ($J-K_s \approx 0.8$ mag) compact source probably not
belonging to the galaxy. The faint (m$_J \approx 20.8$ mag) galaxy {\sf
''G1''} (cf. \cite{papaderos99}) is visible to the SW.
{\bf b),c):} Surface brightness and color profiles. The thick grey
line shows a decomposition fit to the host galaxy by means of a
modified exponential distribution (see Sect. \ref{decomposition})
with $b,q=2.7,0.90$\ .}
\label{ftol65}
\end{figure*}

Star formation in this very metal-deficient BCD (\zsun/24; Kunth \&
Sargent \cite{kunth83}, Masegosa et al. \cite{masegosa94}, Izotov et
al. \cite{izotov01}) is taking place mainly in a chain of five compact
(1\farcs2$\la\diameter\la$2\arcsec) sources located at the NE part of
an irregular, blue LSB envelope (\cite{papaderos99}).
The combination of an extraordinarily blue $J-H$\bg\ ($\sim$0.1 mag)
with a red $H-K$\bg\ ($\approx$0.8 mag) color in the surroundings
of the brightest SF region {\sf e} (Fig. \ref{ftol65}a) points 
to a substantial contribution of ionized gas emission. 
This is also suggested by the combination of a blue $U-B$ ($\sim$--1.05 mag) 
with a moderately red $B-R$ ($\sim$0.4--0.5 mag) color derived by 
\cite{papaderos99} all over the NE half of Tol 65, 
several 100 pc away from the opposite tips of the SF chain. 
Further evidence for intense ionized gas emission is provided by Keck 
spectroscopy by Izotov et al. (\cite{izotov01}), who derived a 
large EW(\ha) of $>$1000 \AA\ in the SF component of the BCD.

Ionized gas contamination appears to be small midway in the SF 
chain. The $B-R$=0.15 mag (\cite{papaderos99}), 
together with $J-H$\bg\ and $H-K_s$\bg\ of 0.44 and 0.2 mag, respectively, 
observed in region {\sf c} are consistent with a single burst stellar 
age of $\la$100 Myr. Note that the $J-H$ color in region {\sf c} is 
barely bluer than the average value for the LSB component (0.48 mag).
This is also true for the $H-K_s$ color, which could be constrained 
at $R^*\ga$\p23\ to $\la$0.2 mag (cf. Fig. \ref{ftol65}c). 

Deep optical surface photometry ($\mu\ga$28.5 $B$ \sbb) for Tol 65 has first
been presented in \cite{papaderos99}. These authors found optical SBPs
to show an outer exponential regime for $R^*>$7\arcsec\ and a
flattening relative to the exponential fit inwards of
\rr$\approx$3\arcsec. In order to adequately decompose 
the optical SBPs, they modelled the LSB component with 
Eq. (\ref{med}) and a depression parameter $q\ga 0.8$.
The \flat\ profile of the LSB component is better visible
in NIR wavelengths, where an even stronger flattening is 
required ($q\approx$0.9) to fit the data. 
The $J$ SBP shows in the radius range $5\farcs0\leq R^{\star} \leq 7\farcs 3$ 
an exponential intensity fall-off with a scale length $\alpha=0.28$ kpc, 
in good agreement with the value derived in \cite{papaderos99}
($\alpha=0.26$ kpc).
The average LSB colors of $J-H$=0.48$\pm$0.17, together with 
the a $B-R \la 0.6$ mag and $B-J\la $1.3 mag, inferable from 
the SBPs in \cite{papaderos99}, are consistent with a stellar age
$\la$1 Gyr, assuming an instantaneous SF process 
(cf. Sect. \ref{lsbcolors}).
\subsection{Tol 1214-277}                                              
\label{tol1214}
\begin{figure*}[!t]
\begin{picture}(18,10.4)
\put(0,3.4){{\psfig{figure=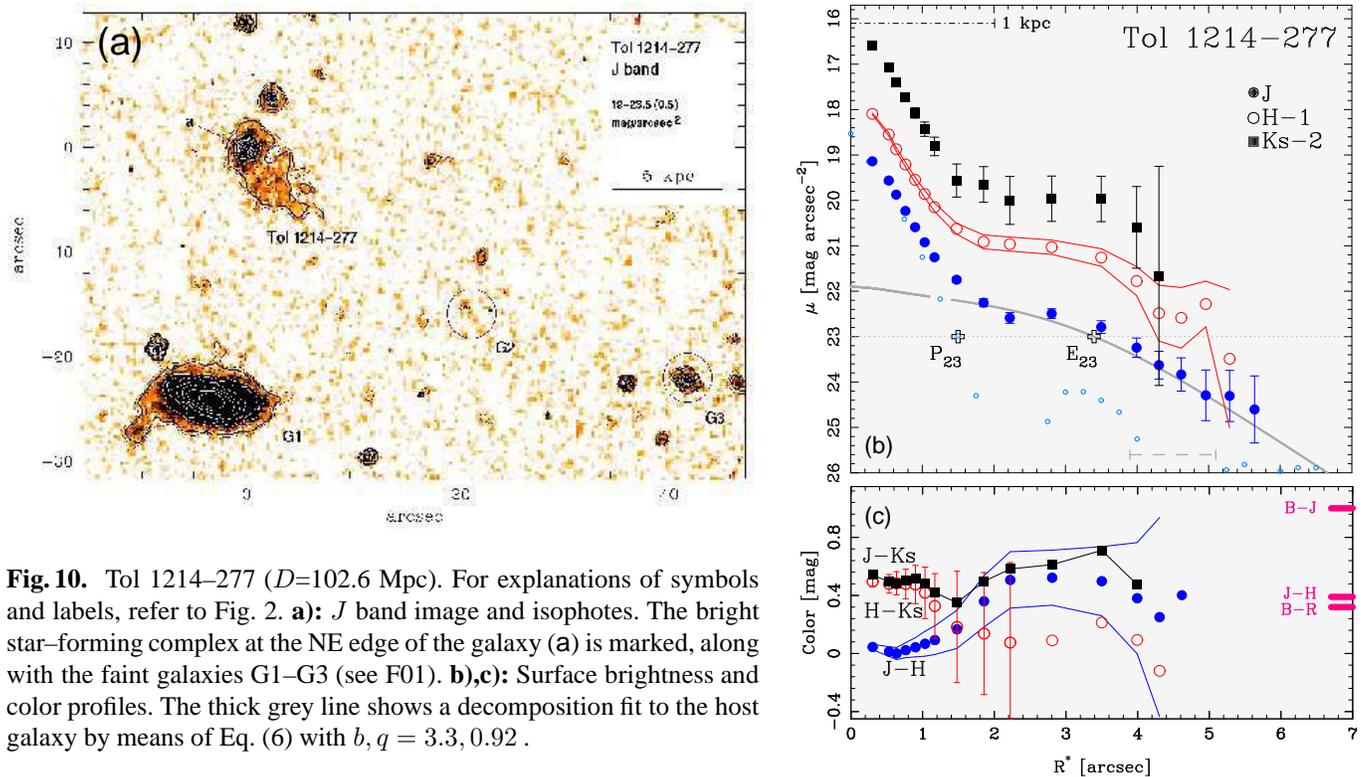,height=7.0cm,angle=0,clip=}}}
\put(10.5,4.){{\psfig{figure=H4029F28.ps,width=7.4cm,angle=-90,clip=}}}
\put(10.55,0.1){{\psfig{figure=H4029F29.ps,width=7.44cm,angle=-90,clip=}}}
\put(1.2,9.7){\Large\sf (a)}
\put(11.4,4.4){\sf (b)}
\put(11.4,3.45){\sf (c)}
\PutWin{0.0}{1.7}{10cm}{\caption[]{\label{ftol1214} 
Tol 1214--277 ($D$=102.6 Mpc). For explanations of symbols and labels, refer to
Fig. \ref{ftol3}.  {\bf a):} $J$ band image and isophotes. The bright
star--forming complex at the NE edge of the galaxy ({\sf a}) is
marked, along with the faint galaxies G1--G3 (see \cite{fricke01}).
{\bf b),c):} Surface brightness and color profiles. The thick grey
line shows a decomposition fit to the host galaxy by means of Eq. (\ref{med})
with $b,q=3.3,0.92$\ .}  }
\end{picture}
\end{figure*}

This metal-poor (\zsun/25; \cite{fricke01}, Izotov et
al. \cite{izotov01}) cometary iI BCD undergoes strong SF activity at
the NE end of an elongated stellar LSB body with an apparent size of
$\approx$6.9$\times$3.4 kpc. The brightest SF complex (labelled {\sf
a} in Fig. \ref{ftol1214}a) contributes nearly one half of the BCD's
optical light within the 25 $B$ \sbb\ isophote (\cite{fricke01}).  The
$J$\bg\ magnitude determined for region {\sf a}, 18.6 mag, shows that
in the NIR the starburst still provides $\sim$1/3 of the galaxy's
total flux.  The colors of this region ($J-H$\bg=--0.02 mag and 
$H-K_s$\bg=0.51 mag) are consistent with a young burst age of $<$7 Myr
for the BCD (see also \cite{fricke01}).

Only the $J$ band SBP of Tol 1214-277 could be observed with a
sufficient quality for a decomposition (Fig. \ref{ftol1214}b). 
Similar to Tol 65, it shows a \flat\ distribution, as already reported from
deep optical VLT FORS\,I data by \cite{fricke01}. 
Fitting Eq. (\ref{med}) to the $J_{\rm LSB}$ SBP, with the ($b$,$q$) parameters
fixed to the values derived in \cite{fricke01} (3.3,0.92), we obtain a scale length
$\alpha$=0.53 kpc, slightly larger than that in the optical 
($\alpha=$0.49 kpc; \cite{fricke01}).

Fricke et al. (2001) reported for Tol 1214-277 nearly constant $U-B$ and $B-R$
colors of respectively --0.42 mag and 0.34 mag over an intensity span of 8 mag (colors  
adapted to the galactic absorption assumed here).
Such colors, together with the $J-H\approx$0.4 mag and $B-J\approx$1.1 mag 
derived here, are slightly bluer that those of Tol 65, supporting the hypothesis 
that Tol 1214-277 is a relatively unevolved dwarf galaxy.

\subsection{Mkn 178 (UGC 6541)}
\label{mk178}
\begin{figure*}[!ht]
\begin{picture}(18,10)
\put(0,0.1){{\psfig{figure=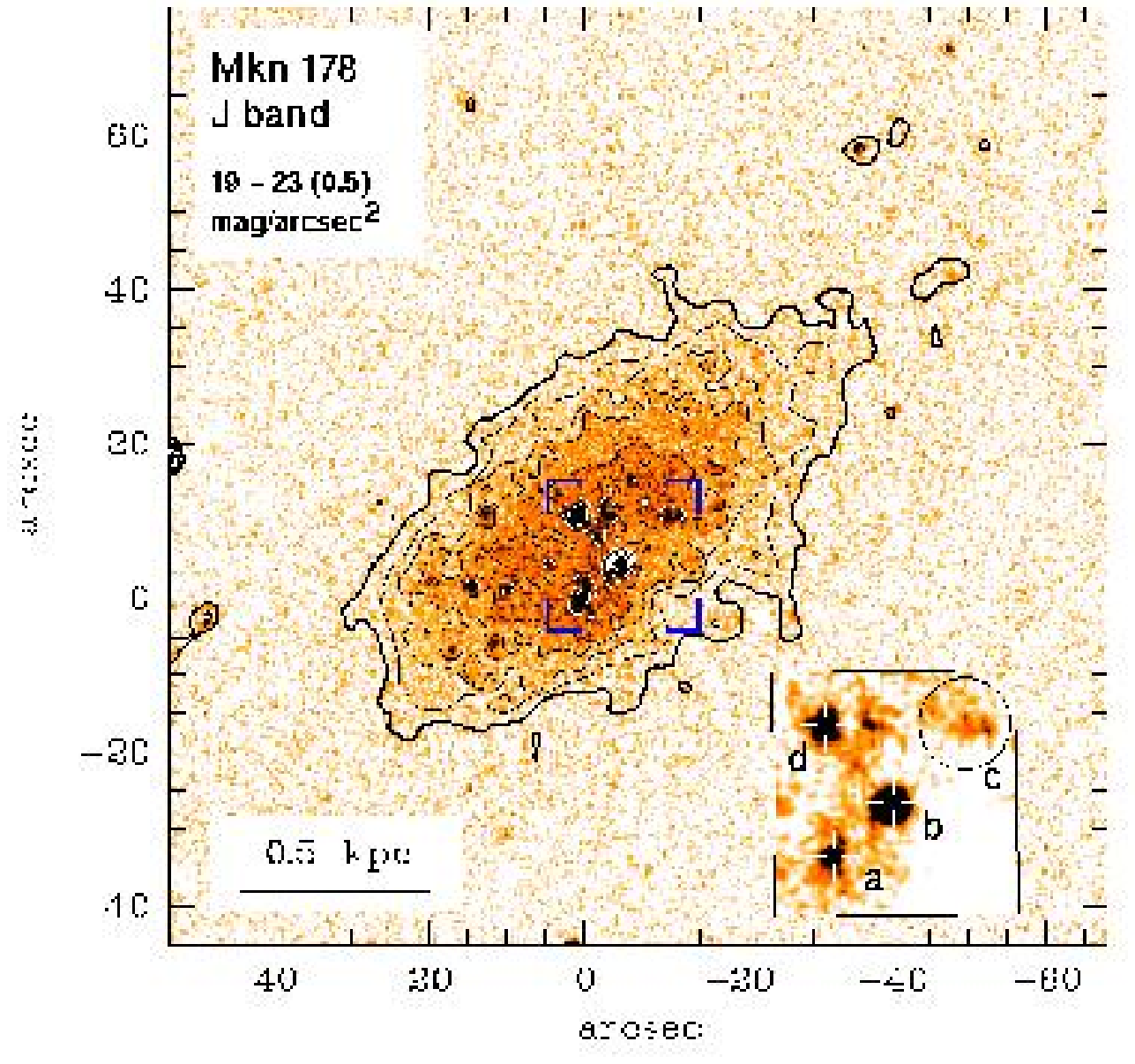,height=10cm,angle=0,clip=}}}
\put(10.9,4.09){{\psfig{figure=H4029F31.ps,width=7.02cm,angle=-90,clip=}}}
\put(10.95,0.29){{\psfig{figure=H4029F32.ps,width=7.02cm,angle=-90,clip=}}}
\put(9.4,9.2){\Large\sf (a)}
\put(11.8,4.3){\sf (b)}
\put(11.8,3.45){\sf (c)}
\end{picture}
\caption[]{Mkn 178 ($D$=4.2 Mpc). 
For explanations of symbols and labels, refer to Fig. \ref{ftol3}.
{\bf a):} $J$ band image and isophotes. The inset shows a close--up
of the star--forming regions (delimited by brackets in the main
image).  The regions {\sf a-d} described by Gonzalez-Riestra et al.
(\cite{gonzalez88}) and Papaderos et al. (\cite{papaderos02}) are
labeled.  {\bf b),c):} Surface brightness and color profiles. }
\label{fmk178}
\end{figure*}

The spectroscopic properties of this intrinsically faint ($M_B$=--13.9
mag) iE BCD, a member of the CVn cloud I of galaxies (Makarova et
al. \cite{makarova98}), have been investigated in
e.g. Gonz\'alez-Riestra et al. (\cite{gonzalez88}) and Guseva et
al. (\cite{guseva00}).
Optical images reveal a complex morphology in the SF component,
notably a pronounced separation of HSB regions into a compact assembly
of bright knots to the SE (regions {\sf a} and {\sf b}) and an
extended arc-like segregation of fainter sources to the NE. NIR
images, on the contrary, show a more regular morphology, with only a
few prominent sources in the central part of Mkn 178 (see
Fig.\ref{fmk178}a).
Region {\sf a} ($m_J$\bg=17.5 mag; cf. inset in Fig. \ref{fmk178}a)
coincides with the optically brightest region in the BCD and is the
main locus of active star formation. The brightest NIR source, {\sf b}
($m_J$\bg=16.5 mag), is optically faint and shows nearly the same $B-R$
color as the surrounding diffuse emission within the plateau component
(cf. Papaderos et al. \cite{papaderos02}).
Sources {\sf c} and {\sf d} are immersed within the northern 
featureless region. The non-detection of an optical counterpart for
region {\sf d}, and the overall optical/NIR morphology of the SF component
are suggestive of inhomogeneous, large-scale dust absorption on
a spatial scale of $\sim$1 kpc. This may cause the apparent 
separation of the SF component in two large detached complexes 
and hide knot {\sf d} in optical wavelengths. 

Gonz\'alez-Riestra et al. (\cite{gonzalez88}) and Guseva et
al. (\cite{guseva00}) have shown ionized gas emission to be negligible
in region {\sf b} (EW(H$\beta$)=24\dots 34 \AA).  By the color
excess of E($B-V$)=0.25 given in Gonz\'alez-Riestra et
al. (\cite{gonzalez88}), its observed $B-R$ color (0.67 mag, Papaderos
et al. \cite{papaderos02}) transforms to 0.28 mag, suggesting a
single-burst age of $\la$100 Myr.  This is consistent with the colors
of $J-H$\bg=0.24, $H-K'$\bg=0.06 and $B-J$\bg=1.6 mag obtained in the
present work, if they are de--reddened adopting the same amount of
intrinsic extinction.
The presence of stellar complexes with an age of the order of 
$\sim$10$^8$yr is in line with the detection of numerous 
luminous AGB stars in NIR color-magnitude diagrams 
(Schulte--Ladbeck et al. \cite{schulte00}), from which 
these authors infer significant SF activity over the 
last few 10$^8$ yr.

NIR SBPs of Mkn 178 (Fig. \ref{fmk178}b) resemble closely
the optical ones (Papaderos et al. \cite{papaderos02}), showing an
extended plateau component ontop a smooth, exponential LSB host
galaxy.  For the latter we derive a $J$ band scale length of 0.28 kpc
and colors ($J-H\approx$0.53, $H-K\arcmin \approx$0.2, $B-J\approx$2.0
and $B-R\approx$1.1 mag) that consistently indicate an old stellar LSB
background.

\subsection{Mkn 1329 (IC 3589/91, UGC 7790, VCC 1699)}                
\label{mk1329}
\begin{figure*}[!ht]
\begin{picture}(18,10)
\put(0.14,0.1){{\psfig{figure=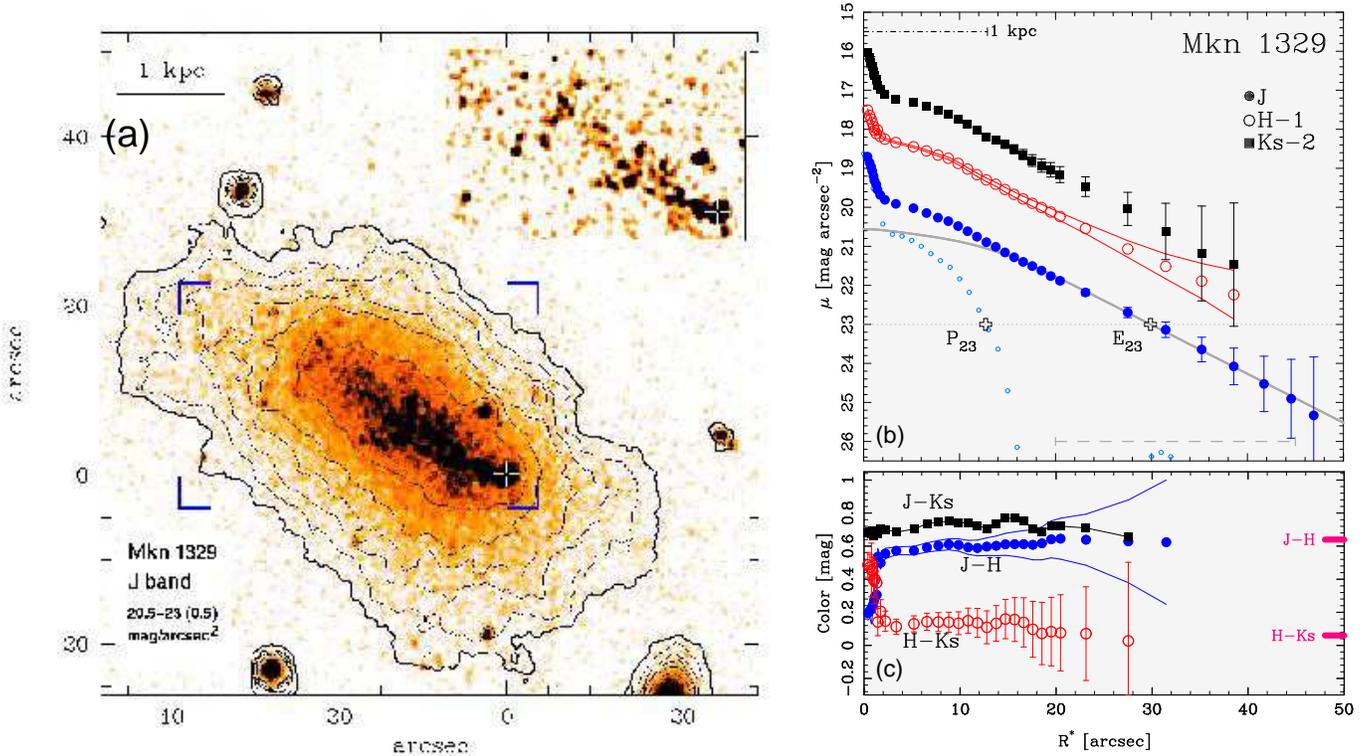,height=9.9cm,angle=0,clip=}}}
\put(10.9,4.01){{\psfig{figure=H4029F34.ps,width=7.1cm,angle=-90,clip=}}}
\put(10.95,.17){{\psfig{figure=H4029F35.ps,width=7.15cm,angle=-90,clip=}}}
\put(1.5,8.2){\Large\sf (a)}
\put(11.75,4.25){\sf (b)}
\put(11.75,1.15){\sf (c)}
\end{picture}
\caption[]{Mkn 1329 ($D$=16 Mpc).
For explanations of symbols and labels, refer to Fig. \ref{ftol3}.
{\bf a):} $J$ band image and isophotes.  {\sf inset:} {\em
hb-}transformed image of the central area (delimited by the brackets
in the main image), showing the distribution of the bright regions
along the major axis. The white cross marks the bright SW star--forming
region for orientation.
{\bf b),c):} Surface brightness and color profiles. The thick grey
line shows a decomposition fit to the host galaxy by means of a
modified exponential distribution, Eq. (\ref{med}),
with $b,q=1.6,0.70$\ .}
\label{fmk1329}
\end{figure*}

This relatively metal-rich (\zsun/5, Guseva et al. \cite{guseva00}) 
cometary iI BCD (M$_{\rm B}\approx$ --16.8, Yasuda et al. \cite{yasuda97})
is associated with a group of 11 galaxies (LGG 296, Garc\'{\i}a \cite{garcia93}) 
within the Virgo Cluster. 

As shown in the contrast-enhanced image (inset in
Fig. \ref{fmk1329}a), numerous irregular concentrations are present in
the inner zone of its elongated stellar LSB component, over a
projected length of $\sim$3.5 kpc.  An H$\alpha$ exposure by
Gallagher \& Hunter (\cite{gallagher89}) shows that SF activity is
almost exclusively confined to a bright \ion{H}{ii} region (marked
with a cross in Fig. \ref{fmk1329}a) at the SW end of the
high-surface brightness component. The \ha\ flux derived 
for that SF region by the latter authors yields, at the
distance adopted here, an H$\alpha$ luminosity of
$\approx$1.5$\times$10$^{40}$ erg\,s$^{-1}$ and a star formation rate
(SFR) of $\sim$ 0.1 M$_{\sun}$ yr$^{-1}$.  The dominant SF region in
Mkn 1329 is identifiable with a compact ($\diameter \la$ 300 pc),
luminous ($M_J$ = --14) NIR source; its colors, ($J-H$)\bg =0.16 and
($H-K_s$)\bg=0.51, are probably significantly influenced by nebular
emission (EW(H$\beta$)=276 \AA, Guseva et al. \cite{guseva00}).

The morphology of the SF regions and the high luminosity of the SW
complex in Mkn 1329 are typical among cometary iI BCDs. The similarity 
of these systems with respect to the morphology of their LSB and SF component
suggests a comparable evolutionary state/history or common mode of star formation.
Propagation of SF activities along those objects' major axes has been
proposed and observationally supported by a number of recent studies
(cf. Noeske et al. \cite{noeske00} and references therein). 

The SBPs of Mkn 1329 (Fig. \ref{fmk1329}b) indicate a moderate central
flattening of the LSB component, which can be approximated both by a
S\'ersic law with $\eta_{\rm LSB}\approx 0.7$, or a \med\ with
$b,q$ = (1.6,0.7) (see Table \ref{tab_phot} and
Sect. \ref{decomposition}). The detection of a moderate central
flattening in the LSB component owes to the comparatively small
contribution of the superposed younger stellar population in Mkn 1329
($\approx$17\% of the total $J$ band light).
The NIR colors in the outskirts of Mkn 1329, ($J-H$)\bg =0.64, ($H-K_s$)\bg =0.06,
suggest an evolved stellar LSB component. Adopting a $B-J$ color characteristic
of a few Gyr old stellar population, we can estimate the $B$ band structural 
parameters from the $\mu _{\rm E,0}$, $\alpha$ and M$_{\rm LSB}$ determined
from $J$ SBPs. We find that the host galaxy of Mkn 1329 shows, similar to other 
cometary BCDs (Noeske et al. \cite{noeske00}), structural properties intermediate 
between extended dIs/dEs and compact iE/nE BCDs.

\subsection{IC4662 (ESO 102-G014, He 2--269)}
\label{ic4662}
\begin{figure*}[!ht]
\begin{picture}(18,18)
\put(0,9){{\psfig{figure=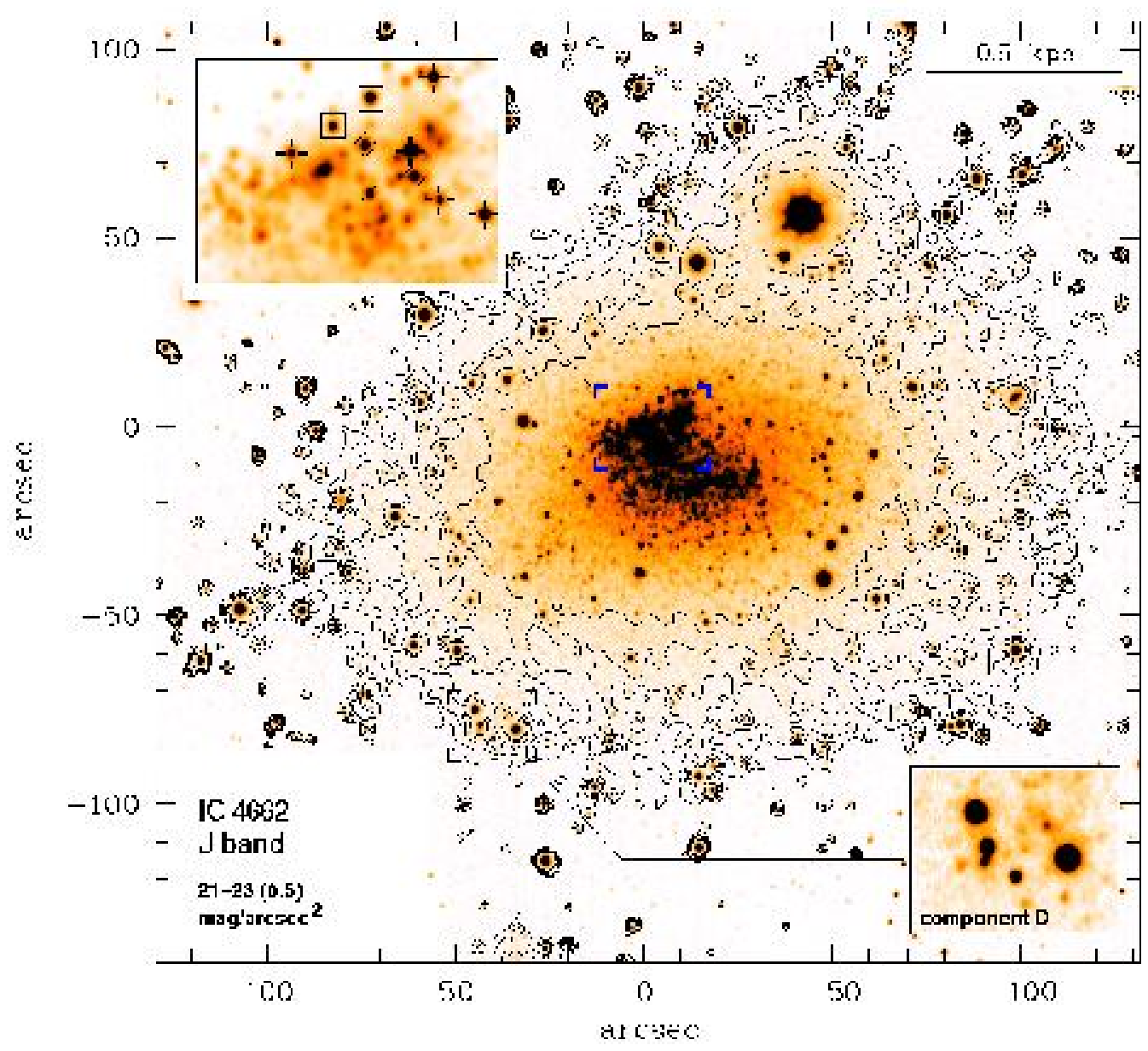,height=9.0cm,angle=0,clip=}}}
\put(10.8,12.04){{\psfig{figure=H4029F37.ps,width=7.1cm,angle=-90,clip=}}}
\put(10.84,8.19){{\psfig{figure=H4029F38.ps,width=7.13cm,angle=-90,clip=}}}
\put(0.17,0){{\psfig{figure=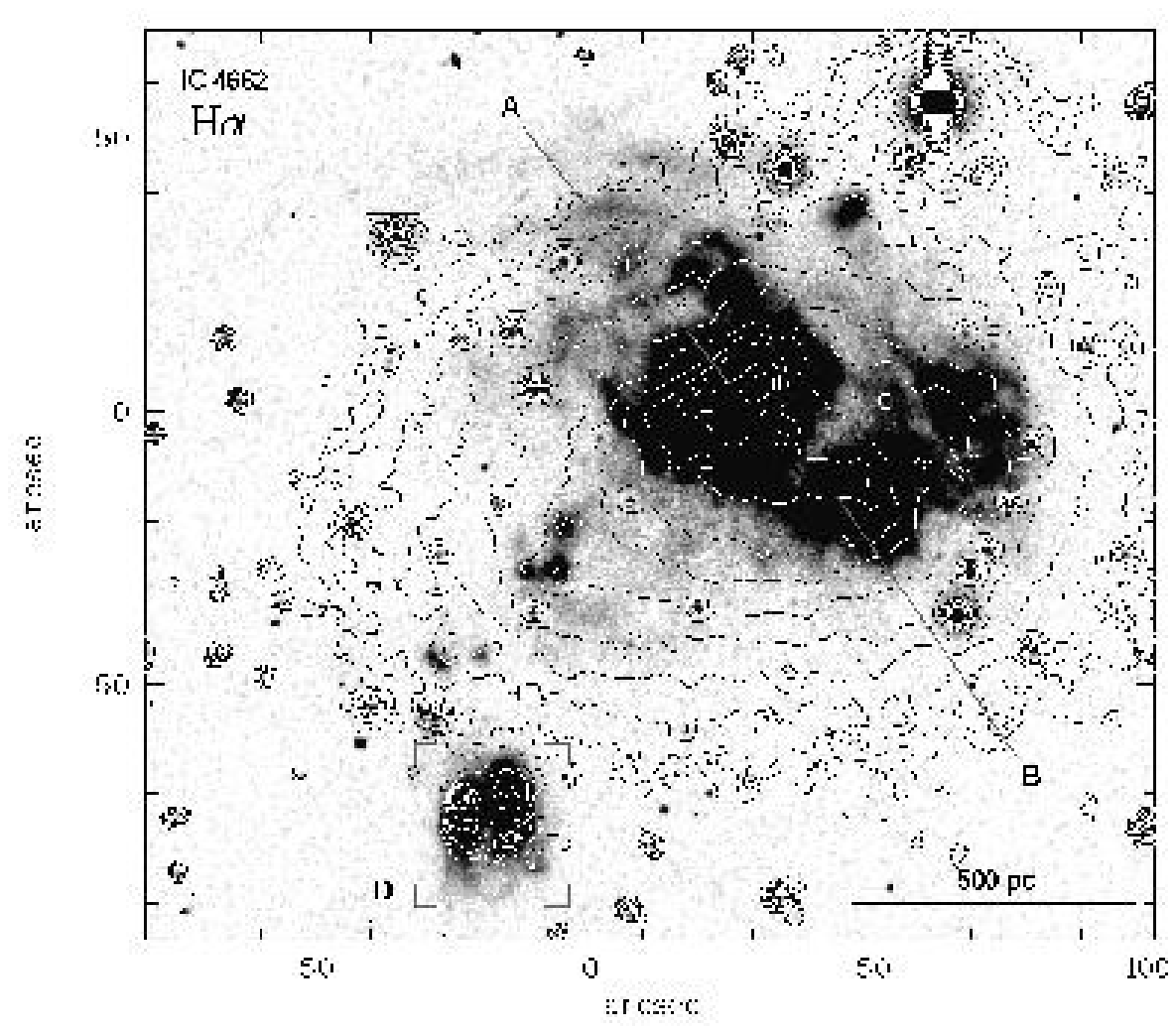,height=8.56cm,angle=0,clip=}}}
\put(2.8,10.75){\Large\sf (a)}
\put(1.55,7.0){\Large\sf (b)}
\put(11.7,12.3){\sf (c)}
\put(11.7,9.15){\sf (d)}
\PutWin{11.0}{3.0}{7.0cm}{\caption[]{\label{fic4662}
IC 4662 ($D$=2 Mpc). For explanations of symbols and labels, refer to
Fig. \ref{ftol3}.
{\bf (a):} 
$J$ band image and isophotes. The insets show magnified
views of the star-forming region A in the central part of the BCD
and of the D component, about 1\farcm 4 from A. Some bright sources in region
A are classified according to Schulte--Ladbeck et
al. (\cite{schulte01}) into red supergiants ($J-H\sim$0.75, crosses),
blue supergiants ($J-H<$0.5, squares), and AGB stars ($J-H>$0.9,
lozenges).  
{\bf (b):} 
Continuum-subtracted \ha\ map (from Papaderos
et al. \cite{papaderos03}), showing that all regions (A--D from
Heydari--Malayeri \cite{heydari90}) are \ha\ emitting.
{\bf (c),(d):} 
Surface brightness and color profiles. The thick grey
line shows a fit to the host galaxy using a modified exponential
distribution, Eq. (\ref{med}), with $b,q=4.8,0.97$ .}}
\end{picture}
\end{figure*}

This nearby ($D$=2 Mpc; Heydari--Malayeri et al. \cite{heydari90}) dwarf
galaxy has been studied spectroscopically by e.g. Pastoriza \& Dottori
(\cite{pastoriza81}), Stasinska et al. (\cite{stasinska86}),
Heydari-Malayeri et al. (\cite{heydari90}) and Hidalgo-G\'amez et
al. (\cite{hidalgo01}).  The latter authors derived the oxygen
abundance in the brighter SF regions A\&B to \zsun/6.5\dots \zsun/7.6,
respectively.  A continuum-subtracted \ha\ map (Fig. \ref{fic4662}b)
from Papaderos et al. (\cite{papaderos02}) reveals an extended and
complex morphology of the ionized gas emission in the upper half of
the galaxy, where the \ha\ emission peaks, and shows a number of
shells extending up to $\sim$0.5 kpc NE.  Another interesting feature
is the extranuclear region D, seen in the outer regions ($\mu \ga$ 22
J \sbb) of IC 4662.  Hidalgo-G\'amez et al. (\cite{hidalgo01}) find
this \ha\ emitting region to be less metal--rich than the central 
SF regions (A), and to show a significant recession velocity 
difference (250 $\pm$ 150 km/s, Hidalgo-G\'amez et
al. \cite{hidalgo02}) to IC 4662. Our deep images show that region D 
is not well detached from the main body of diffuse \ha\ emission in
IC4662 (contrary to Heydari--Malayeri et al. \cite{heydari90}), but
apparently connecting with regions A\&B through a chain of \ha\ sources.
 
Hidalgo-G\'amez et al. (\cite{hidalgo02}) have proposed that region D may be
either a chemically and kinematically distinct complex within IC 4662,
or a close companion object. The latter possibility is particularly
interesting in view of the hypothesis that very close, gas--rich 
dwarf companions might be conceivable triggering agents of
starburst activity in BCDs (Taylor et al. \cite{taylor95}, Pustilnik
et al. \cite{pustilnik01}, Noeske et al. \cite{noeske01}).

From the available data, we resolve a wealth of morphological
information in the central part (Fig. \ref{fic4662}a) of IC
4662, notably the head-tail morphology of the starburst. Luminosities
and colors of the brightest point sources in region A are in agreement
with those of red supergiants, supporting the results by
Heydari-Malayeri et al. (\cite{heydari90}), as well as of blue
supergiants and AGB stars (cf. the upper left inset in
Fig. \ref{fic4662}a; see Schulte--Ladbeck et
al. \cite {schulte01} for color limits separating the latter classes
of giant stars).

As IC 4662 is located at low galactic latitude (--17.8$\degr$),
photometric studies of its LSB component are complicated by the dense 
foreground Galactic stellar field.  Also residuals in the removal of the NW bright star
may affect the photometry for faint isophotal levels. 
A tentative exponential fit to the $J$ SBP
yields for \rr $>$65\arcsec\ ,
i.e. outside significant nebular emission, a scale length of
$\sim$0.21 kpc.  However, inspection of the $J_{\rm LSB}$ profile
shows that the BCD follows a \flat\ distribution, fitted best with
Eq. (\ref{med}) with a central surface brightness of 16.8 $J$ \sbb, a
scale length of $\sim$150 pc and a depression parameter as large as
$q\approx 0.97$. If so, the starburst contributes $\sim$50\% of the
total $J$ light of the BCD. IC 4662 was hitherto classified
as a dwarf irregular galaxy (cf. e.g. de Vaucouleurs et
al. \cite{devaucouleurs91}). However, the structural properties of its LSB
host, as well as its intense and spatially extended SF activity, place
IC 4662 in the range of BCDs, making it probably one of the closest BCDs
known.

\subsection{UM 488 (Mkn 1304, SBS 1139+006, UGC 6665)}  
\label{um448}
\begin{figure*}[!ht]
\begin{picture}(18,10)
\put(0,0.1){{\psfig{figure=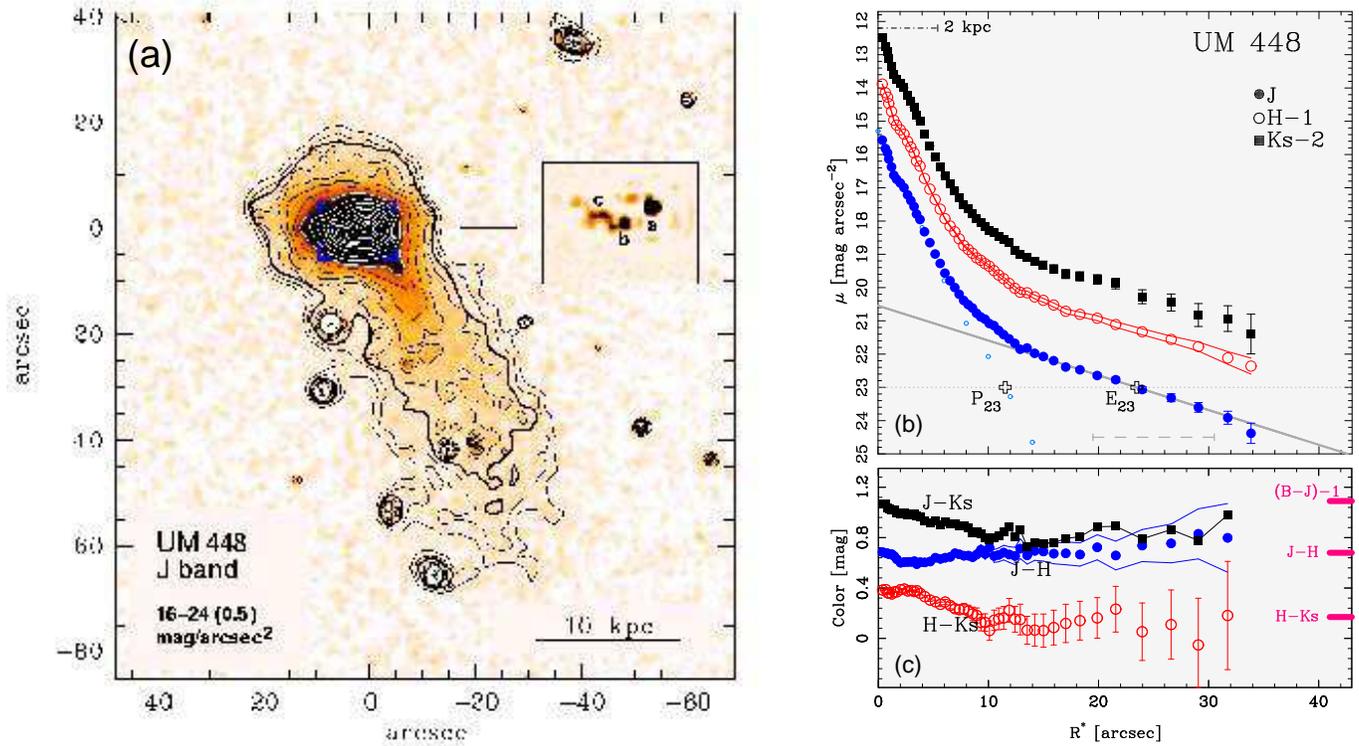,height=10cm,angle=0,clip=}}}
\put(10.9,3.93){{\psfig{figure=H4029F41.ps,width=7.0cm,angle=-90,clip=}}}
\put(10.95,0.23){{\psfig{figure=H4029F42.ps,width=7.02cm,angle=-90,clip=}}}
\put(1.6,9.2){\Large\sf (a)}
\put(11.8,4.35){\sf (b)}
\put(11.8,1.15){\sf (c)}
\end{picture}
\caption[]{UM 448 ($D$=76.1 Mpc). 
For explanations of symbols and labels, refer to Fig. \ref{ftol3}.
{\bf a):} $J$ band image and isophotes.  The inset shows a
contrast--enhanced close--up of the star--forming regions (delimited
by brackets in the main image); individual knots ({\sf a--c}) are
labeled.
{\bf b),c):} Surface brightness and color profiles.}
\label{fum448}
\end{figure*}

This is a distant ($D\approx$ 76 Mpc, Mirabel \& Sanders
\cite{mirabel88}) blue compact galaxy, known to be relatively
metal--deficient (\zsun/8.5; Masegosa et al. \cite{masegosa94}, Izotov
\& Thuan \cite{izotov99}).  Despite its resemblance 
(Fig. \ref{fum448}a) to {\sl cometary} iI BCDs, UM 448 does not
qualify as a dwarf due to its intrinsic luminosity ($M_B=$--19.7 mag)
and large linear extent (major axis length $\sim$25 kpc). Active star
formation is taking place within an extended (diameter $\sim$7 kpc)
high surface brightness region at the NE part of the galaxy, showing
some fainter SW extension.  The starburst component contributes
$\approx$70\% of the $J$ band light.  Guseva et al. (\cite{guseva00})
estimated the starburst luminosity to be powered by $1.6\times 10^6$ O
stars (value referring to the distance assumed here),
0.6\% of which are undergoing their Wolf-Rayet phase. 
Studies by Sage et al. (\cite{sage92}) indicate an ongoing SFR 
of 21 \msun\ yr$^{-1}$, and a large reservoir of molecular gas
amounting to $2.8\times 10^9$ \msun\ or $\sim$1/2 of the
\ion{H}{i} mass ($5.8\times 10^9$ \msun; Mirabel \& Sanders
\cite{mirabel88}) of UM 448. 

On unsharp-masked images, the SF region splits
into 3 high-surface brightness entities (see the inset in
Fig. \ref{fum448}a), the brightest one being the reddest
($J-H$\bg\ =0.7 mag) whereas the fainter regions {\sf b} and {\sf c} are 
slightly bluer ($J-H$\bg$\approx$0.5 mag).
The SBPs (Fig. \ref{fum448}b) show an exponential decay for
\rr$>$15\arcsec\ with a $J$ band scale length of 3.8 kpc, somewhat
smaller than the value $\sim$5.2 kpc (14\arcsec) inferred in Telles et
al. (\cite{telles97}).


\section{Discussion}
\subsection{The shape of NIR surface brightness profiles \label{dis1}}
The SBPs derived for the sample BCDs in Sect. \ref{sample_galaxies}
bear close resemblance to those typically inferred from optical 
broad-band data (cf., e.g., \cite{papaderos96a}, Telles et al. \cite{telles97}, Marlowe
et al. \cite{marlowe97}, Doublier et al. \cite{doublier97},\cite{doublier99}, Cair\'os et
al. \cite{cairos01a}). In most cases, SBPs show in their outer part an
exponential intensity decrease and red, nearly constant colors. This 
outermost SBP part is attributable to the underlying LSB host galaxy 
which, except for a few rare cases, shows clear evidence for a several 
Gyr old stellar population.  At intermediate and small radii the emission 
of a BCD is dominated by the younger stellar component. In the main class
of iE BCDs (\cite{loose86}) its luminosity output is reflected on two 
conspicuous SBP components.

1) At small radii (\rr$\la$100 pc), a feature commonly observed in
optical and NIR SBPs is a central intensity excess. This is seen in
e.g., the $J$ SBP of Tol 3, Mkn 178, Mkn 1329, Haro 14 (our sample) or
Mkn 36 and UM 462 (\cite{c02b}). This component
is due to the brightest and typically youngest stellar assembly. 
In most cases, this narrow innermost excess is due to one and the same region in 
both optical and NIR wavelengths.  An exception is presented by 
Mkn 178, where the brightest region in the NIR is offset from that 
in the optical by $\sim$150 pc, or by the BCD
Mkn 35 in the sample of \cite{c02b}.

2) The second SBP feature has been referred to as {\sl plateau} in
\cite{papaderos96a}. It shows a nearly constant or slowly decaying
intensity, and extends typically out to $\mu\sim$24 $B$ \sbb. 
SBPs with a prominent plateau on top a more extended exponential LSB envelope 
can in general not be adequately fit by a simple function (e.g. a
\ser profile) over their whole intensity span. 
A meaningful decomposition of such SBPs is only possible when 
an extra component (e.g., a \ser distribution with $\eta\leq 1$) 
is introduced in order to fit the plateau 
(\cite{papaderos96a}, Cair\'os et al. \cite{cairos01a}).
There is, so far, no observational support for the plateau being a
dynamically distinct or even interaction-induced stellar entity with
nearly constant ${\cal M/L}$.
Quite contrary, appreciable color gradients ($\la$1.5 $B-R$ \cgg) 
in the HSB part of BCDs indicate that the plateau light is 
mainly due to a young and moderately evolved stellar population.
\cite{papaderos96b} have shown that a conspicuous 
plateau in the SBPs of many iE BCDs can naturally result
from the superposition of diffuse and compact SF sources with varying luminosity and
galactocentric distance on a more extended, exponential LSB component. 
Conversely, the plateau is nearly absent in the intrinsically brighter
nE BCDs, where most of the starburst light originates from the nuclear
region of the BCD. The {\em overall} SBPs of these systems show frequently a concave
shape, fitted satisfactorily by Eq. (\ref{sersic}) with a \ser exponent
$\eta\geq$1, and sometimes resemble a de Vaucouleurs profile.

Discerning the formation history of the stellar populations memorized
in the plateau light poses a challenge for surface photometry and CMD
studies, as nebular line emission (e.g. Tol 65; Papaderos et
al. \cite{papaderos99}, Izotov et al. \cite{izotov01}, Tol 1214-277;
\cite{fricke01}) and patchy dust absorption (Tol 3, Mkn 178; this
paper, Mkn 33 and Mkn 35; \cite{c02b}, I Zw 18; Cannon et
al. \cite{cannon02}) may both hamper standard age-dating techniques.
In addition, an important and mostly overlooked source of systematic
uncertainties in the determination of colors or EWs within the plateau
stems from the unknown line-of-sight contribution of the underlying
old LSB component. As pointed out in Sect. \ref{apphot}, the latter
may redden colors of compact stellar clusters in the SF region by up
to a few tenths of magnitude. Corrections of this order have also been
reported for the iE BCDs Mkn 370 and Mkn 178 by Cair\'os et
al. (\cite{cairos02a}) and Papaderos et al. (\cite{papaderos02}),
respectively. 
Evidently, for such a correction one has to assume a model for 
the intensity distribution of the LSB component just beneath the 
SF regions (i.e., for \rr$\la$\p23). It is a common practice to 
extrapolate the exponential slope of the LSB periphery of BCDs 
all the way to \rr$=$0\arcsec. However, the universal validity 
of this procedure for BCDs has been questioned in, e.g., P96a. 
These authors discussed observational evidence for \flat\ profiles 
in BCDs and proposed the alternative fitting formula Eq. (\ref{med}) to 
approximate such convex distributions. Alternatively, \cite{c02b} discuss 
cases where the LSB host galaxy shows a concave profile which is better 
fit by a \ser model with an exponent $\eta>1$. 
The photometric structure of the LSB host of BCDs will be 
discussed in the light of the present NIR data in the next section.

\subsection{ Hints to a centrally flattening exponential distribution
in the underlying stellar LSB component? \label{dis2}}
%
\begin{figure}
\includegraphics[angle=270,width=8.4cm]{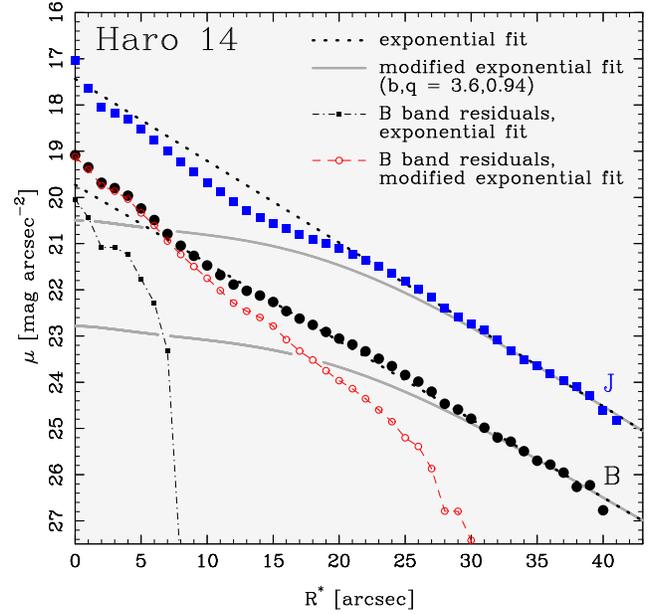}
\caption{Decomposition of the $B$ band SBP of Haro 14 (filled circles)
assuming for its LSB host a pure exponential model (dotted line) or a
modified exponential distribution (Eq. \ref{med}; thick gray curve) 
with the flattening parameters ($b$,$q$)=(3.6,0.9) derived in the $J$ band. The 
surface brightness distribution of the star-forming component, 
as obtained by subtraction of the respective LSB model from the 
$B$ SBP is illustrated by the filled and open interconnected symbols
(see discussion in the text).}
\label{decomp_opt_nir}
\end{figure}
Nine of the BCDs included in our sample (Table \ref{tab_phot}) show
signatures of a \flat\ profile in their LSB hosts.  This is because an
exponential law, or any \ser distribution with $\eta > 1$, fitted to
the outer part of their $J$ SBPs, predicts at intermediate to small
radii a higher intensity than the observed value (cf. Sect. \ref{decomposition}).  
Of course, pure exponential fits to $J_{\rm LSB}$ profiles cannot always be 
definitely ruled out within the 1$\sigma$ uncertainties. 
However, such fits would either overestimate the central intensity of the 
stellar LSB host, hence underestimate the luminosity fraction of the SF 
component (cf. Sect. \ref{dis2}) or systematically overestimate the
$J$ band exponential scale length, thus imply an implausibly large $B-J$ 
color gradient for the old underlying LSB population. 

The limited size of our present sample does hardly allow to estimate the
frequency of \flat\ profiles in BCDs.
This issue will be addressed in a forthcoming paper of this series, 
focussing on the complete NIR sample. However, the evidence gathered 
so far (see also Sect. \ref{decomposition}) strongly supports the idea 
that a  substantial fraction of BCDs shows \flat\ profiles in its LSB host galaxy.
The rare detection of \flat\ profiles in BCDs in previous optical studies 
can be explained by the fact that 
in those wavelengths extended starburst emission overshines the LSB host 
within typically its inner $\sim$2 exponential scale lengths 
(\cite{papaderos96b}, Noeske \cite{noeske99}, Cair\'os
\cite{cairos00}, Cair\'os et al. \cite{cairos01a}).  
As a result, a \flat\ profile in iE/nE BCDs could be detected only when 
a significant depression of the exponential LSB slope occurs  for \rr$>2\alpha$, 
i.e. if $b>2$ in Eq. (\ref{med}). 
NIR observations do not overcome, but alleviate the problem of severe light 
pollution by the burst, proving in many cases crucial for disentangling a pure 
exponential from a \flat\ distribution in the underlying LSB galaxy.

This is illustrated in Fig. \ref{decomp_opt_nir} on the example of the
iE BCD Haro 14. 
Its $B$ SBP is roughly exponential out to \rr$=$40\arcsec\ 
($\mu\sim$27 $B$ \sbb), suggesting that the exponential slope seen 
in the LSB periphery continues all the way to \rr=0\arcsec. 
In that case, one is tempted to conclude that the ongoing burst 
gives rise to a modest luminosity increase, only. 
Indeed, subtraction of the exponential LSB model from the
SBP suggests that SF sources (small interconnected squares in
Fig. \ref{decomp_opt_nir}) are all confined within \rr=8\arcsec\ and
that they account for no more than 7\% ($\geq$16.5 mag) of the $B$
light of the BCD.  Evidently, this conclusion is hardly compatible
with the extended morphology of the SF component and copious \ha\ 
emission of Haro 14 (cf. Marlowe et al. \cite{marlowe97}, Doublier et
al. \cite{doublier99}).

One arrives at a quite opposite conclusion on inspection of the
NIR profiles.
Figure \ref{decomp_opt_nir} shows that a pure exponential law falls
short of fitting the $J$ SBP inwards of \rr$<$20\arcsec; a plausible 
fit to the NIR data must invoke  a significant flattening of the
exponential slope inside $\ga$3 scale lengths of the LSB component.
An adequate fit to the $J_{\rm LSB}$ is best achieved 
with a \med\ (Eq. \ref{med}) with ($b$,$q$)=3.6,0.94.
Such a strongly flattened exponential LSB model yields
a stellar mass by a factor $\sim$3 smaller than that of a purely
exponential profile and implies that the starburst (small 
interconnected circles in Fig. \ref{decomp_opt_nir}) contributes
$\sim$60\% of the $B$ light within the Holmberg radius, i.e. more 
than a factor of eight larger than the value inferred before. 
The isophotal $B$ magnitude within 25 $B$ \sbb\ is then
increased by $\sim$1.1 mag due to the starburst. 
This value is still compatible to the mean value of 0.75 mag deduced 
for BCDs in \cite{papaderos96b} and Salzer \& Norton (\cite{salzer99}).

The considerations above show that the choice of the model
for the LSB emission may have far-reaching implications to our view about
the SF amplitude and photometric fading of BCDs. 
The usual assumption of a purely exponential LSB model 
may, in some cases, strongly underestimate the luminosity
fraction of the young stellar population, and lead to the conclusion
that a ``burst'' is merely a minor luminosity enhancement in the lifetime of a BCD. 
If, on the contrary, the LSB emission is assumed to follow a 
\flat\ profile, then the estimated luminosity of the superimposed young 
stellar population may increase by more than a magnitude. 

Consequently, different LSB models imply different
amounts of photometric fading of a BCD once SF activities have terminated. 
This has to be borne in mind when discussing evolutionary links 
between BCDs and other dwarf galaxies, as well as a possible
fading of luminous blue compact galaxies at medium redshift to 
local spheroidals (cf. e.g. Guzm\'an et al. \cite{guzman98}).
Note that the assumption of a \flat\ profile has no effect on 
the \emph{extrapolated} central surface brightness $\mu_{\rm E,0}$ and the
exponential scale length $\alpha$ of the LSB host galaxy of a BCD.
As a result it is not expected 
to significantly change the systematic difference between 
BCDs and other types of dwarf galaxies on the
$\mu_{\rm E,0}$--$M_{\rm LSB}$ and $\log(\alpha)$--$M_{\rm LSB}$ 
parameter space (\cite{papaderos96b}, Marlowe et al. \cite{marlowe97}, 
Salzer \& Norton \cite{salzer99}, Papaderos et al. \cite{papaderos02}).

A dedicated investigation of the frequency and origin of
systematic deviations from the exponential law 
(e.g. a \flat\ profile, or \ser profile 
with $\eta>1$) in the stellar LSB host of BCDs 
is apparently of great interest.
A first step towards such studies is to test the suitability of 
different empirical functions in approximating the observed LSB 
intensity distribution. 
This issue will be briefly discussed in the following section.
%
\subsection{The \ser\ law vs the \med\ for studies of \flat\ SBPs \label{dis3}}

\begin{figure}
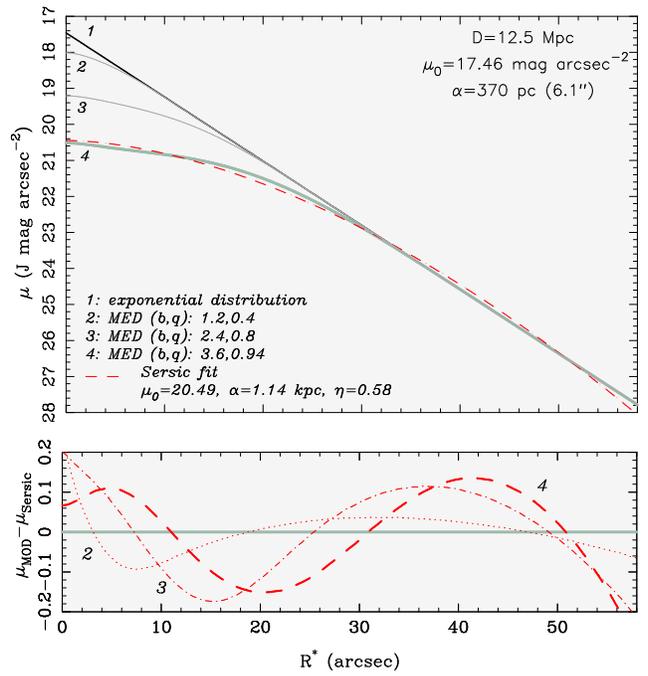

\begin{picture}(8.4,9)
\put(0,3.3){{\psfig{figure=H4029F44.ps,width=8.4cm,angle=-90.,clip=}}}
\put(0.07,0){{\psfig{figure=H4029F45.ps,width=8.32cm,angle=-90.,clip=}}}
\end{picture}
\caption[]{{\bf upper panel:} Fit to the LSB component of Haro 14
(thick gray curve, labelled {\it 4}) assuming an inwards flattened
exponential distribution (\med, Eq. \ref{med}) with the
{\sl extrapolated} central surface brightness $\mu_{\rm E,0}$ and 
exponential scale length $\alpha$ indicated at the upper-right, 
and flattening parameters ($b$,$q$)=(3.6,0.94). 
Distributions {\it 2} and {\it 3} correspond to a \med\ with equal 
$\mu_{\rm E,0}$ and $\alpha$ but different ($b$,$q$), and {\sl 1} describes 
a pure exponential distribution ($q=0$). 
The dashed curve shows a fit to distribution {\sl 4} using a 
\ser fitting law (Eq. \ref{sersic}).
{\bf lower panel:} Residuals between the assumed \med\ distributions
{\sl 2} through {\sl 4} (upper panel) and \ser fits.}
\label{comp1}
\end{figure}

As noted in Sect. \ref{decomposition}, also the \ser\ law, commonly
applied to structurally analyze various extragalactic systems, can
approximate \flat\ SBPs.  In this section, we further discuss why in
this work preference was given to the \med\ model to fit and
quantitatively study such SBPs.

Both the \ser\ law and the \med\ are fitting functions of empirical
origin (Sect. \ref{decomposition}). The choice of either model does
therefore not imply any assumptions on the physical background of
\flat\ SBPs, such as the dynamics or stellar mass distribution of a galaxy.
However, although either function was found to satisfactorily
approximate the \flat\ SBPs found in this work, they principally
differ with respect to their shape.
This difference is illustrated in the comparison shown in
Fig. \ref{comp1}, where we fit for illustrative purposes a \ser
profile (thin dashed line) to the \med\ model (heavy line, labelled
{\sl 4}) obtained for the LSB component of Haro 14 in
Sect. \ref{sample_galaxies}.  Distributions {\sl 2} and {\sl 3}
illustrate \med\ profiles with equal {\sl extrapolated} central
surface brightness $\mu_{\rm E,0}$ (17.46 $J$ \sbb) and exponential
scale length $\alpha$ (0.37 kpc) as {\sl 4}, but differing degrees of
flattening ($q$) with respect to the pure exponential ($q=0$) model
{\sl 1}.  

The \med\ {\sl 4} compares best to a \ser profile with a central
surface brightness $\mu_{\rm S,0}=20.45$ $J$ \sbb, a scale length
$\beta\approx$1.14 kpc and an exponent $\eta=0.57$.  Note that the
deduced $\mu_{\rm S,0}$ compares well to the {\sl actual} central
surface brightness of the LSB host galaxy
\begin{equation}
\mu_{\rm E,0}+2.5\log\left(\frac{1}{1-q}\right)\approx 20.5\,\, {\rm
mag/}\sq\arcsec,
\end{equation}
implied by the \med\ model.

It is evident that the \ser\ law and the \med\ distribution in
Fig. \ref{comp1} closely follow each other over a radius range of
nearly 10 scale lengths $\alpha$.  Nevertheless, the lower panel of
Fig. \ref{comp1} shows that the residuals between either model are
small ($\la$0.2 mag), but systematic.
This difference reflects the fact that a \ser\ law with $\eta <1$
differs from an exponential at all radii, while the \med\
approaches an exponential slope for $R > b \alpha$.

We recall that both a \ser\ and a \med\ model were found to
approximate well the inner portion of a \med\ profile (e.g. within
$\sim$1.5 cutoff-radii $b \alpha $). The question which distribution
gives a more adequate description of the observed \flat\ SBPs can
hence only be assessed through studies of the outer parts of such
profiles. If the \flat\ SBPs show a perfectly exponential fall-off in
their outer parts, continuing several scale lengths beyond a cutoff
radius $b\alpha$ (cf. Eq. \ref{med}), a \ser\ law will fail to correctly
reproduce these light distributions, and no \ser model fit to the LSB
emission can correctly recover the exponential scale length $\alpha$
and will not be free of ambiguity (Sect. \ref{sersic_drawbacks}).
Alternatively, if subsequent studies show that the LSB profiles
deviate systematically from the exponential law at all radii, then a
\ser profile might be the preferable fitting law, albeit the drawbacks
discussed in the following Sect. \ref{sersic_drawbacks}.

Such considerations call for deeper photometric studies of the LSB
component of BCDs, and in dwarf galaxies in general.  An ultimate
check requires deep surface photometry, allowing to derive SBPs out to
$\sim 10\alpha$ with an accuracy better than 0.2 mag (see the
residuals in Fig. \ref{comp1}), not reached by currently available
data.

As long as tests of this kind await to be done, we argue, in agreement
with \cite{c02b}, that a pure (Eq. \ref{exponential}) or modified
exponential (Eq. \ref{med}) fitting formula should be preferably used
to fit the underlying LSB component of BCDs. 
Previous optical studies (e.g. \cite{loose86},  \cite{papaderos96a}, Telles et al.
\cite{telles97}, Marlowe et al.  \cite{marlowe97},  Salzer \& Norton \cite{salzer99}, 
Vennik et al. \cite{vennik00}, Cair\'os et al. \cite{cairos01a}, 
\cite{cairos01b}, Makarova et al. \cite{makarova02})
of BCDs suggest that those systems mostly show no strong deviations
from an exponential law over their outer LSB SBPs. Also if small
deviations from an exponential law which are hardly detectable in
current data should exist, a \med\ gives a robust approximation by
fixing the LSB profile shape.  While a \ser\ law at first view offers
a greater flexibility in this profile region, its application in the
extremely low $S/N$ regime to quantify marginal deviations from an
exponential is problematic, as discussed in the following section.

\subsubsection{A cautionary note on structural analyses of BCDs using the \ser\ law}
\label{sersic_drawbacks}

A drawback of \ser models when fitted to \flat\ profiles is that
the solution can vary considerably, depending on the fitted radius
range.
For instance, a \ser fit to the \med\ {\sl 4} (Fig. \ref{comp1}) for
$<$23.5 $J$ \sbb, or within 5$\alpha$ ($\sim$30\arcsec) yields an
$\eta=0.45$, significantly lower than the value above.  Note a \ser
exponent $\eta<$0.5 reflects a central minimum in the radial
luminosity density distribution of the LSB component (P96a).  The
stability of the shape parameter $\eta$ has been explored in detail in
\cite{c02b}; these authors remarked that a \ser fit can be very
uncertain for concave SBPs, where the exponent $\eta$ can vary by up
to an order of magnitude, depending on the fitted range and the radial
sampling of points in the profile.

Another objection to the view that \ser models offer a robust tool to
systematize the structural properties of dwarfs comes from the
degeneracy of the shape parameter $\eta$ vs. the pseudo-scale length
$\beta$. A fit of Eq. (\ref{sersic}) to the \med\ distribution {\sl 2}
yields an $\eta$=0.94 and a $\beta\approx$0.44 kpc ($>\alpha$).  As
for profile {\sl 3}, it is best approximated with an $\eta=0.75$ and
$\beta\approx 0.72$ kpc ($2\alpha $).  We see that while the \ser
exponent $\eta$ decreases with increasing degree of flattening,
i.e. as we go from distribution {\sl 2} towards {\sl 4}, the
corresponding scale length $\beta$ increases from 0.44 kpc to 1.15
kpc. Actually, the exponential scale length $\alpha$ of the \med\
profiles considered here is nowhere recovered by fitting a \ser
model. Instead, one is left with the pseudo-scale length $\beta$,
which by itself alone carries no quantitative information on the
structural properties of the LSB component.
 
Obviously, the scale length $\beta$ is only meaningful in connection
with $\eta$. However, these two quantities are rendered impractical
for a systematic study of the LSB component by their strong non-linear
coupling (see discussion in e.g. Young \& Currie \cite{young94},
Cellone \& Buzzoni \cite{cellone01}), and dependence on the fitting
procedure (see above).  Furthermore, observational uncertainties
connected with, e.g., the sky subtraction (see e.g. Cellone \& Buzzoni
\cite{cellone01}, \cite{c02b}), filtering of images prior to surface
photometry, and imperfect removal of background sources may also skew
the $\beta$ vs. $\eta$ parameters in a hardly predictable manner,
making \ser fits to the outer part of SBPs a hazardous procedure
(\cite{c02b}).  These problems are probably not worrisome in studies
of early type galaxies, i.e. systems with little morphological
distortions and a nearly constant ${\cal M/L}$, where also the
central, high $S/N$ regions are accessible to constrain the global
photometric structure.  In irregular SF galaxies, however, the derived
\ser\ parameters depend on subjective choices in the data processing
(see above), profile fitting and, equally important, on the profile
extraction methods themselves. This is particularly important in the
case of BCDs, where mostly only the outer, low $S/N$ part of the LSB
host is accessible to structural studies.

It is conceivable that the latter circumstances imprint the \ser
parameters ($\mu_{\rm S,0}$,$\beta$,$\eta$) inferred recently for a
sample of luminous BCGs by Bergvall \& \"Ostlin (\cite{bergvall02}).
We believe that the extremely large $\eta_{\rm LSB}$ exponents they
derive (up to 20), in conjunction with central surface brightnesses
$\ll 0$\,\sbu , should be regarded formal only, rather than a
manifestation of an extraordinarily dense Dark Matter halo which, as
advocated by these authors, dominates the stellar dynamics of a BCD.

\subsection{The colors of the stellar LSB host galaxy \label{dis4}}
\begin{figure*}[!ht]
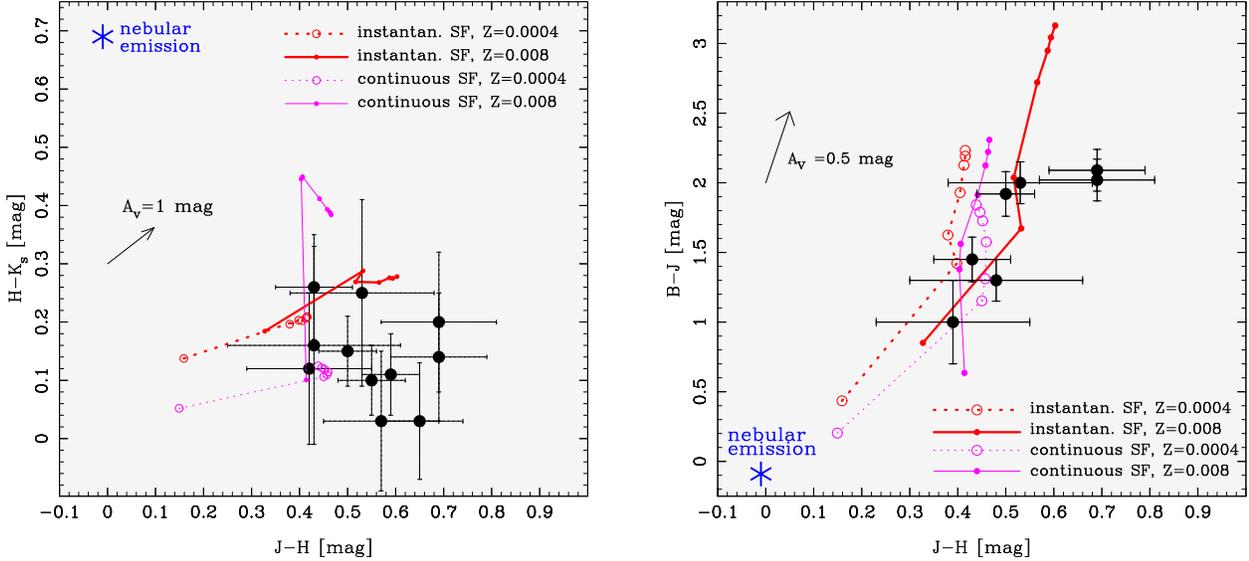

\centerline{\includegraphics[angle=270,width=7.75cm,clip=]{H4029F46.ps}\hspace{1cm}\includegraphics[angle=270,width=7.75cm,clip=]{H4029F47.ps}}
\caption{Colors of the stellar LSB host galaxies (filled circles) as
listed in Table \ref{tab_lsbcolors}.  For comparison, we show the
color evolution of synthetic stellar populations, calculated with the
GALEV evolutionary synthesis model (Schulz et al. \cite{schulz02};
Anders, Bicker \& Fritze--v.Alvensleben 2002, priv. comm.).  
An initial mass function with a Salpeter slope and respective lower 
and upper stellar mass cutoffs of 0.08 and 100 M$_{\odot}$ was assumed. 
We show single--age stellar populations, formed in an instantaneous 
burst, as well as in a continuous process with a constant SFR, both 
calculated for metallicities of \zsun/50 and \zsun/2.5.
For clarity, only population ages of 0.1, 0.5, 1, 3, 7, 10 and 14 Gyr 
are shown. The youngest (0.1 Gyr) point of each evolutionary track 
corresponds to the bluest $H-K_s$ and $B-J$ color.
All NIR colors shown are transformed to the 2MASS photometric system
(see Sect. \ref{transformations}), while the $B$ band refers to the
Johnson system.  Note that the error bars indicate cumulative
uncertainties due to all major error sources (see
Sect. \ref{lsbcolors}).  The colors of ionized gas emission from a
metal--deficient (\zsun/20) \ion{H}{ii} region ($T_e=10^4$ K,
$n_e=$100\,cm$^{-3}$, Case B recombination; cf. Kr\"uger
\cite{krueger92}) are indicated by the asterisk.  Arrows show
reddening vectors.}
\label{fig_lsbcolors}
\end{figure*}
%
The 
available NIR data allowed not only for the
detection and structural analysis of the LSB component for all 
sample BCDs, but also for the determination of at least one 
NIR color in the LSB host (Sect. \ref{lsbcolors}). 

The $B-J$ vs. $J-H$ and $H-K_s$ vs. $J-H$ LSB colors (Table
\ref{tab_lsbcolors}), whenever available, are displayed in
Fig. \ref{fig_lsbcolors}.  For comparison, color predictions from the
GALEV evolutionary synthesis model (Schulz et al. \cite{schulz02};
Anders, Bicker \& Fritze -- v. Alvensleben 2002, priv. comm.) are
overlaid in both panels, calculated for stellar populations with 
metallicities of \zsun/50 and \zsun/2.5, formed either in an
instantaneous burst or continuously with a constant SFR.
As evident from Fig. \ref{fig_lsbcolors}, NIR colors ($J-H$
vs. $H-K_s$) show little evolution for old stellar populations
($\la$0.1 mag for an age $>$0.5 Gyr). However, optical--NIR colors,
such as, e.g., $B-J$ (right panel of Fig. \ref{fig_lsbcolors}), change
by up to $\sim$1.5 mag for ages $>$0.5 Gyr. Within the above discussed
errors, such colors will allow to constrain the age of the stellar LSB
host of old iE/nE BCDs with a precision of 1--2 Gyr, once
precise spectroscopic measurements of its metallicity become available
(see the discussion of Gil de Paz et al. \cite{gildepaz00b}).
Fig. \ref{fig_lsbcolors} shows that the observed colors are generally
compatible with the GALEV predictions within their 1$\sigma$
uncertainties. This is also the case for model predictions from 
PEGASE (Fioc \& Rocca--Volmerange \cite{fioc97}).

Five BCDs in our sample, Pox4, Pox4B, Tol 1400-411, Tol 65 and
Tol 1214-277 (Table \ref{tab_lsbcolors}), show blue ($J-H<$0.5,
$B-J<$1.5) colors in their LSB component. This may be partly due to an
appreciable contribution of extended nebular emission, as suggested by
previous optical spectrophotometric work. The blue optical--NIR LSB
colors of Tol 65 and Tol 1214-277 may be attributed to a comparatively
young ($\la$1 Gyr) photometrically dominant stellar population
(cf. Papaderos et al \cite{papaderos99}, \cite{fricke01}).  The
optical--NIR colors of the LSB host of the remaining sample BCDs can
be reconciled with an old stellar population of subsolar metallicity.
This result is in line with most published optical--NIR photometry on
stellar hosts of BCDs (Gil de Paz et al. \cite{gildepaz00b}, Vanzi et
al. \cite{vanzi02}, Papaderos et al. \cite{papaderos02}), which
indicates a comparable range of colors for the LSB population of BCDs.

We can compare the NIR colors of the LSB component with data from the 
literature for three sample BCDs, only (Tol 3, Haro 14 and UM 461).
The colors of Tol 3 compare well with those reported by Vanzi et al. (\cite{vanzi02}) 
over the whole radius range of (\rr=0\arcsec\dots 50 \arcsec). 
Our photometry for Tol 3 and Haro 14 is also in agreement with
Doublier et al. (\cite{doublier01}) for small radii 
(for \rr$\la$15\arcsec\ and \rr$\la$10\arcsec, respectively).
For larger radii, however, the NIR colors by Doublier et al. 
are not compatible with our results within the 1$\sigma$ uncertainties.
Outside the star-forming component the color profiles by Doublier et al. approach
values of $J-H>2$ for Tol 3, and $J-H\approx 1.0 \dots >2.5$,
$H-K\approx 0 \dots -0.6$ for Haro 14. For UM 461, their color profiles
vary between $\approx 1.1 \dots 2.3$ ($J-H$) and $\approx -0.6 \dots
0.5$ ($H-K$). 
\section{Summary and Conclusions}

We have analyzed deep Near-Infrared (NIR) $J$, $H$, $K$ images of 12
Blue Compact Dwarf (BCD) Galaxies and one luminous Blue Compact
Galaxy.  These objects, together with those studied in an accompanying
paper (Cair\'os et al. 2003), constitute the first part of a sample
of 40 BCDs for which deep NIR images were obtained in the framework of
a large-scale multi-wavelength study.  The limiting surface
brightnesses of our data, $\sim$ 23.5 to 25.5 \sbb\ in $J$ and
$\sim$22 to 24 \sbb\ in $K$, allow for the detection and study of the
NIR structural properties and colors of the underlying stellar
low-surface brightness (LSB) component in all sample galaxies. This
evolved LSB host, underlying the star-forming regions is known to
exist in the majority of BCDs and to contain the bulk of the stellar
mass in these systems.
Consequently, a systematic determination of its structural properties 
(e.g., radial stellar surface density distribution, central 
surface brightness, exponential scale length) and of the 
gravitational potential it forms may prove crucial for the 
understanding of the starburst activity, dynamics and evolution 
of BCDs. Other than in optical wavelengths, where extended 
starburst emission hides the LSB host inside its inner 2--3 
exponential scale lengths, NIR studies allow to extend surface 
photometry of the underlying old stellar background to a smaller 
galactocentric distance, thereby better constrain its overall 
intensity distribution.
Our results can be summarized as follows:
\begin{enumerate}
\item Surface brightness profiles (SBPs) of our sample BCDs can 
at large galactocentric radii \rr\ be well approximated by an 
exponential fitting law, in agreement with previous observational
evidence based on deep optical surface photometry. Also, the 
exponential scale lengths derived in the optical and NIR spectral 
domain are in mutual agreement, implying minor optical--NIR color 
gradients in the LSB component.
\item
The LSB host galaxy of several sample BCDs shows on a galactocentric
radius comparable to the size of star-forming component (within its inner 
1--3 exponential scale lengths) a conspicuous intensity depression 
with respect to the purely exponential slope, observed for larger radii.
This type of an inwards flattening exponential SBP, classified \flat\ in 
Binggeli \& Cameron (\cite{binggeli91}), has been frequently derived 
in dwarf irregular and dwarf elliptical galaxies.
The rare detection of \flat\ profiles in BCDs from previous optical studies 
can be attributed to the fact that in those wavelengths extended 
starburst emission overshines the underlying LSB population within 
typically its inner $\sim$2 exponential scale lengths. 
A possible high frequency of \flat\ profiles among dwarf galaxies 
will not have a notable effect on the 
structural difference between BCDs and dwarf irregulars
on the parameter space defined by the central surface brightness, exponential
scale length and absolute magnitude of their LSB host galaxy.
However, it may have important implications for our view about BCDs, 
as it would significantly increase the estimated starburst-to-LSB 
luminosity fraction, and therefore the amount of 
photometric fading of these systems, once the starburst activity has 
terminated. This information is crucial for, e.g., establishing or 
discarding the hypothesis of faint dwarf spheroidals being the 
evolutionary endpoints of BCDs.
In the same way, a \flat\ intensity distribution would impose 
new observational constraints to the derivation of the total 
stellar mass and its radial density distribution within the 
underlying LSB host galaxy of BCDs.
\item
The exact shape and physical origin of \flat\ SBPs in 
dwarf galaxies are still to be investigated.
We find that such SBPs can be well approximated
by a modified exponential fitting formula proposed in Papaderos 
et al. (1996a). Alternatively, 
a S\'ersic law can also yield good fits to 
\flat\ profiles, albeit small systematic residuals. 
However, the practical applicability of the \ser\ law 
to the LSB emission of BCDs is limited by the strong 
non-linear coupling of its free parameters, and the
extreme sensitivity of the achieved solutions to, 
e.g., small uncertainties in the sky subtraction and
SBP derivation.
\item For the majority of the LSB host galaxies in our sample 
we derive optical-NIR colors indicative of an evolved stellar 
population of subsolar metallicity. 
The metal-deficient BCDs Tol 65 and Tol 1214-277 show blue colors 
($B-J<$1.3 mag) in their underlying stellar LSB component.
\item 
NIR images, in combination with optical data reveal for some 
BCDs signatures of appreciable and non-uniform dust absorption 
on a spatial scale as large as $\sim$1 kpc.
The NIR images, being less affected by dust extinction and widespread
ionized gas emission than optical data, allow for the detection of a 
variety of morphological features within the star-forming component, 
as e.g. coherent assemblies of compact stellar clusters, that might 
be remnants from previous episodes of collective star formation. 
An elaborate multiwavelength investigation of such sources
holds the promise of a better understanding of the history and
spatial progression of star-forming activity in BCDs.
\end{enumerate}
%


\begin{acknowledgements}
Research by K.G.N. has been supported by the Deutsche
Forschungsgemeinschaft (DFG) Grants FR325/50-1 and FR325/50-2. LMC
acknowledges support from the European Community Marie Curie Grant
HPMF--CT--2000--00774.  We thank U. Fritze -- v. Alvensleben,
P. Anders, J. Bicker and J. Schulz for kindly providing the GALEV
models.  This research has made use of the NASA/IPAC Extragalactic
Database (NED) which is operated by the Jet Propulsion Laboratory,
CALTECH, under contract with the National Aeronautic and Space
Administration.  This publication makes use of data products from the
Two Micron All Sky Survey, which is a joint project of the University
of Massachusetts and the Infrared Processing and Analysis
Center/California Institute of Technology, funded by the National
Aeronautics and Space Administration and the National Science
Foundation.
\end{acknowledgements}


\end{document}